\documentclass[]{svjour3}                
\smartqed  
\usepackage{graphicx}
\usepackage{mathptmx}

\journalname{The Astronomy \& Astrophysics Review}

\def\lesssim{\mathrel{\hbox{\rlap{\hbox{\lower4pt\hbox{$\sim$}}}\hbox{$<$}}}}
\def\gtrsim{\mathrel{\hbox{\rlap{\hbox{\lower4pt\hbox{$\sim$}}}\hbox{$>$}}}}

\def\pg{\mbox{PG~0122+200}}
\def\pp{\mbox{PG~1159$-$035}}
\def\pr{\mbox{PG~2131+066}}
\def\pt{\mbox{PG~1707+427}}
\def\rxj{\mbox{RX~J2117.1+3412}}
\def\v4334{\mbox{V4334 Sgr}}
\def\ngc{\mbox{NGC 1501}}
\def\vv{\mbox{VV 47}}

\begin{document}

\title{Evolutionary and pulsational properties of white dwarf stars}

\titlerunning{Evolutionary and pulsational properties of white dwarfs}

\author{Leandro G. Althaus \and
        Alejandro H. C\'orsico \and
        Jordi Isern \and
        Enrique Garc\'\i a--Berro}

\institute{Leandro G. Althaus and Alejandro H. C\'orsico \at 
           Facultad de Ciencias Astron\'omicas y Geof\'{\i}sicas,
           Universidad Nacional de La Plata,
           Paseo del  Bosque s/n,
           (1900) La Plata, Argentina\\
           Instituto de Astrof\'{\i}sica La Plata, IALP, CONICET-UNLP\\
           \email{althaus,acorsico@fcaglp.unlp.edu.ar}
           \and
           J. Isern \at 
           Institut de Ci\`encies de l'Espai, CSIC,
           Campus de la UAB,  
           Facultat de Ci\`encies, Torre C-5, 
           08193 Bellaterra, Spain\\
           Institut d'Estudis Espacials de Catalunya,
           c/Gran Capit\`a 2--4, Edif. Nexus 104,   
           08034 Barcelona, Spain\\
           \email{isern@ieec.fcr.es}
           \and
           E. Garc\'\i  a--Berro \at
           Departament de F\'\i sica Aplicada, 
           Universitat Polit\`ecnica de Catalunya, 
           c/Esteve Terrades 5, 
           08860 Castelldefels, Spain\\
           Institut d'Estudis Espacials de Catalunya,
           c/Gran Capit\`a 2--4, Edif. Nexus 104,   
           08034 Barcelona, Spain\\
           \email{garcia@fa.upc.edu}}

\date{Received: \today / Accepted: \today}

\maketitle

\begin{abstract}
White  dwarf  stars are  the  final  evolutionary  stage of  the  vast
majority of stars, including our  Sun.  Since the coolest white dwarfs
are very old objects, the  present population of white dwarfs contains
a wealth of information on the evolution of stars from birth to death,
and  on  the  star  formation  rate  throughout  the  history  of  our
Galaxy. Thus, the study of  white dwarfs has potential applications to
different fields  of astrophysics. In particular, white  dwarfs can be
used  as independent  reliable  cosmic clocks,  and  can also  provide
valuable  information  about  the  fundamental parameters  of  a  wide
variety of stellar populations, like  our Galaxy and open and globular
clusters.    In  addition,   the  high   densities   and  temperatures
characterizing  white  dwarfs  allow  to  use these  stars  as  cosmic
laboratories for studying  physical processes under extreme conditions
that  cannot be achieved  in terrestrial  laboratories.  Last  but not
least, since many white dwarf stars undergo pulsational instabilities,
the  study  of  their  properties  constitutes  a  powerful  tool  for
applications beyond stellar astrophysics.  In particular, white dwarfs
can  be  used  to   constrain  fundamental  properties  of  elementary
particles such as axions and  neutrinos, and to study problems related
to the variation of fundamental constants.

These potential  applications of  white dwarfs have  led to  a renewed
interest  in  the  calculation   of  very  detailed  evolutionary  and
pulsational  models for  these stars.   In  this work,  we review  the
essentials  of the  physics of  white dwarf  stars.  We  enumerate the
reasons that  make these stars excellent chronometers  and we describe
why white  dwarfs provide  tools for a  wide variety  of applications.
Special emphasis is placed on  the physical processes that lead to the
formation of white  dwarfs as well as on  the different energy sources
and processes  responsible for  chemical abundance changes  that occur
along their evolution.  Moreover, in  the course of their lives, white
dwarfs cross different  pulsational instability strips.  The existence
of  these  instability  strips  provides astronomers  with  an  unique
opportunity to peer into their internal structure that would otherwise
remain  hidden from  observers.   We  will show  that  this allows  to
measure with  unprecedented precision the stellar masses  and to infer
their envelope thicknesses, to probe the core chemical stratification,
and to  detect rotation rates  and magnetic fields.   Consequently, in
this work, we  also review the pulsational properties  of white dwarfs
and the most recent applications of white dwarf asteroseismology.

\end{abstract}


\section{Introduction}
\label{intro}

White  dwarf  stars are  the  final  evolutionary  stage of  the  vast
majority of stars.  Indeed, more than 97\% of all stars, including our
Sun, are expected to ultimately end their lives passively, getting rid
of their outer layers and  forming white dwarfs.  For this reason, the
present population of white dwarfs contains valuable information about
the evolution  of individual stars from  birth to death  and about the
star  formation rate  throughout  the history  of  our Galaxy.   These
stellar remnants are the  cores of low- and intermediate-mass hydrogen
burning  stars, and  have no  appreciable sources  of  nuclear energy.
Hence, as  time passes by, white  dwarfs will slowly  cool and radiate
the stored thermal energy, becoming dimmer and dimmer.

The interest  in the study  of white dwarf  stars in recent  years has
been  triggered in  part by  their  importance relative  to some  open
questions in other  fields of astrophysics --- see  the recent reviews
of Winget \& Kepler (2008),  Fontaine \& Brassard (2008), and also the
valuable early  reviews on the physics and  evolutionary properties of
white dwarfs of Koester \& Chanmugam (1990) and D'Antona \& Mazzitelli
(1990).  The  huge amount of high-quality observational  data of these
stars,   obtained  from  both   space-borne  observatories   and  from
ground-based  telescopes, has  made possible  to use  the  white dwarf
cooling sequence as independent age and distance indicators for a wide
variety  of stellar  populations.  As  a matter  of fact,  an accurate
knowledge  of  the rate  at  which  white  dwarfs cool  constitutes  a
fundamental  issue  and  provides   an  independent  cosmic  clock  to
constrain  the age and  previous history  of the  Galactic populations
including  the disk  (Winget et  al.  1987;  Garc\'\i a--Berro  et al.
1988a,b; Hernanz et  al.  1994; Torres et al.   2002) and globular and
open clusters (Richer et al.  1997; Von Hippel \& Gilmore 2000; Hansen
et al.  2002; Von Hippel et  al.  2006; Hansen et al.  2007; Winget et
al.  2009; Garc\'\i a--Berro et al. 2010).  In addition, because white
dwarf  progenitors lose  their  outer layers  ---  which are  carbon-,
nitrogen-  and oxygen-rich  --- at  the  top of  the asymptotic  giant
branch  (AGB),  they  are  significant contributors  to  the  chemical
evolution   of  the  Galaxy.    More  than   10,000  spectroscopically
identified  white   dwarfs  with  determined   effective  temperatures
($T_{\rm eff}$) and gravities  ($\log g$) have been currently detected
(Kleinman  et  al.   2004;  Kepler   et  al.   2007),  giving  us  the
opportunity  to  explore  the  white dwarf  mass  distribution,  which
ultimately provides  insights into mass-loss  processes during stellar
evolution and the mass budget of the Galaxy.

White  dwarfs usually play  a key  role in  some types  of interacting
binaries, such  as cataclysmic variables,  where mass transfer  from a
companion  star  onto  the  white  dwarf gives  rise  to  very  strong
energetic events.  In addition, white dwarfs in binary systems are the
candidate progenitors of  Type Ia supernovae, which are  thought to be
the  result  of   the  explosion  of  a  white   dwarf  exceeding  the
Chandrasekhar  limit due  to  accretion from  a mass-losing  companion
star.  Understanding  the evolutionary  and  structural properties  of
white dwarfs  helps improving our konwledge of  supernova events, with
the  important  underlying  implications  for  cosmology.   Also,  the
presence  of white dwarfs  in binary  systems with  millisecond pulsar
companions  allows to  constrain the  age of  millisecond  pulsars and
thereby  the  timescale for  magnetic  field  decay  (van Kerkwijk  et
al. 2005).

Surveys carried out  by the MACHO team (Alcock et  al. 1995; Alcock et
al. 1997; Alcock  et al. 2000) suggest that  a substantial fraction of
the halo dark  matter could be in the form of  very cool white dwarfs.
Since  then, the EROS  (Lasserre et  al.  2000;  Goldman et  al. 2002;
Tisserand et al. 2006), OGLE (Udalski et al. 1994), MOA (Muraki et al.
1999)  and  SuperMACHO (Becker  et  al.   2005)  teams have  monitored
millions of  stars during several  years in both the  Large Magellanic
Cloud  (LMC)  and the  Small  Magellanic  Cloud  (SMC) to  search  for
microlensing events.  Most of them  have challenged the results of the
MACHO  experiment  --- see,  for  instance,  Yoo  et al.   (2004)  and
references therein.   In addition, there  have been claims  that white
dwarfs could be the stellar  objects reported in the Hubble Deep Field
(Ibata et al. 1999; M\'endez  \& Minniti 2000).  However, these claims
remain  inconclusive for  lack of  spectroscopic  identifications. The
Hubble Deep Field  South has provided another opportunity  to test the
contribution  of  white  dwarfs  to  the  Galactic  dark  matter.   In
particular,  three white  dwarf  candidates among  several faint  blue
objects  which exhibit  significant proper  motion have  been recently
found  (Kilic  et  al. 2005).   They  are  assumed  to belong  to  the
thick-disk  or  halo  populations.   If  these  are  spectroscopically
confirmed it would imply that white dwarfs account for $\lesssim 10\%$
of the  Galactic dark matter,  which would fit comfortably  within the
results of  the EROS team.  All in  all, the study of  the white dwarf
population has  important ramifications  for our understanding  of the
structure and evolution of the Milky Way (D\'\i az--Pinto et al. 1994;
Isern et al. 1998; Torres et  al. 1998; Garc\'\i a--Berro et al. 1999;
Torres et al. 2001; Torres et al. 2002; Garc\'\i a--Berro et al. 2004;
Camacho et al. 2007; Torres et al. 2008).

Besides the  implications of  the white dwarf  cooling theory  for the
study of  stellar populations, white dwarfs  constitute also extremely
interesting objects in their own  right.  Typically, the mass of white
dwarfs is about half that of the Sun, and their size resembles that of
a  planet.   Their  compact   nature  translates  into  large  average
densities,  large  surface gravities  and  low  luminosities.  At  the
extremely high densities of  white dwarfs, electrons become degenerate
and  quantum mechanics ---  the Pauli  principle ---  determines their
equation of  state, their structure,  and the existence of  a limiting
mass above which no stable white dwarf can exist (Chandrasekhar 1939).
These properties  turn white dwarfs into  extremely attractive objects
as  cosmic laboratories  to  study numerous  physical processes  under
extreme   conditions   that   cannot   be  achieved   in   terrestrial
laboratories.  Just to indicate a few ones, we mention the equation of
state, crystallization and physical  separation processes at very high
density  (Garc\'{\i}a-Berro et  al.  1988a;  Garc\'{\i}a-Berro  et al.
1988b; Isern et al.  1991; Segretain et al.  1994; Isern et al.  1997;
Winget et  al 2009;  Garc\'{\i}a-Berro et al.   2010), the  physics of
neutrino  emission (Winget  et  al.  2004),  the  existence of  axions
(Isern  et  al.  2008)  and  the  variation  of fundamental  constants
(Garc\'{\i}a--Berro  et al.  1995).   The study  of the  structure and
evolution  of  white  dwarfs  thus  allows  us  to  test  our  current
understanding  of the  behavior  of matter  at  extreme densities  and
pressures.

In  the  course  of   their  evolution,  white  dwarfs  cross  several
pulsational instability  phases.  The pulsational  pattern of variable
white  dwarfs has  become  a  powerful tool  for  probing their  inner
regions   that   would  otherwise   remain   unaccessible  to   direct
observations (Fontaine \& Brassard  2008; Winget \& Kepler 2008).  The
recent  development  of  white dwarf  asteroseismological  techniques,
combined with  the improvements in  the evolutionary models  for these
stars have allowed to face  some key problems in astrophysics, such as
the core  chemical composition of low-mass stars  and the implications
for thermonuclear reaction rates, and the theory of crystallization at
high densities. In addition, pulsating white dwarfs are powerful tools
which have  applications beyond stellar  astrophysics.  In particular,
pulsating  white dwarfs  can be  used to  constrain the  properties of
elementary particles, such as axions and neutrinos (Isern et al. 1992;
C\'orsico et al.  2001b; Isern et al. 2010).

All these potential applications of white dwarfs have led to a renewed
interest  in the  calculation of  full evolutionary  models  for these
stars. In this work we discuss  the essentials of the physics of white
dwarf evolution  in order  to understand the  reasons that  make these
stars excellent  chronometers and potential tools with  a wide variety
of  applications.   In Sect.   \ref{essentials},  we  will describe  a
simple model  for white dwarf evolution,  making use of  the fact that
these stars  are strongly degenerate stellar  configurations, and thus
taking  advantage  of  the   fact  that  the  mechanical  and  thermal
structures can be treated separately.  This simple model will allow us
to  understand  the evolution  of  white  dwarf  as a  simple  cooling
process.  In  Sect.  \ref{detailed}, we will  describe the predictions
obtained  from the  full theory  of  stellar evolution,  based on  our
present  understanding  of the  physical  processes  occurring in  the
interior of these stars.  In Sect.  \ref{hdeficient}, special emphasis
will be  placed on  the formation and  evolution of  H-deficient white
dwarfs.    We   will    discuss   the   pulsational   properties   and
asteroseismological    tools    in    Sects.    \ref{pulsations}    to
\ref{tools}.  Furthermore, we  will discuss  as well  the  most recent
asteroseismological  results for  the different  classes  of pulsating
white   dwarfs.  This   is   done  in   Sect.   \ref{families}.    The
observational aspects  of white dwarfs  are briefly touched  upon. The
interested  reader  is referred  to  the  review  articles of  Koester
(2002),  Hansen \&  Liebert (2003)  --- for  general overviews  of the
field ---  Barstow (2006) --- for  hot white dwarfs ---  and Winget \&
Kepler (2008) and Fontaine \&  Brassard (2008) --- for recent accounts
of  the  observational aspects  and  applications  of pulsating  white
dwarfs.   However, we  will  discuss in  some  detail the  statistical
properties of the white dwarf population, and in particular, the white
dwarf luminosity function.


\section{The essentials of white dwarfs}
\label{essentials}

\subsection{A historical perspective}

\begin{figure}
\centering
\includegraphics[width=0.9\columnwidth]{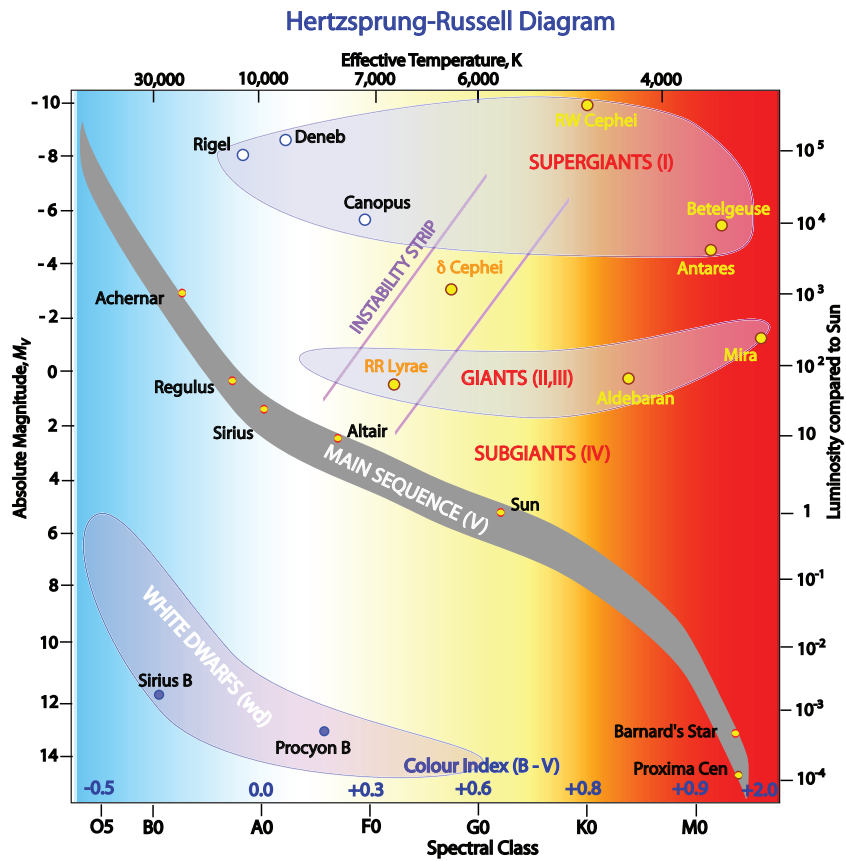}
\caption{Hertzsprung-Russell diagram.  The location of white dwarfs is
         clearly seen below the main sequence. From R. Hollow, CSIRO.}
\label{hr}
\end{figure}

White  dwarf stars have  captured the  attention of  astronomers since
their early discovery in 1914, when H.  Russell (Russell 1914) noticed
that the  star now known as 40  Eridani B, was located  well below the
main   sequence   on  the   Hertzsprung-Russell   diagram,  see   Fig.
\ref{hr}. He  rapidly realized that  this star was characterized  by a
very small radius (hence the ``dwarf'' in the name), comparable to the
size of the Earth.  But a major puzzle was posed by the discovery that
the companion  star to Sirius, Sirus  B, was also a  white dwarf.  The
determination  of the  spectral type  (Adams 1915)  and  luminosity of
Sirius B provided the first  estimate of its radius.  Because the mass
of this star was known to be about $1\, M_{\odot}$, it was possible to
infer the mean  density of a white dwarf for  the first time: $5\times
10^4$  gr/cm$^3$.  The  measurement of  the gravitational  redshift by
Adams (1925) resulting from the strong gravity of Sirius B provided an
independent test of  the high density of this white  dwarf, and at the
same  time a confirmation  of general  relativity.  It  is nonetheless
interesting to note that nowadays  the accepted value of the effective
temperature of the star is  much larger than the originally estimated,
thus implying a much larger average density of about $10^6$ gr/cm$^3$.

It was clear for astronomers that  these stars were indeed a new class
of star  quite different from  ordinary stars.  But  the extraordinary
value of the density derived for  Sirius B was difficult to explain at
that time,  and the  answer had  to wait until  the advent  of quantum
mechanics.   As  a matter  of  fact,  the  existence of  white  dwarfs
provided one of the first tests  of the quantum theory of matter and a
demonstration  of the  Pauli exclusion  principle for  electrons.  The
realization that completely degenerate  electrons should be the source
of internal pressure in the interiors of white dwarf (Fowler 1926) led
to  the conclusion  that white  dwarfs should  be stable  objects, and
established   the  study   of  the   zero-temperature  configurations.
Subsequent  investigations  by  Anderson  (1929),  Stoner  (1930)  and
Chandrasekhar (1931), demonstrated the fundamental conceptual coupling
of relativity and quantum statistical mechanics, which yielded another
key   breakthrough,  the   existence   of  a   limiting  mass,   where
gravitational forces overwhelm degenerate electron pressure, and above
which no stable white dwarf can exist.  This limiting mass, also known
as  the  Chandrasekhar limiting  mass,  constitutes  one  of the  most
important concepts in stellar  astrophysics, and it is responsible for
the differences  in the  evolutionary properties of  low and  and high
mass stars an in their final outcomes.

\subsection{Mass distribution}

\begin{figure}
\centering
\includegraphics[width=0.9\columnwidth]{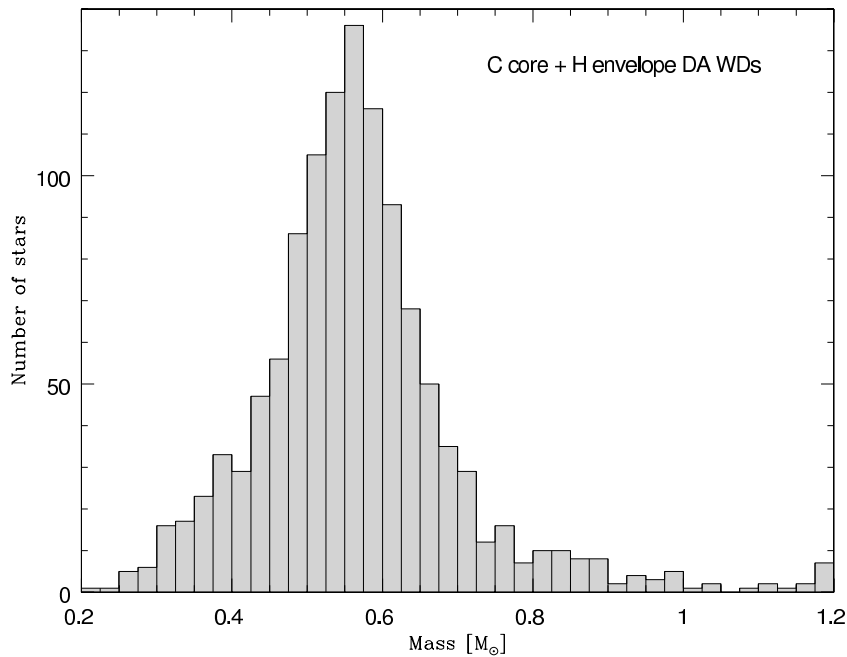}
\caption{Mass distribution of H-rich  white dwarfs with $T_{\rm eff} >
         12,000$  K.   The  peak  mass  is located  at  $M=  0.562  \,
         M_{\odot}$. From Madej et al. (2004). Reproduced by permision
         of ESO.}
\label{massdist}
\end{figure}

The number of known white  dwarfs has increased significantly from one
in 1914 in the first Herztsprung-Russell diagram to about 1,000 in the
1970s as a  result of targeted searches for  high proper motion stars.
More  recently, imaging  surveys  provided by  the  Sloan Digital  Sky
Survey (SDSS)  have largely  expanded the number  of spectroscopically
identified  white dwarfs to  about 10,000  (Eisenstein et  al.  2006),
including the  detection of  ultracool white dwarfs,  pulsating H-rich
white  dwarfs and  H-deficient  white dwarfs.   This  increase in  the
number  of  detected white  dwarfs  has  extraordinarily improved  our
knowledge  of  the white  dwarf  luminosity  and  mass functions.   In
addition,  spectroscopy  of  this  much  enlarged sample  led  to  the
discovery  of   new  and  interesting  species   of  degenerate  stars
(Eisenstein et al. 2006; Koester et al. 2009).

It is now accepted that white dwarfs constitute the end-product of all
stars with initial  masses up to $\sim 11  \, M_{\odot}$ (Siess 2007).
Most of  the mass of a typical  white dwarf is contained  in its core,
which  is  made of  the  products of  He  burning,  mostly carbon  and
oxygen. Small  amounts of H and  He are left over  after the mass-loss
phases  have ended.   Taking into  account the  previous thermonuclear
history and  the efficiency of gravitational settling,  it is expected
that the structure  of a typical white dwarf corresponds  to that of a
compositionally  stratified  object  with  a  mass of  about  $0.6  \,
M_{\odot}$ consisting  of a carbon-oxygen  core surrounded by  a thin,
He-rich envelope  --- of  at most $0.01  \, M_{\odot}$  --- surrounded
itself   by    a   thinner    H-rich   layer   of    $\sim   10^{-4}\,
M_{\odot}$. Although very thin,  the outer layers are extremely opaque
to  radiation  and regulate  the  energy  outflow  of the  star,  thus
playing, as we saw, a crucial role in the evolution of a white dwarf.

An  important property  of the  white dwarf  population is  their mass
distribution,  which conveys  information about  the evolution  of our
Galaxy.  From the mass distribution  of white dwarfs it is possible to
constrain the  late stages of  stellar evolution since it  reveals the
amount of  mass lost  during the  lifetime of the  star from  the main
sequence (Liebert et al.   2005).  The surface gravities and effective
temperatures  of  white  dwarfs  are  usually  determined  from  model
atmosphere  fits to  spectral lines.   On average,  it turns  out that
white dwarfs are characterized by surface gravities $\log g \simeq 8$.
Coupled  with theoretical mass-radius  relations, this  yields average
masses of $M\approx  0.6 \, M_{\odot}$.  In passing,  we note that for
cool white  dwarfs, H and  He spectral lines  are not visible  and the
stellar  radius   (and  mass)  have  to  be   inferred  from  accurate
trigonometric  parallaxes and  mass-radius  relations.  Typical  white
dwarf  mass distributions  (see  Fig.  \ref{massdist})  show that  the
values of  the masses of most  white dwarfs cluster  around this value
(Kepler  et al.   2007), with  a tail  extending towards  high stellar
masses.   The rather  narrow mass  distribution of  white dwarfs  is a
remarkable characteristic  of these stars.  Massive  white dwarfs have
spectroscopically determined  masses within 1.0  and $1.3\, M_{\odot}$
and are  believed to harbor cores  composed mainly of  oxygen and neon
--- at  least for  non-rotating  stars (Dom\'\i  nguez  et al.   1996;
Ritossa et al. 1996) --- in contrast to average-mass white dwarfs, for
which carbon-oxygen cores are expected.  The existence of such massive
white dwarfs has been suggested to  be the result of the merger of two
averaged-mass white  dwarfs in close  binaries (Guerrero et  al. 2004;
Lor\'en-Aguilar  et al.   2009) or  of the  evolution  of heavy-weight
intermediate-mass   single  stars   that  have   experienced  repeated
carbon-burning  shell flashes  (Ritossa et  al.  1999).   Finally, the
white  dwarf  mass distribution  comprises  a  population of  low-mass
remnants.  Because  low-mass progenitors would  need exceedingly large
ages to reach  the white dwarf stage, these  low-mass white dwarfs are
mostly  produced in binary  systems, where  the stellar  evolution has
been truncated by mass transfer (Sarna et. al 1999).

\subsection{The white dwarf luminosity function}

\begin{figure}[t]
\centering
\includegraphics[width=0.9\textwidth,clip]{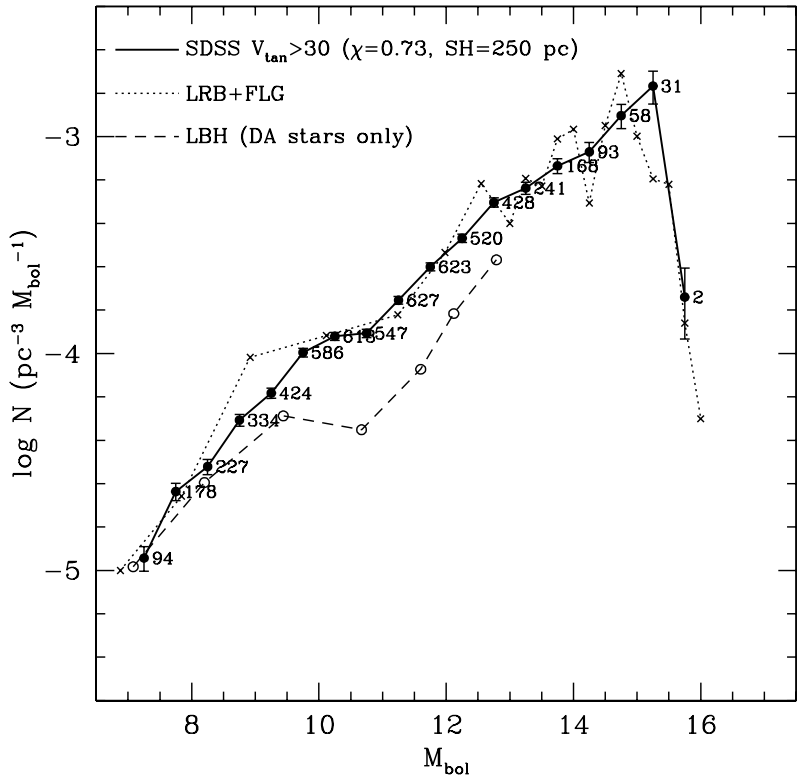}
\caption{Luminosity  function of  disk white  dwarfs derived  from the
         SDSS, from Harris et  al. (2006). Reproduced by permission of
         the AAS.}
\label{wdlf}
\end{figure}

Fifty  years ago  (Schmidt  1959)  it was  first  recognized that  the
coolest and  faintest white  dwarfs are the  remnants of  the earliest
stars formed  in the Solar  neighborhood, and that the  cooling theory
which we will  review below could be very useful  to estimate the time
elapsed  since  star formation  started  in  the  Galactic disk.   The
fundamental  tool  for studying  the  properties  of  the white  dwarf
population is the white dwarf luminosity function, which is defined as
the  number of  white  dwarfs  per cubic  parsec  and unit  bolometric
magnitude (or  luminosity) as a  function of the  bolometric magnitude
(or  luminosity).  The white  dwarf luminosity  function not  only can
provide valuable information about the age, structure and evolution of
our Galaxy but  it also provides an independent test  of the theory of
dense  plasmas (Isern et  al.  1997;  Isern et  al.  1998).   Also, it
directly   constrains   the   current   death   rate   of   low-   and
intermediate-mass  stars in  the  local neighborhood  which, in  turn,
provides  an  important  tool  to  evaluate  pre-white  dwarf  stellar
evolutionary sequences.   Furthermore, the relative  simplicity of the
structure and evolution of white dwarfs makes them extremely useful as
astroparticle  physics laboratories  (Raffelt 1996;  C\'orsico  et al.
2001b).  However, in order to  use the white dwarf luminosity function
to attack these  astrophysical problems, it is necessary  to have good
observational   data,  good   evolutionary  models   and   good  input
physics. Three decades later,  the white dwarf luminosity function has
helped  in  dealing  with  all  these problems  and  many  more.   The
observational white  dwarf luminosity function, its  uses and inherent
limitations  have  been  discussed  over  the years  in  a  number  of
excellent papers  (Weidemann 2000; M\'endez \& Ruiz  2001; Bergeron et
al. 2001; Hansen \& Liebert 2003).

At the present time, the most promising source of large numbers of new
white  dwarfs in  both the  disk and  halo is  the SDSS  (York  et al.
2000).  Thus  far, about 10,000  white dwarfs have been  identified in
the SDSS (Harris  et al.  2006; Eisenstein et  al.  2006; DeGennaro et
al. 2008) using  the reduced proper motion diagram  first described in
Luyten  (1922).  For  this  sample, the  disk  white dwarf  luminosity
function  was  constructed  using  improved proper  motions  based  on
comparison  of  positions between  the  SDSS  and  USNO surveys,  high
quality  $ugriz$  photometry, and  improved  atmospheric models.   The
resulting white  dwarf luminosity function is  surprisingly smooth and
drops off abruptly at $M_{\rm bol} = 15.3$ --- see Fig. \ref{wdlf}.

\subsection{Spectroscopic classification}
\label{spectro}

\begin{figure}
\centering
\includegraphics[width=0.9\columnwidth,clip]{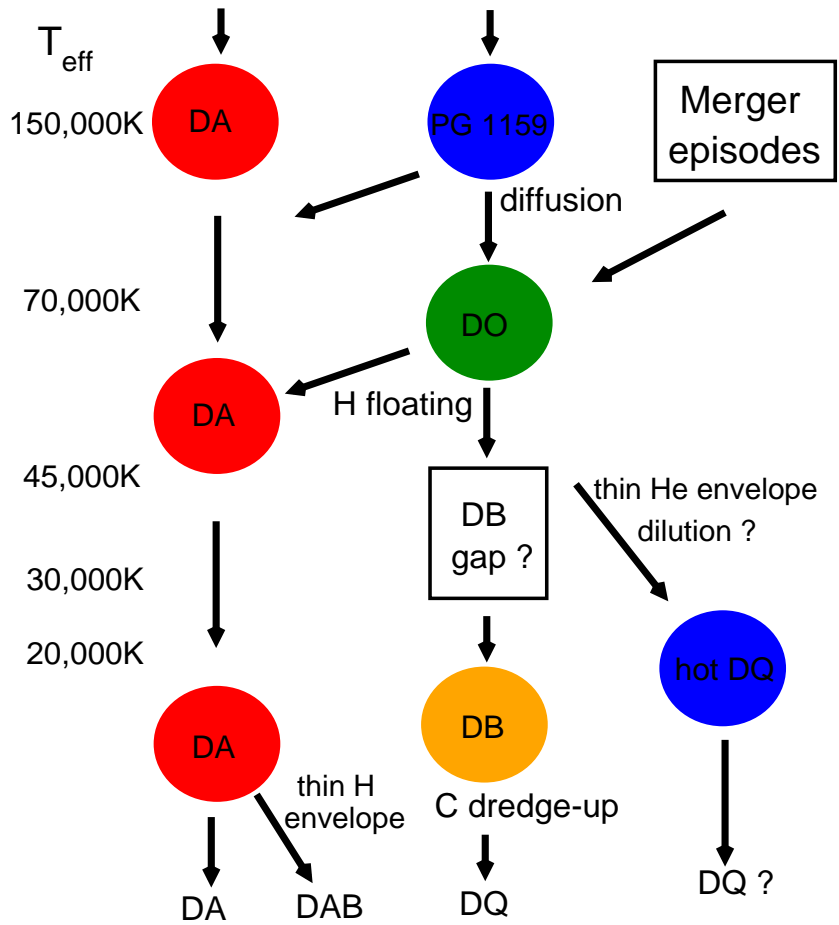}
\caption{A  scheme of the  several evolutionary  paths that  hot white
         dwarfs may follow as they  evolve.  The left column gives the
         effective  temperature.  Most  white dwarfs  are  formed with
         H-rich  envelopes  (DA spectral  type),  and  remain as  such
         throughout   their    entire   evolution   (second   column).
         H-deficient white dwarfs like DOs may follow different paths,
         either from the hot and He-, carbon-, and oxygen-rich PG 1159
         stars or  from post-merger events.   Traces of H in  PG 1159s
         and  DOs may  lead to  other white  dwarf varieties.  PG 1159
         stars are also believed to be the predecesors of the recently
         discovered  new class  of  white dwarfs:  the  hot DQs,  with
         carbon-rich atmospheres. Accretion  of metals by cool He-rich
         white dwarfs  from interstellar medium  or from circumstellar
         matter may lead to DZ white dwarfs.}
\label{spectral}
\end{figure}

Traditionally,  white dwarfs  have been  classified into  two distinct
families  according   to  the  main  constituent   of  their  surface.
Spectroscopic observations reveal that the surface composition of most
white  dwarfs consists almost  entirely of  H with  at most  traces of
other  elements. These  are the  so-called  DA white  dwarfs and  they
comprise  about 85\% of  all white  dwarfs ---  see Eisenstein  et al.
(2006)  and  references  therein.   To  the other  family  belong  the
H-deficient white  dwarfs with  He-rich atmospheres, usually  known as
non-DA  white  dwarfs, which  make  up to  almost  15\%  of the  total
population.  H-deficient white dwarfs are  thought to be the result of
late thermal flashes experienced  by post-AGB progenitors or of merger
episodes.  The  non-DA white dwarfs  are usually divided  into several
different   subclasses:   the  DO   spectral   type  (with   effective
temperatures 45\,000 K $\leq T_{\rm  eff} \leq 200\,000$ K) that shows
relatively strong lines of singly ionized He (He{\sc ii}), the DB type
(11\,000 K $\leq T_{\rm eff}  \leq 30\,000$ K), with strong neutral He
(He{\sc  i}) lines,  and the  DC,  DQ, and  DZ types  ($T_{\rm eff}  <
11\,000$ K) showing traces of  carbon and metals in their spectra.  As
a DO white dwarf evolves, the He{\sc ii} recombines to form He{\sc i},
ultimately transforming into a DB white dwarf.  The transition from DO
to the cooler  DB stage is interrupted by the  non-DA gap (that occurs
at 30\,000  K $  < T_{\rm eff}  < 45\,000$  K) where few  objects with
H-deficient atmospheres  have been observed (Eisenstein  et al. 2006).
To  this  list,  we  have  to  add  those  white  dwarfs  with  hybrid
atmospheres or peculiar abundances, and  the recent discovery of a new
white dwarf spectral type with carbon-dominated atmospheres, the ``hot
DQ'' white  dwarfs, with  $T_{\rm eff}\sim 20\,000$  K (Dufour  et al.
2007,  2008a).  Hot  DQ  white dwarfs  are  thought to  be the  cooler
descendants of some PG 1159 stars, and the result of convective mixing
at smaller  effective temperatures (Dufour  et al.  2008a;  Althaus et
al. 2009b).

Although this  classification is in  line with our  understanding that
most  giant stars  will evolve  into white  dwarfs with  either H-rich
atmospheres or H-deficient composition, the existence of some of these
white  dwarfs  poses  a  real  challenge  to  the  theory  of  stellar
evolution,  which cannot  adequately explain  their  origin.  Finally,
there  is ample  observational evidence  that individual  white dwarfs
undergo spectral  evolution, i.e., the surface composition  of a given
white  dwarf  may  change as  it  evolves  as  a result  of  competing
processes such as convection, mass-loss episodes, accretion, radiative
levitation  and gravitational  settling. The  interplay  between these
processes may help to understand the different evolutionary paths that
white dwarfs may follow as the surface temperature decreases, see Fig.
\ref{spectral}.  For  instance, the empirical evidence  that the ratio
of DA  to non-DA white  dwarfs changes with effective  temperature and
the  existence of  the non-DA  gap are  interpreted as  the  result of
changes in  the surface  composition from He-dominated  to H-dominated
and vice versa as evolution proceeds.  Also, the presence of traces of
H in  the outer  layers of  the hot H-deficient  white dwarfs  like PG
1159s or  DOs can turn  the spectral type  of these white  dwarfs into
that of a DA type as a result of gravitational settling.

White  dwarfs span  a wide  range of  both effective  temperatures and
luminosities. Values of $T_{\rm eff}$  range from about 150\,000 K for
the hottest  members to  4\,000 K for  the coolest  degenerate dwarfs.
The stellar  luminosity ranges from roughly $10^3$  to about $10^{-5}$
$L_{\odot}$ for  the faintest observed white dwarfs.   The majority of
known white dwarfs have temperatures higher than the Sun and hence the
``white''  in their  name.  Because  the intrinsic  faintness  of most
white dwarfs, quantitative studies of these stars, traditionally based
on  photometric  and  spectroscopic  observations, are  restricted  to
nearby objects.  Hence, the vast majority of observed white dwarfs are
representative  of   the  solar  neighborhood.    Other  observational
techniques and the advent of large-scale ground-based surveys and deep
Hubble Space Telescope exposures,  have revealed the presence of white
dwarf populations located well beyond our own neighborhood, such as in
distant open and globular clusters and, most probably, in the Galactic
halo, thus  enabling us  to extract information  and to  constrain the
properties of such populations.

\subsection{Simple treatment of white dwarf structure and evolution}

As previously  mentioned, white  dwarfs have numerous  applications in
different  fields of  astrophysics. In  particular, the  use  of white
dwarfs  as  accurate cosmic  clocks  has led  over  the  years to  the
development of detailed evolutionary  models of these stars, including
sophisticated  constitutive   microphysics.   Before  describing  such
models,   we  will  discuss   a  simple   treatment  of   white  dwarf
evolution. At the  high densities of white dwarf  interiors, matter is
completely pressure-ionized and the chemical potential of the electron
gas  is much  larger than  the thermal  energy.  This  means  that the
Fermi-Dirac  distribution that  characterizes electrons  reduces  to a
step-like function.   In particular, all  the energy states up  to the
Fermi energy ($\varepsilon_{  \rm F}$) are occupied.  Thus,  to a very
good  approximation, electrons are  almost completely  degenerate, and
those  electrons with energies  close to  $\varepsilon_{ \rm  F}$ will
make  the largest  contribution to  the pressure.   Hence, to  a first
approximation,  the  mechanical  structure  of  white  dwarfs  can  be
described in terms of a Fermi gas at zero temperature.

A key feature  to understand the evolution of white  dwarfs is that in
degenerate configurations, the  mechanical and thermal structures are,
to a good  approximation, decoupled from each other  and, thus, can be
treated separately. This basic  property of degenerate structures will
allow us  to derive a simple  picture of how white  dwarfs evolve with
time  and  to  discuss  the  role played  by  the  different  physical
processes  in the  evolution of  these stars.   Indeed,  the remaining
thermal  reservoir  (which  implies  $T\neq  0$)  is  responsible  for
radiation and, thus,  for the evolution of white  dwarfs. Instead, for
the    mechanical   structure,    the   limit    $T=0$    is   usually
assumed. Accordingly,  in what follows,  we will treat  the mechanical
and  thermal  structures  separately.   We  begin  by  describing  the
structure  of   white  dwarfs,   as  given  by   the  zero-temperature
approximation.  Then  in Sect. \ref{mestel} we will  describe a simple
model  to understand  the  basic thermal  properties  of white  dwarfs
responsible for their  evolution.  We will show that  evolution can be
described essentially  as a rather  simple cooling process,  where the
source of star  luminosity is approximately provided by  the change in
the internal energy stored in the ions.

\subsubsection{Zero-temperature approximation: Chandrasekhar's theory}
\label{chandra}

In this approximation, the structure of a white dwarf is determined by
the pressure  of non-interacting completely  degenerate electrons.  In
the limit of complete degeneracy,  the electron pressure can be easily
derived by considering only  the energy states up to $\varepsilon_{\rm
F}$.  The pressure and mass density have the well-known form:

\begin{eqnarray}
\label{eos}
P&=&\frac{8 \pi}{3 h^3}
\int_{0}^{p_{\rm F}} {{(p^4/m_e) dp}\over{\sqrt{1+(p/m_ec)^2}}} \nonumber\\
&=&A\ [x (x^2+1)^{1/2} (2 x^2-3) + 3\sinh^{-1}(x)]\nonumber\\ 
\ \\
\varrho&=&C\ x^{3}\,, \nonumber 
\end{eqnarray}

\noindent where the dimensionless Fermi momentum is given by $x=p_{\rm
F}/m_{\rm e}c$,  $A= \pi  m_{\rm e}^{4} c^{5}/3\  h^{3}$ and  $C=8 \pi
m_{\rm e}^{3} c^{3} \mu_{\rm e} M_{\rm u} / 3\ h^{3}$, and the rest of
the symbols  have their  usual meaning. Note  that this  expression is
valid for any relativistic degree.

For  a spherically  symmetric  white dwarf  at  zero temperature,  the
equilibrium  structure is  specified by  the equations  of hydrostatic
equilibrium and mass conservation

\begin{eqnarray}
\label{eh}
\frac{d P}{d r}&=&- \frac{G m \varrho}{r^2}\nonumber\\
\ \\
\frac{d m}{d r}&=&4 \pi r^2 \varrho\ \,,\nonumber
\end{eqnarray}

\noindent together with the equation of state given in Eq. (\ref{eos})
and the usual boundary conditions.  Here, $r$ is the radial coordinate
and $m$ the mass inside a sphere of radius $r$.

Chandrasekhar (1939) showed that  the problem can be reformulated into
a second order differential equation for a dimensionless function with
essentially  two parameters that  have to  be specified:  the chemical
composition via the molecular  weight per electron ($\mu_{\rm e}$) and
the central density. To see  this, note first that Eqs. (\ref{eh}) can
be casted in the form

\begin{eqnarray}
\frac{d}{d r}\left(\frac{r^2}{\varrho}\frac{d P}{d r}\right)&=&
-4 \pi G\, r^2 \varrho\ \,.
\end{eqnarray}

\noindent    Introducing    the    variables   $\eta=r/\alpha$,    $y=
\sqrt{1+x^{2}}$, and $\Phi= y/y_{0}$, it is straightforwardly found

\begin{eqnarray}
\frac{1}{\eta^2}\frac{d}{d \eta}\left(\eta^2\frac{d \Phi}{d \eta}\right)&=&
-\left(\Phi^2 - \frac{1}{y_0^2}\right)^{3/2} \,,
\label{chandraeq}
\end{eqnarray}

\noindent  where  $y_{0}^{2}=x_0^{2}+1$  is  related  to  the  central
density,  $\varrho_{0}$, via  Eq.  (\ref{eos}),  that  is, $\varrho_0=
C({y_0}^2-1)^{3/2}$ and $\alpha$ is given by

\begin{eqnarray}
\alpha&=& \left( {{2A}\over{\pi G}} \right)^{1/2}  {{1}\over{C
y_{0}}} \, .
\end{eqnarray}

\noindent Eq.  (\ref{chandraeq}) describes the structure of completely
degenerate  configurations  based  on  the  equation  of  state  of  a
degenerate electron gas. It takes into account the gradual change from
non-relativistic to  relativistic electrons, with  increasing density.
Eq.   (\ref{chandraeq})  is   solved  numerically  with  the  boundary
conditions:   $\Phi(0)=1$   and   $\Phi(\eta_{\rm  I})=1/y_0$,   where
$\eta_{\rm I}$  defines the radius  of the stellar  configuration, $R=
\alpha\ \eta_{\rm  I}$.  For a fixed  value of $\mu_{\rm  e}$, i.e., a
given chemical  composition, a one-parameter family of  models is thus
obtained  integrating Eq.  (\ref{chandraeq})  for different  values of
the central density, thus  defining implicitly a relation between mass
and  radius such  that  the more  massive  the star,  the smaller  its
size. This is  the famous mass-radius relation for  white dwarf stars.
In   the  limit   $\varrho  \rightarrow   \infty$,   the  relativistic
``softening'' of the equation of state (for relativistic electrons, $P
\sim \varrho^{4/3}$) causes  the radius $R$ to become  zero, while the
mass approaches a limiting value $5.826 / \mu_{\rm e}^2 \, M_{\odot}$,
the   Chandrasekhar   limiting  mass.    According   to  our   present
understanding of stellar  evolution, a core made of  carbon and oxygen
is expected for most white dwarfs.  In this case, $\mu_{\rm e} \approx
2$, and the Chandrasekhar mass becomes $1.45\, M_{\odot}$.  For larger
masses, gravitational forces  overwhelm the electron pressure support,
and no stable white dwarf can exist.

The theoretical  mass-radius relation for  white dwarfs, which  can be
improved  by  including  corrections  to  the equation  of  state  and
finite-temperature  effects, is an  accepted underlying  assumption in
nearly  all studies of  white dwarfs  properties, including  the white
dwarf  mass distribution and  the luminosity  function, which  in turn
convey important  information about  Galactic evolution and  about the
late stages of stellar evolution.  Because of this, it is essential to
provide  observational support  for this  theoretical  relation.  This
means that it is necessary  to obtain independent measures of the mass
and radius of individual white dwarfs.  The best way to do this is for
white  dwarfs  in visual  binaries,  where  the  stellar mass  can  be
inferred from Kepler's  law and the radius can be  obtained from a fit
to the  spectrum of the white  dwarf atmosphere, once  the distance to
the binary  system is known. Independent determinations  of radius and
mass   can  also  be   obtained  from   white  dwarfs   with  measured
gravitational redshifts in  binary systems.  Although independent mass
and radius determinations  are available only for a  few white dwarfs,
they  strongly support  the theoretical  mass-radius relation  and the
existence of a limiting mass.  Another remarkable consequence of these
observational  determinations is  that they  discard the  existence of
white dwarfs with H-rich cores. Note that for white dwarfs with H-rich
cores   ($\mu_{\rm  e}\simeq   1$),  the   mass  of   the  equilibrium
configuration  is much  larger for  a given  stellar radius  than that
resulting from a carbon-oxygen composition.  This is in agreement with
the results of the stellar  evolution theory, which predicts that H is
burnt  into heavier elements  in evolutionary  stages previous  to the
white dwarf formation.

The Chandrasekhar limiting mass is  one of the most important concepts
in  astrophysics.   The  existence  of  a  limiting  mass  has  strong
implications not only for the  theory of white dwarfs itself, but also
for stellar evolution in general.  This is particularly true regarding
the occurrence of mass-loss episodes: low-mass stars must lose a large
fraction of their  mass at some point of their  evolution to end-up as
white dwarfs with masses smaller than $1.4\, M_{\odot}$.  Finally, the
concept  of the  Chandrasekhar limiting  mass is  intimately connected
with  the most  energetic events  known in  the  Universe: supernovae.
Those supernovae  which do not  show H in  their spectra ---  known as
type I  supernova (SNIa), or thermonuclear supernovae  --- are thought
to be the result  of a instability of a white dwarf  with a mass close
to  the Chandrasekhar  mass. One  of  the two  proposed mechanisms  to
explain  the  properties  of  SNIa  involves  a  massive  white  dwarf
progenitor  in  a  binary   system  accreting  from  a  non-degenerate
companion --- see Branch et al.   (1995) and Kotak (2008) for a recent
review. The second  mechanism involves the merger of  two white dwarfs
with sufficiently large stellar masses due to the secular radiation of
gravitational waves in a  close binary system --- see Lor\'en--Aguilar
et  al.  (2009)  and references  therein for  recent three-dimensional
calculations of this phenomenon.  In the first scenario, mass transfer
from  an accretion disk  onto the  massive white  dwarf may  cause the
white  dwarf to  exceed  the  Chandrasekhar mass,  thus  leading to  a
thermonuclear  runaway and  a explosion  that completely  disrupts the
white dwarf.  In the second scenario, the coalescence process may lead
to high  enough temperatures to ignite the  degenerate merger remnant,
leading  again to  the same  final output.   On the  other  hand, core
collapse supernovae arise from stars with degenerate cores with masses
larger than the Chandraskhar limiting mass.

\subsubsection{Mestel's model of white dwarf evolution}
\label{mestel}

The zero-temperature  model we have described in  the previous section
provides  an appropriate  description of  the mechanical  structure of
white dwarfs. However, white dwarfs are not zero temperature stars. In
fact, they are observed to  have high surface temperatures and to lose
energy,  which implies  the  existence of  temperature gradients,  and
consequently of higher interior  temperatures.  Thus, white dwarfs are
not  static  objects and  are  expected  to  experience some  kind  of
evolution.   As we already  mentioned, the  advantage of  dealing with
degenerate   configurations  is  that   the  mechanical   and  thermal
properties are effectively decoupled  from each other.  This allows us
to capture  the main features of  white dwarf evolution by  means of a
very simple  model, which is often  referred to as  the Mestel's model
(Mestel 1952).   Here, the  problem of the  evolution of  white dwarfs
reduces essentially to  two main aspects. The first  of these consists
in the  determination of the total  energy content, $E$,  of the star,
whilst the second  one is the determination of the  rate at which this
energy is radiated away.

The first problem involves  a detailed knowledge of the thermodynamics
of the  degenerate core of the  white dwarf, which  contains more than
99$\%$ of  the mass, and  the structure of  which is specified,  as we
saw, by completely degenerate electrons ($T=0$). An important property
of the  core, which simplifies  the evaluation of  $E$, is that  it is
characterized by an almost constant temperature, $T_{\rm c}$.  This is
because  degenerate  electrons  are  very  efficient  at  transporting
energy,  and thus  a very  small temperature  gradient is  required to
transport the energy  flux in the core. Although  the assumption of an
isothermal  core is not  valid in  young and  hot white  dwarfs, where
neutrinos constitute a main energy sink, it is completely justified in
evolved  and  cool white  dwarfs.   The  second  problem involves  the
solution of the equation of energy transfer through the non-degenerate
envelope above  the core.  In  the envelope, energy is  transferred by
radiation and/or convection, which  are less efficient mechanisms than
electron  conduction. In  fact, the  extremely large  temperature drop
between  the  core  and  the  surface  occurs  mainly  in  the  opaque
envelope. Because  of this, the envelopes  of white dwarfs  play a key
role in  white dwarf evolution, since  they control the  rate at which
energy is transferred from the core into space.

In  what follows, a  simple approach  to solve  the problem  of energy
transfer in  the outer layers of  a white dwarf will  be explained. By
making reasonable simplifications,  simple power law relations between
the surface  luminosity, the stellar mass and  the central temperature
of   the  white   dwarf  will   be  derived.    The  first   of  these
simplifications  concerns  the  way  in which  energy  is  transported
throughout  the envelope. We  will assume  that energy  is transferred
only by radiation,  which is true except for  cool white dwarfs, where
convection is the dominant mechanism of transport in the outer layers.
In  this case,  the  structure of  the  envelope is  specified by  the
equation of hydrostatic equilibrium,  Eq. (\ref{eh}), and the equation
of  radiative transfer,  as  given by  the  photon diffusion  equation
(Kippenhahn \& Weigert 1990)

\begin{eqnarray}
\frac{d T}{d r}& =& -\frac{3}{4 a c}
~\frac{\kappa \varrho}{T^3}~\frac{L_r}{4 \pi r^2} \,,
\label{foton}
\end{eqnarray}

\noindent  where $L_r$ is  the luminosity  or the  net flux  of energy
transported through the sphere of radius $r$ (erg /s), and $\kappa$ is
the  radiative  opacity.  A  common approximation  for  the  radiative
opacity  is  to  adopt  Kramers's  law for  bound-free  and  free-free
processes: $\kappa  = \kappa_0\ \varrho\ T^{-3.5}$,  where $ \kappa_0$
depends  on the  chemical composition  of the  envelope. We  will also
assume  that  the envelope  is  thin  enough  that $m\approx  M$  (low
density), and  that any source or  sink of energy can  be neglected in
the envelope,  so that  $L_r \approx L$,  where $L$ means  the surface
luminosity.   Under these  assumptions, the  equation for  the thermal
structure of the  envelope can be derived by  dividing the hydrostatic
equilibrium equation by Eq. (\ref{foton}):

\begin{eqnarray}
\frac{d P}{d T}&=&\frac{16 \pi a c }{3 \kappa_0}
~\frac{G M T^{6.5}}{\varrho L} \,.
\label{dpdt}
\end{eqnarray}

To integrate this equation the equation of state for the envelope must
be  specified.   Because  white  dwarfs  are  observed  to  have  high
effective   temperatures,  and   considering  the   smaller  densities
characterizing the  very outer  layers, we expect  the envelope  to be
non-degenerate.  Thus,  we will assume  that the equation of  state is
that of  a non-degenerate,  ideal gas (ions  and electrons),  which is
given by

\begin{eqnarray}
\label{eq:eos_g_ideal}
P&=&\frac{\Re}{\mu} \varrho T \quad \quad \mbox{with} \quad \mu^{-1}=\sum_{i} 
(1+Z_i)\frac{X_i}{A_i} \,. \\ 
\nonumber
\end{eqnarray}

Here $\Re$  is the gas constant,  $\mu$ is the  mean molecular weight,
$A_i$  the atomic  weight,  $Z_i$  the charge  number,  and $X_i$  the
abundance  by  mass  fraction  of  the  chemical  species  $i$.  Thus,
Eq. (\ref{dpdt}) becomes now

\begin{eqnarray}
P\ d P&=&\frac{16 \pi a c }{3 \kappa_0}
~\frac{\Re G M} {L \mu}\ T^{7.5}\ d T \,.
\end{eqnarray}

We integrate this equation from the surface, where we assume $P=0$ and
$T=0$, to the base of the envelope as defined by the transition layer.
This  layer is an  idealized representation  of the  abrupt transition
that separates the inner degenerate core from the thin, non-degenerate
envelope.  The integration yields

\begin{eqnarray}
\varrho_{\rm tr}&=&\left(\frac{32 \pi a c}{8.5\ 3 \kappa_0} \frac{\mu}{\Re} 
\frac{G M}{L}\right)^{1/2}\ T_{\rm tr}^{3.25} \,,
\label{densi}
\end{eqnarray}

\noindent  which   provides  a   relation  between  the   density  and
temperature at  the transition  layer ($\varrho_{\rm tr}$  and $T_{\rm
tr}$, respectively)  for a white  dwarf characterized by  stellar mass
$M$ and surface luminosity $L$.  At the transition layer, the pressure
of  degenerate electrons  is equal  to the  pressure of  an  ideal gas
(pressure is  continuous across the transition layer).   This gives us
another  relation between  density and  temperature at  the transition
layer.   Note that  we  consider non-relativistic  electrons.  In  the
non-relativistic limit, $x\rightarrow  0$ and Eqs.  (\ref{eos}) reduce
to $P \sim (\varrho/\mu_e)^{5/3}$.  Thus, we have

\begin{eqnarray}
10^{13} \left(\frac{\varrho_{\rm tr}}{\mu_e}\right)^{5/3}=\frac{\Re}{\mu_e} 
\varrho_{\rm tr} T_{\rm tr}\quad \Longrightarrow\quad \varrho_{\rm tr}=
2.4 \times 10^{-8}\mu_{\rm e}\ T_{\rm tr}^{3/2} \,.\\
\nonumber
\end{eqnarray}

Because we  have assumed  that the entire  core is isothermal,  we can
replace $T_{\rm  tr}$ by the  core temperature, $T_{\rm  c}$. Finally,
from the last  two equations we arrive at a  simple power law relation
between  the surface  luminosity,  the stellar  mass  and the  central
temperature of the white dwarf:

\begin{eqnarray}
L&=&5.7 \times 10^{5}\ \frac{\mu}{\mu_{\rm e}^2}\ \frac{1}{Z(1 + X)}\ 
\frac{M}{M_{\odot}}\ T_{\rm c}^{3.5}\quad \mbox{erg /s} \,,
\label{ltc}
\end{eqnarray}

\noindent where $Z$ and $X$  are, respectively, the metallicity of the
envelope and the H abundance by mass.

Although some of the assumptions made in the treatment of the envelope
are not usually verified in white dwarfs, the simple relation given by
Eq.  (\ref{ltc}) enables us to  infer some important features of these
stars.  Specifically,  for typical compositions and a  stellar mass of
$M=1\,  M_{\odot}$, we  find that  for $L=  10\, L_{\odot}$  and  $L =
10^{-4}\,  L_{\odot}$, the central  temperatures are  of the  order of
$T_{\rm c}  \approx 10^8$  K and $T_{\rm  c} \approx 4\times  10^6$ K,
respectively.   The    high   central   temperatures    expected   for
high-luminosity  white  dwarfs  discard  the  presence  of  H  in  the
degenerate core.  In fact, simple stability arguments show that at the
high densities  of the core,  stable nuclear burning is  not possible,
because  the white  dwarf  would  be destroyed  by  a thermal  runaway
(Mestel 1952).  The fact that no nuclear sources can be present in the
core  of white  dwarfs at  their  birth, constrains  their origin  and
suggests that white dwarfs must be the result of complete H exhaustion
occurred at some point in the evolution of their progenitors.

Before proceeding forward, we can make a simple estimate of the radial
extent of the  envelope of a typical white dwarf. To  this end, we use
Eqs.  (\ref{foton}) and (\ref{densi}), and Kramers' law to obtain

\begin{eqnarray}
d T&=&-\frac{1}{4.25}~\frac{G M \mu} {\Re}\ \frac{1}{r^2}\ d r\,.
\end{eqnarray}

\noindent We  integrate this  equation over the  radial extent  of the
envelope

\begin{eqnarray}
\int_{0}^{T_{\rm tr}}\ d T&=&-\frac{1}{4.25}~\frac{G M \mu} {\Re}\  
\int_{R}^{R_{\rm tr}}\ \frac{1}{r^2}\ d r\,,
\end{eqnarray}

\noindent to obtain

\begin{eqnarray}
T_{\rm tr}&=&\frac{1}{4.25}~\frac{G \mu} {\Re}\  \frac{M}{R}\  
\left(\frac{R}{r_{\rm tr}} - 1 \right)
\end{eqnarray}

For  a low-luminosity  white dwarf  with $M=1\,  M_{\odot}$, $R=0.01\,
R_{\odot}$ and  $T_{\rm c}= 4\times  10^6$ K, we  get $R -  r_{\rm tr}
\approx 30$  km.  Thus, a  very thin envelope isolates  the degenerate
core from the outer space. Through this thin envelope, the temperature
drops appreciably  from the central  to surface values.   Although the
envelope controls  the rate  at which energy  is transferred  from the
core  to  the space,  it  has  nevertheless  little influence  on  the
structure of the white dwarf. In  fact, the radius of a white dwarf is
well approximated by the  calculations that assume complete degeneracy
throughout.

The absence  of nuclear  energy sources in  the white  dwarf interiors
brings us to our first point, namely the total energy content, $E$, of
a white  dwarf. The precise question  to be answered  is the following
one. If there  is no stable nuclear burning,  which energy sources are
involved when a normal white  dwarf radiates energy?  It is clear that
some energy source or reservoir must be present, since if this was not
the  case it  may not  shine.  We  will focus  on this  issue  in what
follows.  The  first point to be  noted is that  although white dwarfs
cannot contract appreciably, because  of electron degeneracy --- under
the conditions of strong degeneracy, pressure is almost independent of
temperature --- the residual  contraction resulting from the gradually
decreasing ion pressure  plays an important role in  the energy budget
of the  star.  This is because  white dwarfs are  compact objects, and
hence a  small change  of the radius  releases an important  amount of
gravitational energy.   For instance, for  a white dwarf with  $M= 1\,
M_{\odot}$, a  change of 1\%  in the radius  produces a change  in the
gravitational energy

\begin{eqnarray}
\mid\Delta E_{\rm grav}\mid\ \propto \frac{G M^2}{R}\ \frac{\Delta R}{R} \sim 
10^{48}\ {\rm erg}.
\end{eqnarray}

\noindent At  a typical luminosity of  $L=0.1\,L_{\odot}$, this energy
would  be lost  in  about $10^8$  yr.   Thus, the  energy released  by
contraction  is by  no means  negligible.  It  can be  shown  from the
virial  theorem for degenerate  configurations that  the gravitational
energy released during  contraction is of the same  order of magnitude
of the stellar luminosity $L$.

If we neglect  neutrino losses, the total energy  content of the white
dwarf results from contributions  of ions, electrons and gravitational
energy,  i.e., $E= E_{\rm  ion} +  E_{\rm elec}  + E_{\rm  grav}$. The
luminosity is given by the temporal decrease of the total energy $E$:

\begin{eqnarray}
L&=&-\frac{dE}{dt}=-\frac {d}{dt}(E_{\rm ion} + 
                                  E_{\rm elec} + 
                                  E_{\rm grav} ).  
\end{eqnarray}

According  to the  virial theorem  for degenerate  configurations, the
release of gravitational  energy is used to increase  the Fermi energy
of  the  electrons,  $\varepsilon_{   \rm  F}$  ---  specifically  the
density-dependent  contribution of  the energy  of  electrons (Koester
1978).  This behavior is different  from that of a non-degenerate gas,
where  the  released gravitational  energy  is  used  to increase  the
internal energy  of both  ions and electrons.   This gives rise  to an
important simplification  of the treatment, since in  order to compute
the  temporal change of  $E$, only  the changes  in the  {\sl thermal}
contribution to $E$  need to be considered. Indeed,  if we assume that
the core is  isothermal, the luminosity of the star  can be written in
terms of the decrease of the central temperature as

\begin{eqnarray}
L= -\frac{\partial}{\partial T_{\rm c}} (E_{\rm ion}^T + E_{\rm elec}^T)\ 
\frac{dT_{\rm c}}{dt}.  
\end{eqnarray}

\noindent where $E_{\rm ion}^T$ and $E_{\rm elec}^T$ are the total ion
and electron thermal energies. Thus,  in terms of the specific heat of
ions and electrons, the luminosity equation becomes

\begin{eqnarray}
L=  -\langle C_V\rangle M  \frac {dT_{\rm c}}{dt},  
\end{eqnarray}

\noindent where  $\langle C_V\rangle  = C_{V}^{\rm elec}  + C_{V}^{\rm
  ion}$  and $M$  is the  white dwarf  mass.  Thus,  we have  a simple
relation between the luminosity and  the rate of change of the central
temperature. Clearly,  the source  of luminosity of  a white  dwarf is
essentially the decrease in the  thermal energy of ions and electrons.
Because  of this,  white dwarf  evolution  is described  as a  cooling
process.   However,   at  low  temperatures,   $C_{V}^{\rm  elec}  \ll
C_{V}^{\rm   ion}$,  since   strongly   degenerate  electrons   barely
contribute  to the specific  heat.  In  fact, for  strongly degenerate
electrons, we have

\begin{eqnarray}
C_V^{\rm elec} \propto \frac {k_{\rm B}T}{\varepsilon_F} \rightarrow 0 \,.  
\end{eqnarray}

\noindent  Thus,  we  arrive  at  the  core  feature  of  white  dwarf
evolution: the source of luminosity of a cool white dwarf is basically
the change of the internal energy stored in the ions.

Using the relation between luminosity and central temperature given by
Eq. (\ref{ltc}), which reads $L = C M T_{\rm c}^{3.5}$, where $C$ is a
constant, we get

\begin{eqnarray}
C M  T_{\rm c} ^{3.5}&=&-\langle C_V^{\rm ion}\rangle M 
\frac {dT_{\rm c}}{dt}\,.
\label{eqcool}   
\end{eqnarray}

\noindent We  now assume an  ideal gas for  a single ion. That  is, we
adopt  $ C_{V}^{\rm  ion}= 3  \Re /  2 A  $, where  $A$ is  the atomic
weight, and integrate Eq. (\ref{eqcool}) from an initial time $t_0$ to
the present  time $t$.   We define the  {\sl cooling time}  as $t_{\rm
  cool}= t - t_0$, and the following cooling law is obtained

\begin{eqnarray}
t_{\rm cool} \approx \frac{10^8}{A}\ \left(\frac{M/M_\odot}{L/L_\odot}\right)^{5/7} 
{\rm years},  
\end{eqnarray}   

\noindent where  it has been  assumed that the central  temperature at
the beginning of white dwarf formation is much larger than the present
central temperature of the white dwarf.  This is the Mestel's model of
white dwarf  evolution (Mestel 1952),  and it provides a  simple power
law  relation between  the  cooling time  of  a white  dwarf, and  its
stellar mass, surface luminosity,  and core chemical composition.  For
$A=12$ (a  pure carbon  core), typical cooling  ages for  the faintest
observed white dwarfs ($L \sim 10^{-4.5}\, L_{\odot}$) of $10^{10}$ yr
are derived. These evolutionary  timescales are sufficiently long that
white dwarfs  remain visible  for a long  time. Thus, the  white dwarf
phase constitutes a major phase in stellar life. Cool white dwarfs are
very  old objects,  and  considering the  relative  simplicity of  the
physical processes  responsible for their evolution,  these stars turn
out to be robust tools to date stellar populations.

Although Mestel's  model is the  simplest picture of how  white dwarfs
evolve,  it captures  the essentials  of  the physics  of white  dwarf
evolution.  It shows that the cooling time of a white dwarf is related
essentially  to  its  core  chemical  composition, its  mass  and  its
luminosity.  The following major features emerge:

\begin{enumerate}
\item The cooling times depend  on the core chemical composition ($A$)
      of the  white dwarf:  oxygen white dwarfs  are expected  to cool
      faster than  carbon white dwarfs  because the specific  heat per
      gram of oxygen is smaller than that of carbon --- there are more
      ions in a gram of carbon than in a gram of oxygen.

\item Because  of their larger thermal content,  in principle, massive
      white    dwarfs   are    characterized    by   longer    cooling
      times. Moreover,  it has to be  taken into account  as well that
      these white dwarfs also have  much smaller radii and, thus, they
      have smaller luminosities.

\item The cooling  times increase as the luminosity decreases.
\end{enumerate}

\subsubsection{Improvements to Mestel's model}

\begin{figure}
\centering
\includegraphics[width=0.9\columnwidth,clip]{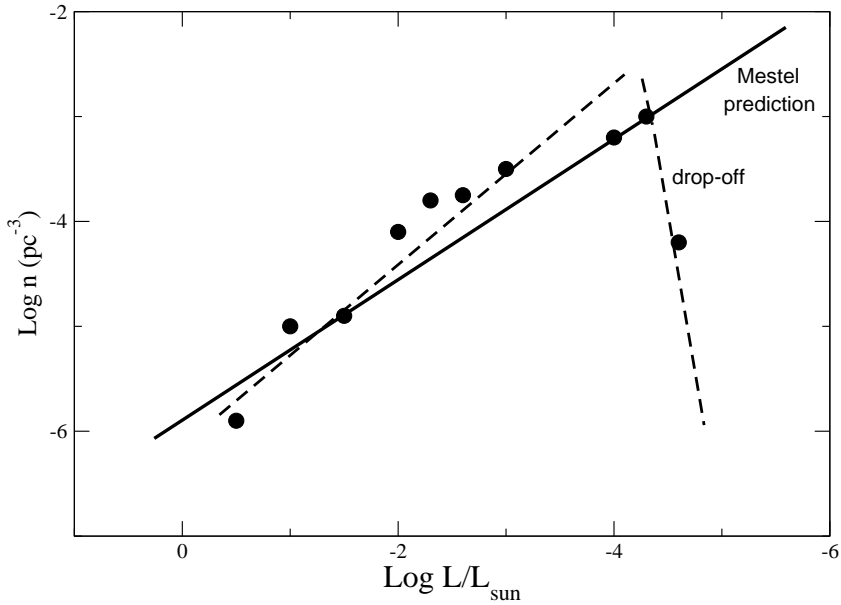}
\caption{A  schematic  view of  the  observed  white dwarf  luminosity
         function,  as  a  function  of the  luminosity  (dots).   The
         prediction  of  the  Mestel  law  is  shown  as  solid  line.
         Note the sharp cutoff at low luminosities.}
\label{lf}
\end{figure}

As previously mentioned, the white dwarf luminosity function gives the
number of white dwarfs per unit volume and interval of magnitude, as a
function of  the magnitude. To  a first approximation, this  number is
proportional to  the characteristic  cooling time, $\tau_{\rm  cool} =
{\rm  d}\,  t_{\rm cool}  /  {\rm  d}\log(L/L_{\odot})$. According  to
Mestel's law, more and more white dwarfs should be observed at fainter
luminosities,  as  it  is  indeed  the  case,  see  Fig.   \ref{wdlf}.
Nonetheless,  as clearly shown  in Fig.   \ref{wdlf}, the  white dwarf
luminosity  function   presents  an   abrupt  drop-off  at   very  low
luminosities  ($L\sim  10^{-4.5}\,  L_{\odot}$).   This  steep  cutoff
cannot  be  explained  within  the  frame of  Mestel's  theory.   Fig.
\ref{lf}  depicts  a  schematic  view  of  the  observed  white  dwarf
luminosity function  together with the prediction given  by the Mestel
cooling law.  Clearly, the  theoretical model predicts that the number
of  white   dwarfs  should  increase   monotonically  with  decreasing
luminosity.  This discrepancy between theory and observation motivated
the  exploration of  physical effects  that might  be  responsible for
shortening  the  cooling times  at  very  low luminosities.   However,
nowadays  we know  that  this shortage  of  white dwarfs  at very  low
luminosities  is a  consequence  of  the finite  age  of the  Galactic
disk. Namely, we know that the origin of the cutoff is due entirely to
the fact  that the coolest  white dwarfs have  not had time  enough to
cool down to luminosities lower than that of the observed drop-off.

However, let  us go  step by  step. The Mestel  model is  a reasonable
description of the behavior of  real white dwarfs only at intermediate
luminosities, where the assumptions of ideal gas, isothermal core, and
non-degenerate   radiative   envelope   are   more  or   less   valid.
Nevertheless, it  requires major improvements  if we want to  use cool
white  dwarfs as  reliable cosmic  clocks.  In  what follows,  we will
comment on the main basic improvements to be considered in the physics
of white dwarfs and on the expected impact on the cooling times.

\subsubsection{Coulomb interactions and Debye cooling} 

At the high densities  characteristic of white dwarf interiors, matter
is completely pressure-ionized.   As early recognized (Kirshnitz 1960;
Abrikosov  1960;  Salpeter  1961),  Coulomb  interactions  modify  the
thermodynamical properties of the  ion gas, in particular the specific
heat.  This, in turn, modifies  the cooling times of white dwarfs, see
Eq.   (\ref{eqcool}).   The   strength  of  Coulomb  interactions  ---
relative  to the  thermal  kinetic  energy ---  is  determined by  the
Coulomb  coupling parameter  $\Gamma= (Ze)^2  /  a k_{\rm  B} T=  2.26
\times 10^5  Z^{5/3} \varrho^{1/3} /T$,  where $a$ is  the inter-ionic
separation  and $k_{\rm  B}$  is the  Boltzmann  constant.  For  small
values of  $\Gamma$, Coulomb forces are of  minor importance (relative
to thermal motions) and the  ions behave like an ideal non-interacting
gas.  But, once $\Gamma$ becomes of order unity, ions begin to undergo
short-range  correlations,  eventually behaving  like  a liquid.   For
large  enough $\Gamma$  ($\sim 180$)  they form  a  lattice structure,
experiencing  a first  order phase  transition with  the corresponding
release of latent heat. This results  in a new source of energy, which
introduces an extra delay in the cooling of white dwarfs.  In the case
of Coulomb plasmas, the latent heat  is small, of the order of $k_{\rm
B}  T_{\rm  s}$ per  ion,  where $T_{\rm  s}$  is  the temperature  of
solidification.  Its  contribution to  the total luminosity  is small,
$\sim 5$\%, but  not negligible (Shaviv \& Kovetz  1976). In addition,
there is an increase in the  specific heat of ions in the crystallized
region (according to the Dulong-Petit  law) due to the extra degree of
freedom associated  with lattice vibration.  But  most importantly, as
the white dwarf cools further, fewer modes of the lattice are excited,
and the heat capacity drops  according to the Debye law.  This reduced
heat capacity starts to manifest  once the temperature drops below the
Debye  temperature ($\theta_{\rm  D}= 4  \times  10^3 \varrho^{1/2}$),
resulting  in   a  fast  cooling   phase  ($C_V  \propto   T^3$)  once
$\theta_{\rm D}/T > 15$. As a result, the thermal content of the white
dwarf gets  smaller.  This fast cooling  is expected to  take place in
very  cool white  dwarfs, that  is,  those characterized  by very  low
surface luminosities.

We can estimate  the importance of the Debye  cooling for white dwarfs
as follows.  From Eq.   (\ref{eqcool}), and approximating the specific
heat of  ions in the  Debye regime by  $ C_V^{\rm ion} \sim  3.2 \pi^4
(T_{\rm c}/\theta_{\rm  D })^3 \Re /A  $, to a  first approximation we
obtain:

\begin{eqnarray}
C \int_{t_0}^{t} dt' = -\frac{16 \pi^4}{5}\ \frac{\Re}{A}\ \frac{1}
{\theta_{\rm D}^3}\int_{T_0}^{T}\  T_{\rm c}^{-1/2}\ dT_{\rm c},
\end{eqnarray}

\noindent  where $T_0$  is the  initial temperature  from  which Debye
cooling  begins  ($T_0  < \theta_{\rm  D}$)  and  $T<T_0$.   The
integration yields the cooling time during the Debye phase

\begin{eqnarray}
t_{\rm D} \sim \frac{32 \pi^4}{5}\ \left(\frac{T}{\theta_{\rm D}}\right)^3\ 
\left[\left(\frac{T_0}{T}\right)^{1/2} - 1\right]\
\frac{TM}{L}\ \frac{\Re}{A}.  
\end{eqnarray}

Note that the cooling time during  the Debye phase is shorter than the
result obtained from Mestel´s law  (given by the last factor) when $T$
is smaller  than about  $0.1 \theta_{\rm D}$.   For instance, if  $T =
0.05 \theta_{\rm  D}$ then $t_{\rm D}$  is a factor of  4 smaller than
that obtained using Mestel's law.  It was hoped that this reduction in
the cooling times  would explain the deficit of  observed white dwarfs
below  $\sim  10^{-4.5}  L_{\odot}$.   However,  it  is  difficult  to
attribute to the onset of Debye cooling the deficiency of white dwarfs
observed at low luminosities. This  can be easily shown computing, for
a  given white  dwarf  mass,  the luminosity  at  which Debye  cooling
becomes relevant  using again the relation between  the luminosity and
the  central  temperature  \mbox{($L=C  M  T_{\rm  c}  ^{3.5}$)},  and
adopting the  central temperature corresponding to the  onset of rapid
Debye cooling (of about $0.1  \theta_{\rm D}$, as we have just shown).
We  use the  Chandrasekhar's model  to estime  the central  density to
compute  $\theta_{\rm  D}$.   Table  \ref{debye} lists,  for  a  given
stellar  mass,  the  Debye  temperature,  $\theta_{\rm  D}$,  and  the
luminosity at the  onset of the rapid Debye  cooling phase. Note that,
for a typical white dwarf,  fast Debye cooling occurs at extremely low
luminosities, and thus no observable consequences are expected in this
case.  Debye  cooling is relevant at observable  luminosities only for
very  massive white  dwarfs, which  are much  less abundant,  see Fig.
\ref{massdist}.

\begin{table}
\caption{Onset of Debye cooling.}
\label{debye}
\begin{center}
\begin{tabular}{lll}
\hline
\hline
$M\; (M_{\odot})$  & $\theta_{\rm D}$ (K) & $L\; (L_{\odot})$  \\
\hline
0.6  &$ 7.0 \times 10 ^6 $ & $10 ^{-7} $   \\
1.0  &$ 2.2 \times 10 ^7 $ & $10 ^{-5.1} $ \\
1.2  &$ 4.0 \times 10 ^7 $ & $10 ^{-4.1} $ \\
\hline
\hline
\end{tabular}
\end{center}
\end{table}

>From these simple  estimates it can be concluded  that the shortage of
dim white dwarfs is difficult to  explain in terms of a rapid cooling.
The absence of very low  luminosity white dwarfs suggests instead that
the Galaxy is not sufficiently  old to contain cooler (and thus older)
white dwarfs. This means that the cutoff of the white dwarf luminosity
function is  the result of the finite  age of the Galaxy  --- see, for
instance, Winget et al.  (1987), Garc\'\i a-Berro et al.  (1988b), and
references  therein,   for  a  discussion.   This   feature  has  been
extensively and quantitatively  explored by numerous investigations to
construct detailed  white dwarf luminosity  functions on the  basis of
sophisticated evolutionary models in order to constrain the age of the
disk of our Galaxy and to understand crucial aspects of the history of
the Galactic disk, like the past star formation rate.

\subsubsection{Physical separation processes}
\label{phasesep}

As already mentioned, when  the Coulomb coupling parameter reaches the
critical value  $\Gamma \simeq 180$, crystallization at  the center of
the white dwarf sets in.  Coulomb interactions lead quite naturally to
crystallization, and the subsequent release of latent heat affects the
evolution of  white dwarfs,  as shown in  the pioneering works  of Van
Horn (1968) and Lamb \&  Van Horn (1975).  The typical luminosities at
which  this  occurs  are  $\log (L/L_{\odot})  \lesssim  -3$.   Later,
Stevenson  (1980)  and  Mochkovitch  (1983) examined  the  release  of
gravitational   energy  associated  with   changes  in   the  chemical
composition  induced  by  crystallization in  carbon-oxygen  mixtures.
Because of  the spindle  shape of the  phase diagram  of carbon-oxygen
mixtures, the  solid formed upon  cyrstallization is richer  in oxygen
than the liquid.  As the solid (oxygen-rich) core  grows at the center
of  the white  dwarf, the  lighter carbon-rich  liquid left  behind is
efficiently redistributed  by Rayleigh-Taylor instabilities  (Isern et
al.   1997).  In  addition, the  effects of  sedimentation  during the
liquid phase  of minor  species such as  $^{22}$Ne, the  most abundant
impurity  present in  the central  regions of  a white  dwarf,  and of
$^{56}$Fe, the second most  important impurity, have also been studied
--- see Isern et  al.  (1991) and Xu \&  Van Horn (1992) respectively,
and also Deloye \& Bildsten (2002) and Garc\'\i a--Berro et al. (2008)
for  the impact  of $^{22}$Ne  sedimentation on  the cooling  times of
white dwarfs.  All  these works show that a  change in the composition
of  the  cores  of  white  dwarfs, either  by  phase  separation  upon
crystallization   or  by  sedimentation   during  the   liquid  phase,
introduces   an  additional   source   of  energy:   the  release   of
gravitational energy due to chemical differentiation.

Strong  empirical  evidence  for  the  occurrence  of  these  physical
separation processes  in the cores  of white dwarfs has  been recently
presented  in Garc\'\i a--Berro  et al.   (2010).  These  authors have
convincingly shown that these  processes constitute an important piece
of physics that has to be considered to date stellar populations using
white dwarf  cooling times.  In particular, the  contribution of phase
separation of carbon-oxygen mixtures upon crystallization on the white
dwarf  cooling times  can be  assessed as  follows.  The  local energy
budget of the white dwarf can be written as:

\begin{eqnarray}
\frac{dL_r}{dm}= -\epsilon_\nu -P\,\frac{dV}{dt}-\frac{dE}{dt},
\end{eqnarray}

\noindent where  $L_{\rm r}$  is the local  luminosity, $\epsilon_\nu$
corresponds to  the energy  per unit mass  per second due  to neutrino
losses, $V=1/\varrho$, and  $E$ is the internal energy  per unit mass.
If the white dwarf is made of two chemical species with atomic numbers
$Z_0$ and $Z_1$, mass numbers  $A_0$ and $A_1$, and abundances by mass
$X_0$ and $X_1$, respectively $(X_0+X_1=1)$, where the suffix 0 refers
to the heavier component, this equation can be written as:

\begin{equation}
-\left(\frac{dL_r}{dm}+\epsilon_{\nu}\right)=C_{\rm v}\frac{dT}{dt}+
T\left(\frac{\partial P}{\partial T}\right)\frac{dV}{dt}
-l_{\rm s}\frac{{dM}_{\rm s}}{dt}\delta(m-M_{\rm s})+
\left(\frac{\partial E}{\partial X_0}\right)\frac{dX_0}{dt},
\end{equation}  

\noindent where $l_{\rm s}$ is  the latent heat and $\dot{M}_ {\rm s}$
is  the  rate at  which  the solid  core  grows.   The delta  function
indicates that the latent heat is released at the solidification front
(Isern  et al.  1997;  Isern et  al. 2000).   Chemical differentiation
contributes  to the  luminosity not  only through  compressional work,
which  is  negligible,  but  also  via  the  change  in  the  chemical
abundances, which leads to the  last term of this equation. We mention
that, to  a first  order, the largest  contribution to $L_r$  from the
change in  $E$ cancels  the $P\,dV$ work  for any  evolutionary change
(with or without  a compositional change). This is,  of course, a well
known  result that  can be  related  to the  release of  gravitational
energy.   Integrating over  the  whole star,  the  next expression  is
obtained:

\begin{eqnarray}
L+L_{\nu}&=&- \int^{M}_0 C_{\rm v}\frac{dT}{dt}\,dm
            - \int^{M}_0 T\left(\frac{\partial P}{\partial
             T}\right)\frac{dV}{dt}\,dm \nonumber\\
          &&+\;\; l_{\rm s} \frac{dM_{\rm s}}{dt}
            - \int^{M}_0 \left(\frac{\partial E}{\partial
             X_0}\right)\frac{dX_0}{dt}\,dm
\end{eqnarray}
 
The first term  of the equation is the well  known contribution of the
heat capacity  of the  star to the  total luminosity. The  second term
represents the  contribution to  the luminosity due  to the  change of
volume.  It  is in general  small since only  the thermal part  of the
electronic pressure, the ideal part of the ions, and the Coulomb terms
contribute. However, when the white dwarf enters into the Debye regime
this term provides about the  80\% of the total luminosity, preventing
the sudden  disappearence of the  star (D'Antona \&  Mazzitelli 1990).
The third term  represents the contribution of the  latent heat to the
total  luminosity at  freezing, Since  the  latent heat  of a  Coulomb
plasma is  small, its contribution  to the total luminosity  is modest
although not negligible.  The last  term is the energy released by the
chemical readjustment of  the white dwarf, that is  the release of the
energy stored in the form of chemical potentials.  In normal stars the
last term  is usually  negligible, since it  is much smaller  than the
energy  released by  nuclear  reactions,  but it  must  be taken  into
account when  all other energy sources  are small.  It  can be further
expanded and related to  the difference between the chemical abundance
of the liquid and the solid as (Isern et al. 1997):

\begin{equation}
\int^{M}_0 \left(\frac{\partial E}{\partial X_0}\right)
\frac{dX_0}{dt}\,dm=
(X_0^{\rm sol}-X_0^{\rm liq})\,\left[\left(\frac{\partial E}
{\partial X_0}\right)_{M_{\rm s}}-\left\langle\frac{\partial E}
{\partial X_0}\right\rangle\right]\frac{dM_{\rm s}}{dt}
\label{prom}
\end{equation}

\noindent where $\left(\partial E  / \partial X_0 \right)_{M_{\rm s}}$
is evaluated at the boundary of the solid core, and

\begin{eqnarray}
\left\langle\frac{\partial E}{\partial X_0}\right\rangle=\frac{1}{\Delta
M}\int_{\Delta M}\left(\frac{\partial E}{\partial X_0}\right) dm
\end{eqnarray}

\noindent The  first term in  the square bracket in  Eq.  (\ref{prom})
represents the  energy released  in the crystallizing  layer ---  as a
result of  the increasing concentration  of oxygen --- and  the second
term  is the  energy  absorbed  on average  in  the convective  region
($\Delta  M$)  driven by  the  Rayleigh-Taylor  instability above  the
cyrstallization front, as a  result of the decreasing concentration of
oxygen.  Because $(\partial E/\partial  X_0)$ is negative --- since it
is dominated  by the ionic  contributions, which are negative  --- and
essentially depends  on the density,  the square bracket  is negative,
and thus the  process of phase separation results in  a net release of
energy.  It is clear that the  energy released by this process will be
dependent on the initial oxygen  profile at the beginning of the white
dwarf  phase, resulting  in  a  smaller contribution  in  the case  of
initially higher oxygen abundances. Note  that a change in the initial
chemical profile  may also affect the  degree of mixing  in the liquid
layers  and thus  the energy  absorbed there,  hence altering  the net
energy released by the process.

Furthermore, it  is possible to  define the total energy  released per
gram of crystallized matter due  to the change in chemical composition
as:

\begin{eqnarray}
\epsilon_{\rm g}=-(X_0^{\rm sol}-X_0^{\rm liq})
\left[\left(\frac{\partial E}{\partial X_0}\right)_{M_{\rm s}}-
\left\langle\frac{\partial E}{\partial X_0}\right\rangle\right]
\end{eqnarray}

>From this,  the decrease  in the cooling  rate introduced  by chemical
differentiation upon solidification can  be easily estimated to a good
approximation if it is assumed  that the luminosity of the white dwarf
only depends on the temperature of the nearly isothermal core. In this
case:

\begin{equation}
\Delta t = \int^{M}_0 \frac{{\epsilon}_{\rm g}(T_{\rm c})}
{L(T_{\rm c})} \;dm
\end{equation}

\noindent where $\epsilon_{\rm g}$ is  is the energy released per unit
of crystallized  mass and $T_{\rm c}$  is the temperature  of the core
when the crystallization front is located  at $m$.  For the case of an
otherwise typical  $0.6 \, M_  {\odot}$ white dwarf, the  total energy
released by chemical differentiation  amounts to $\Delta E\sim 2\times
10^{46}$  erg and  the corresponding  time delay  at $L  = 10^{-4.5}\,
L_{\odot}$ is $\Delta t\sim 1.8$ Gyr. Of course, the total increase in
the cooling times  depends on the initial chemical  profile and on the
transparency of the  envelope as well. The importance  of the envelope
is crucial for two reasons. Clearly, more transparent envelopes result
in faster energy losses and  shorter cooling times.  The second reason
is  more  subtle  and  has  been  largely  overlooked  until  recently
(Fontaine et al. 2001).   At approximately the same evolutionary stage
at  which  crystallization  sets  in,  the  external  convection  zone
penetrates the region where thermal conduction by degenerate electrons
is very efficient.  Such an occurrence (known as convective coupling),
initially produces  a further decrease  in the cooling rate,  but this
will be the subject of our next section.

\subsubsection{Convection}  

In the envelope  of cool white dwarfs, energy  is transferred not only
by radiation,  but also by convection,  which is a  consequence of the
recombination of  the main atmospheric  constituents.  With decreasing
temperature, the  base of the  convection zone gradually  moves deeper
into the star, following the  region of partial ionization. As long as
convection does not reach the  degenerate core, the convergence of the
envelope  temperature profile  to the  radiative  temperature gradient
causes the central temperature to be independent of the outer boundary
conditions, namely the chemical  stratification of the envelope or the
detailed treatment of convection.  When this is the case the heat flow
depends basically on  the opacity at the edge  of the degenerate core,
as  it  is assumed  in  the  Mestel  approximation.  However,  at  low
luminosities convection reaches the  degenerate core and the radiative
gradient convergence  is lost.   The thermal profile  of an  old white
dwarf  becomes thus  strongly constrained,  with the  consequence that
changes in the atmospheric parameters  are reflected in changes in the
core  temperature.  In fact,  the degenerate  core is  isothermal, and
convection --- which extends from  the atmosphere to the outer edge of
the  core ---  is  essentially adiabatic.   This so-called  convective
coupling modifies the relationship between the luminosity of the white
dwarf and its core temperature, and, thus, the rate of cooling of cool
white dwarfs (D'Antona \& Mazzitelli  1989; Fontaine et al. 2001).  In
particular,  the core  temperature becomes  smaller ---  convection is
much  more efficient  than  radiation at  transporting  energy ---  as
compared with  the predictions of  purely radiative envelopes.  Due to
the lower  central temperature,  the star has  initially an  excess of
thermal energy that  has to get rid of. This causes  a decrease in the
cooling  rate   at  low  luminosities.   Depending   on  the  chemical
stratification  of  the outer  layer,  convective  coupling occurs  at
different times in  the lifes of white dwarfs  (D'Antona \& Mazzitelli
1989).  This is critical to understand the different cooling speeds of
cool  white  dwarfs with  different  envelope compositions.   Finally,
convection, and the corresponding mixing episodes, plays a key role in
the  interpretation  of spectral  evolution  (the  observed change  in
surface composition  of a white dwarf  as it evolves)  of these stars.
This is particularly  true if the white dwarf is formed  with a thin H
envelope,  and  convection  penetrates  beyond the  surface  H  layers
(Fontaine et al. 2001).

\begin{figure}[t]
\centering
\includegraphics[width=0.9\columnwidth,clip=true]{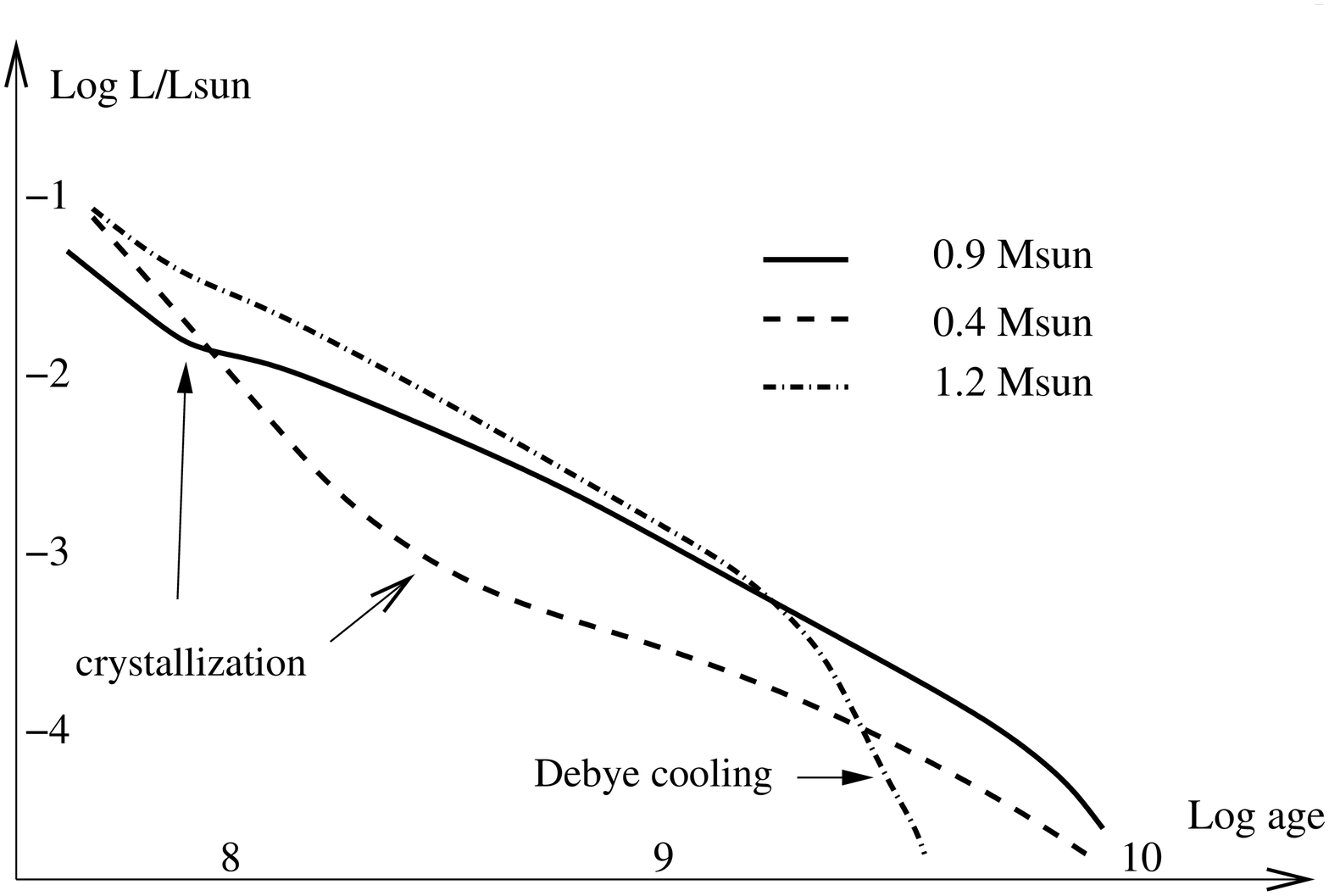}
\caption{Surface  luminosity as  a function  of the  cooling  time for
         white  dwarfs  with  a  carbon  core  and  different  stellar
         masses. The  onset of  crystallization and Debye  cooling are
         indicated.   These  cooling  curves  are  based  on  reliable
         relationships   between  the   luminosity  and   the  central
         temperature, and  better input physics than  those assumed in
         the Mestel's model. See text for details.}
\label{edad}
\end{figure}

\subsubsection{Specific heat of electrons}  

The Mestel model assumes  that electrons are completely degenerate and
so  $C_{V}^{\rm  elec}  \ll   C_{V}^{\rm  ion}$.   This  is  true  for
low-luminosity (low $T_{\rm c}$) white dwarfs, where the specific heat
of strongly degenerate electrons is $C_{V}^{\rm elec}\propto k_{\rm B}
T  / \varepsilon_F  \rightarrow 0  $.   But at  high luminosities  and
particularly for low-mass white dwarfs, the energy of electrons is not
completely  independent of  temperature. Hence,  the specific  heat of
electrons must be considered in  the calculation of the thermal energy
content of  the white dwarf.  For  instance, for a  $0.5 \, M_{\odot}$
carbon-rich white  dwarf at $T_{\rm c}  \sim 10^7$ K  ($L \sim 10^{-3}
L_{\odot}$), $C_{V}^{\rm elec} \sim 0.25 C_{V}^{\rm ion}$.

\subsubsection{Changes in the Mestel's cooling law}

All  the previously  described  physical processes  alter the  cooling
times  of white  dwarfs  given  by the  simple  Mestel's cooling  law.
Improved cooling  times can be  obtained by considering  $L-T_{\rm c}$
relations that take into account  energy transfer by convection in the
envelope of cool  white dwarfs, and a better  treatment of the opacity
and of the equation of state  in the envelope than that assumed in the
Mestel model.  These improved $L-T_{\rm c}$ relations  together with a
more realistic treatment of  the thermal energy content that considers
the effects of  Coulomb interactions (plus the release  of latent heat
on crystallization) and Debye cooling, as well as the specific heat of
electrons  can  be used  to  calculate  improved  cooling times.   The
results  of these calculations  are depicted  in Fig.   \ref{edad} ---
see, for instance, Koester (1972).  The main conclusions are:

\begin{enumerate}

\item The dependence of cooling times on the stellar mass is different
      from  that   predicted  by  the  Mestel   model  ($\tau  \propto
      M^{5/7}$).  This  is especially true at  high luminosities where
      the contribution of electrons to the specific heat increases the
      cooling  times of  low-mass white  dwarfs ---  because  of their
      smaller electron  degeneracy ---  and at low  luminosities where
      massive white dwarfs cool faster as a result of Debye cooling.

\item The impact  of crystallization on the cooling  times is strongly
      dependent  on  the  stellar   mass  and  core  composition.   In
      particular,  massive  white  dwarfs  crystallize  at  very  high
      luminosities  because  of  their  larger densities.   Thus,  the
      corresponding delay introduced by the release of latent heat has
      a smaller effect on the cooling times.
 
\item The cooling times  strongly depend on core chemical composition,
      as predicted by the Mestel's model ($\tau \propto 1/A$).

\item At  observable luminosities, Debye cooling is  relevant only for
      $M>1M_{\odot}$.

\end{enumerate}


\section{The progenitors of white dwarfs}

\subsection{Which stars become white dwarfs?} 

It is  generally accepted that  the immediate progenitors of  the vast
majority of  white dwarfs are  nuclei of planetary nebulae,  which are
the  products  of  intermediate-   and  low-mass  main  sequence  star
evolution.  Stars that  begin their lives with masses  less than about
$10\pm 2\, M_{\odot}$ are expected  to become white dwarfs (Ritossa et
al.   1999;   Siess  2007).   Comparing  this  limit   with  the  mass
distribution on the main sequence  at birth, this means that certainly
more than 95$\%$  of all stars will become  white dwarfs. Typically, a
$8\, M_{\odot}$  progenitor seems  to lead to  a remnant of  $\sim 1\,
M_{\odot}$  and  a  $1\,  M_{\odot}$  star  to  one  of  about  $0.5\,
M_{\odot}$. Because the exceedingly large time required for a low-mass
main sequence  star (less  than $0.8\, M_{\odot}$)  to become  a white
dwarf,  most white  dwarfs  with stellar  masses  smaller than  $0.4\,
M_{\odot}$  are  expected not  to  be  the  result of  single  stellar
evolution, but instead, the result of mass transfer in binary systems.

The  fact that the  maximum mass  of a  white dwarf  is about  $1.4 \,
M_{\odot}$  ---  the Chandrasekhar  limiting  mass  ---  hints at  the
occurrence of  strong mass loss during the  progenitor star evolution,
particularly  during the  AGB stage,  in agreement  with  the existing
observations for this evolutionary  phase. For instance, in the famous
Hyades cluster, stars with  about $2\, M_{\odot}$, the turn-off point,
are still on the main sequence, but the cluster contains several white
dwarfs.  These white dwarfs  have typical masses of $0.6\, M_{\odot}$,
but on the main sequence their progenitors must have been more massive
than the turn-off  point, or they would not have  evolved --- see, for
instance,  Weidemann (2000)  and references  therein.   The difference
between the turn-off  age of the cluster and the  cooling age of white
dwarfs  determines the  time that  the progenitor  of the  white dwarf
spent  on  the  main  sequence  and  on the  giant  phase  and,  thus,
determines  the original  progenitor  mass.  In  this  way a  relation
between initial and final masses can be constructed --- see, Catal\'an
et al.  (2008), Casewell et al.  (2009) and Salaris et  al. (2009) for
recent determinations of the initial and final masses as inferred from
white dwarfs in stellar clusters.

\begin{figure}
\centering
\includegraphics[width=0.9\columnwidth,clip]{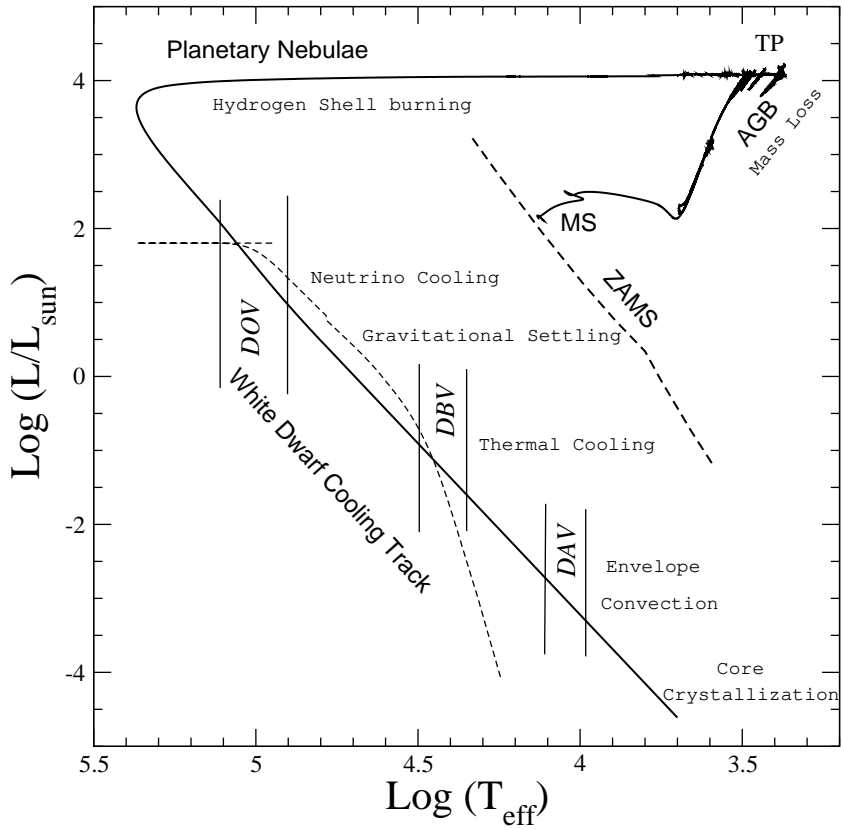}
\caption{Hertzsprung-Russell diagram for the  full evolution of a $3.5
         \,  M_{\odot}$  star  from   the  ZAMS  to  the  white  dwarf
         domain.  Mass-loss  episodes  at  the thermally  pulsing  AGB
         reduce the  stellar mass to $0.66 \,  M_{\odot}$. The various
         physical processes  which occur as white dwarfs  cool as well
         as  the domain of  the pulsating  instability strips  for the
         DOV, DBV  and DAV are  indicated.  Thin dashed  line displays
         the neutrino luminosity.}
\label{hrtrack}
\end{figure}

\subsection{Prehistory of a typical white dwarf} 

The main  phases in the life  of a typical white  dwarf progenitor are
visualized    in   the    Hertz\-sprung-Russell   diagram    of   Fig.
\ref{hrtrack}.  After the long-lived stage of central H burning during
the main sequence phase, the  progenitor star evolves to the red giant
region  to  burn  He  in  its  core.   Here,  the  carbon-oxygen  core
composition that will characterize the emerging white dwarf remnant is
built up. After the end of  core He burning, evolution proceeds to the
AGB.   There, the  He  burning  shell becomes  unstable  and the  star
undergoes recurrent thermal instabilities commonly referred to as {\sl
  thermal pulses}.  As  evolution proceeds along the AGB,  the mass of
the  carbon-oxygen  core  increases  considerably  by  virtue  of  the
outward-moving He burning shell.  Also,  during this stage most of the
remaining  H-rich envelope  is ejected  through very  strong mass-loss
episodes.  When the mass fraction of the remaining envelope is reduced
to $\sim  10^{-3}\, M_{\odot}$ the  remnant star moves rapidly  to the
left in the Hertzsprung-Russell diagram to the domain of the planetary
nebulae.  If  the departure  from the AGB  takes place at  an advanced
stage in the He shell flash cycle, the post-AGB remnant may experience
a last  He thermal pulse on  its early cooling  branch, and eventually
totally  exhausts  its residual  H  content,  thus  giving rise  to  a
H-deficient white dwarf (see below).  When the remaining H envelope is
reduced  to  $\sim  10^{-4}\,  M_{\odot}$, nuclear  energy  generation
becomes virtually extinct.   The surface luminosity decreases rapidly,
and the star enters the terminal phase of its life as a white dwarf.

The newly  formed white dwarf  is left mostly with  only gravitational
and thermal  energy sources  available.  In fact,  during most  of its
final evolution, the  gravothermal (gravitational plus thermal energy)
contribution drives  the evolution, see  Sect.  \ref{detailed}.  Since
electrons are  already degenerate in the interior,  the stellar radius
is not far from the  equilibrium radius of the zero-temperature model,
and   the   remaining  contraction   is   small,   but  not   entirely
negligible. Hence, the star evolves  almost at constant radius along a
diagonal   straight  line   in   the  white   dwarf   region  of   the
Hertzsprung-Russell diagram.

\subsection{The formation of white dwarfs with low H content}
\label{lowH}

An  important fraction  of white  dwarf  stars is  characterized by  a
H-deficient composition.  In agreement with their DA counterparts, the
mass distribution  of these stars exhibits  a sharp peak  near $0.6 \,
M_{\odot}$,  but  it lacks  almost  entirely  the  low- and  high-mass
components  --- see Sect.   \ref{hdeficient}.  The  accepted mechanism
for the formation  of most H-deficient white dwarfs  is the born-again
scenario, that is  the occurrence of a very  late thermal pulse (VLTP)
during the  early stages of white  dwarf evolution when  H burning has
almost ceased  (Fujimoto 1977; Sch\"onberner 1979; Iben  et al.  1983;
Herwig  et al.  1999;  Althaus et  al.  2005a).   During the  VLTP, an
outward-growing  convection  zone  powered  by the  He  burning  shell
develops  and  reaches the  H-rich  envelope  of  the star,  with  the
consequence that  most of  the H content  of the remnant  is violently
burned in  the He  flash-driven convection zone  (Herwig et  al. 1999;
Miller Bertolami  et al.   2006). The resulting  H-burning luminosity,
due mainly to proton captures  by $^{12}$C, may reach about $10^{11}\,
L_{\odot}$ in  a matter of  a few hours.   The star is then  forced to
evolve rapidly back to the AGB and finally into the domain of the He-,
carbon-, and oxygen-rich PG 1159 stars (Werner \& Herwig 2006) at high
$T_{\rm  eff}$.  Afterwards, the  remnant evolves  to the  white dwarf
domain,  as a  H-deficient star.   We postpone  the discussion  of the
evolutionary   properties  of  H-deficient   white  dwarfs   to  Sect.
\ref{hdeficient}.

\begin{figure}
\centering
\includegraphics[width=0.9\columnwidth,clip]{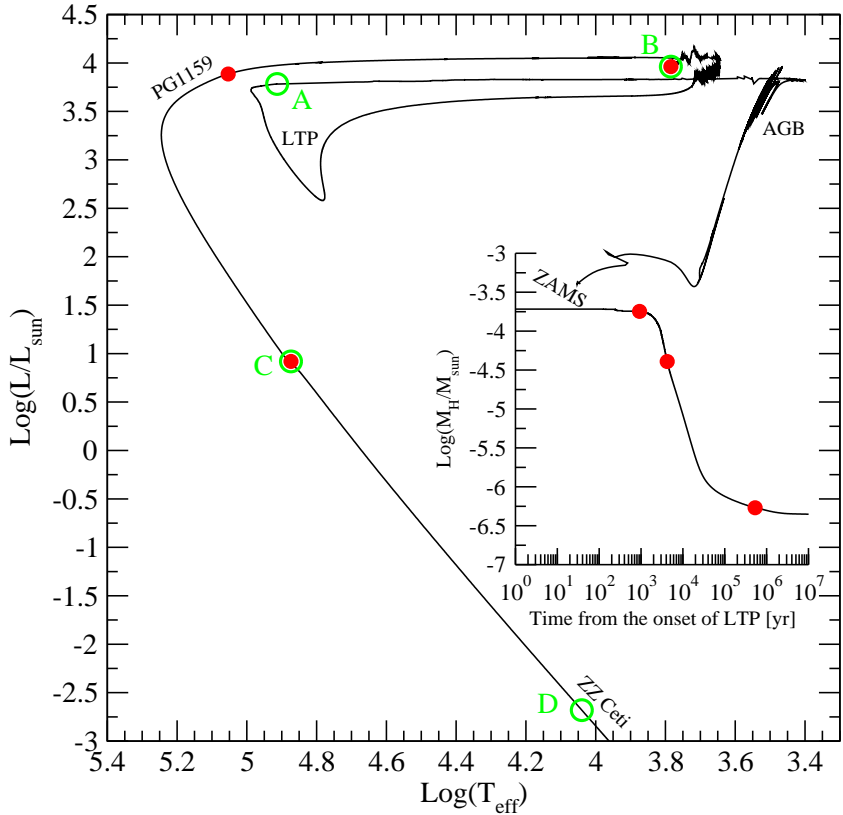}
\caption{Evolutionary scenario  for the formation of a  DA white dwarf
         with a thin H  envelope.  It is shown the Hertzsprung-Russell
         diagram for  the complete evolution of the  initially $2.7 \,
         M_{\odot}$ stellar model from the ZAMS to the ZZ~Ceti domain.
         The  star undergoes  a LTP  after  leaving the  AGB at  $\log
         T_{\rm eff}= 4.9$ (A).  As a result, the remnant evolves back
         to the  red-giant domain where  the H envelope is  diluted by
         surface convection  (B).  Inset: temporal evolution  of the H
         mass in solar units.   Filled dots correspond to those stages
         indicated  along  the  evolutionary  track.   Most  of  H  is
         processed by  nuclear burning during the  PG~1159 stage. From
         Miller Bertolami et al. (2005).}
\label{ltp}
\end{figure}

Late  thermal pulses (LTP)  may also  lead to  the formation  of white
dwarfs  with small  H-rich envelopes.  In contrast  to a  VLTP,  a LTP
occurs before  the remnant star  reaches the white dwarf  stage, while
the H burning  shell is still active. This has  the consequence that H
will not be completely and violently burnt. Specifically, in a LTP the
low  H  content results  from  the  dilution  of the  residual  H-rich
envelope  as a  result of  a dredge-up  episode ---  when  the remnant
returns to the giant phase  --- and following stable proton burning at
high effective temperatures.  After a LTP episode, a large fraction of
the residual  H envelope is burnt.   This is an  important point since
there  is mounting evidence  that some  DA white  dwarfs appear  to be
characterized  by H-rich  envelopes many  orders of  magnitude smaller
than those predicted by the standard theory of stellar evolution.  The
hypothesis  that LTPs  may be  responsible for  the existence  of such
white  dwarfs   has  been   quantitatively  explored  by   Althaus  et
al.  (2005b), who have  shown that  a large  fraction of  the original
H-rich  material of  a post-AGB  remnant is  indeed burned  during the
post-LTP evolution, with  the result that, on the  white dwarf cooling
track  the remaining  H envelope  becomes $10^{-6}  \,  M_{\odot}$, in
agreement with  asteroseismological inferences for some  ZZ Ceti stars
--- see section \ref{ZZceti} in this  review and also Winget \& Kepler
(2008) and Castanheira \& Kepler (2009).

This evolutionary  scenario is illustrated in  Fig. \ref{ltp}. Shortly
after the  occurrence of the LTP  (A), the stellar  remnant returns to
the region  of giant stars, where  H is diluted  by surface convection
and  mixed  inwards with  the  underlying  intershell region  formerly
enriched in He, carbon  and oxygen (B). Afterwards, evolution proceeds
into  the domain of  the central  stars of  planetary nebulae  at high
$T_{\rm eff}$ to become a hybrid  PG~1159 star.  In the meantime, H is
reignited  and, after the  point of  maximum effective  temperature is
reached on the  early white dwarf cooling track (C),  the H content is
reduced to $8 \times 10^{-7}  \, M_{\odot}$, that is, about two orders
of magnitude  smaller than the amount  of H predicted  by the standard
theory of stellar evolution.  This value is the amount of H with which
the star enters the white dwarf domain.  The temporal evolution of the
H content is illustrated in the inset of Fig.  \ref{ltp}.  Most of the
residual  H material  is burnt  over  a period  of roughly  100,000~yr
during the PG~1159 stage.  On  the cooling track, H diffuses outwards,
turning the  white dwarf into one  of the DA  type with a thin  pure H
envelope of a few $10^{-7} \, M_{*}$.

\subsection {The formation of low-mass white dwarfs} 

About 10\% of  the white dwarf population is  characterized by stellar
masses below $0.4 \, M_{\odot}$  (Liebert et al.  2005).  A handful of
such  white dwarfs  have masses  below  $0.2 \,  M_{\odot}$ (Kawka  \&
Vennes  2009).   These low-mass  white  dwarfs  constitute a  separate
sequence of white dwarfs with He cores (because the core mass is below
that required for He ignition), necessarily resulting from binary star
evolution.  Indeed, the Galaxy is  not old enough for these objects to
have  formed  via single-star  evolution.   Instead,  a companion  has
stripped  the envelope  from  the now-visible  white  dwarf before  it
completed  its  red  giant  evolution.   The binary  nature  has  been
confirmed  by the  numerous  discoveries of  periodic radial  velocity
variations of the  H$\alpha$ lines of low-mass DA  white dwarfs (Marsh
et  al.   1995).  Low-mass  white  dwarfs  are  usually found  as  the
companions to  millisecond pulsars (van  Kerkwijk et al.   2005). This
offers  the possibility  of  constraining the  ages  and initial  spin
periods  of the  millisecond pulsars  (Driebe et  al. 1998;  Hansen \&
Phinney 1998;  Althaus et al.  2001;  Benvenuto \& De  Vito 2005).  In
fact, millisecond pulsars  are thought to be recycled  during the mass
transfer stage.   When mass transfer  ends, the pulsar begins  to spin
down, and  from then on,  the ages of  the white dwarf and  the pulsar
component as  inferred from  its rate of  period variation  (spin down
age) should be the same. Finally, the existence of some low-mass white
dwarfs  ($\sim 0.3\,  M_{\odot}$) with  carbon-oxygen  cores resulting
from single-star evolution is not discarded (Prada Moroni \& Straniero
2009).


\section{Detailed models of white dwarf evolution}
\label{detailed}

Mestel's  model with  the  improvements we  have previously  described
provides a  reasonable description of white  dwarf evolution. However,
if we want to use white  dwarfs as reliable cosmic clocks to infer the
age  of  stellar populations,  we  need  to  consider more  elaborated
treatments that include  all the potential energy sources  and the way
energy  is  transported  within  the  star.  The  main  points  to  be
considered are:

\begin{enumerate}

\item The  thermal  and   hydrostatic  evolution have  to  be  treated
      simultaneously. In fact, the  structure of a white dwarf changes
      with time  in response to  the energy radiated. This  means that
      pressure is a function of density {\sl and temperature}.

\item The   thermal energy  is  not the  only  source  of white  dwarf
      luminosity.   There  are additional  energy  sources and  sinks:
      residual  contraction  of  the  outer layers,  nuclear  burning,
      neutrino  losses,  and the  gravitational  energy released  from
      physical  separation processes  in the  core  like carbon-oxygen
      phase  separation upon  crystallization and  diffusion  of minor
      species in the liquid phase.

\item The  core  of  a  white dwarf  is   never  strictly  isothermal,
      particularly in hot white dwarfs.

\item Changes in chemical composition during white dwarf evolution due
      to convective mixing, diffusion processes, nuclear reactions and
      accretion are expected to influence the cooling of white dwarfs.

\end{enumerate}

A proper  treatment of these  issues requires a detailed  knowledge of
the previous  history of the  white dwarf.  This is  particularly true
regarding the calculation of the early phases of white dwarf evolution
at high luminosities where contraction is not negligible. In addition,
the mass  and chemical  composition of the  core and outer  layers ---
which, as we saw, have a marked influence on the white dwarf evolution
--- are  specified  by  the  evolutionary history  of  the  progenitor
star. Thus,  an accurate treatment  of white dwarf  evolution requires
the  use  of complete  stellar  evolutionary  codes  to calculate  the
evolutionary history of the progenitor stars all the way from the main
sequence,  to the  mass-loss  and planetary  nebulae  phases. In  what
follows, we  will comment on  the main aspects  to be considered  in a
detailed treatment of white dwarf evolution.

\subsection{Chemical abundance distribution}
\label{chemical}

\begin{figure}[t]
\centering
\includegraphics[width=0.9\columnwidth,clip]{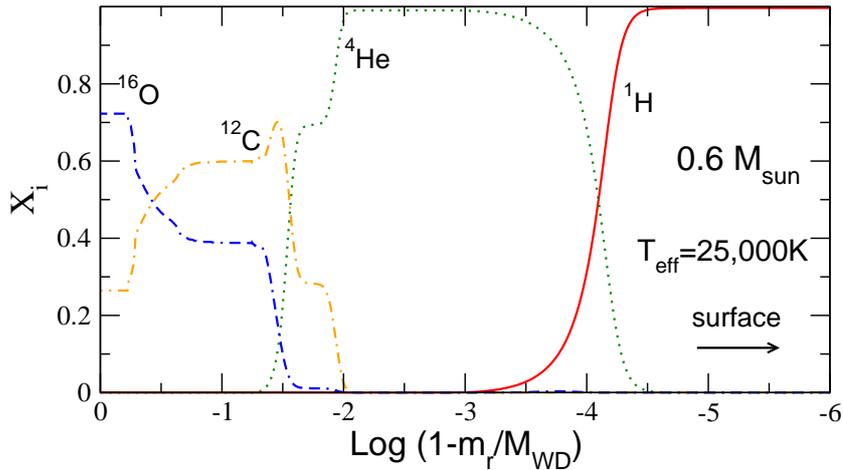}
\caption{$^1{\rm H}$, $^4{\rm He}$, $^{12}{\rm C}$, and $^{16}{\rm O}$
         distribution within a typical DA  white dwarf in terms of the
         outer mass fraction \mbox{$q=1-m/M$}.}
\label{quimi}
\end{figure}

The chemical composition of a white dwarf is determined by the nuclear
history  of  the white  dwarf  progenitor.   The chemical  composition
expected in a typical DA  white dwarf is displayed in Fig. \ref{quimi}
in  terms  of the  outer  mass fraction  $q$.   The  election of  this
coordinate strongly  emphasizes the outer layers of  the model.  Three
different regions  can be distinguished: the  carbon-oxygen core (95\%
of the  mass), which is the  result of convective He  core burning and
the subsequent  steady He burning shell of  prior evolutionary stages.
Because of  the larger temperatures  in the helium burning  shell, the
oxygen abundance  decreases in the outer regions  of the carbon-oxygen
core.  Possible extra-mixing episodes  during core He burning --- like
overshooting and  semiconvection (Straniero et al.  2003)  --- and the
existing  uncertainties in  the  $^{12}{\rm C}(\alpha,\gamma)^{16}{\rm
  O}$  reaction rate turn  the carbon-oxygen  composition of  the core
into  one  of the  main  sources  of  uncertainty weighting  upon  the
determination  of  white  dwarf  cooling  times.   The  core  chemical
composition depends also on the initial stellar mass of the progenitor
star. Lower  oxygen abundances are generally expected  in more massive
white dwarf progenitors. In the  case of massive white dwarfs --- that
is, those with  $M \gtrsim 1.05 M_{\odot} $  --- oxygen-neon cores are
expected (Ritossa et al.  1996), while very low mass white dwarfs, are
characterized by  He cores.  On top  of the carbon-oxygen  core is the
He- and  C- rich intershell, which  is built up by  mixing and burning
episodes during the  last thermal pulse on the  AGB.  The abundance of
carbon  in the  intershell region,  which stems  from  the short-lived
convective mixing  that has driven the carbon-rich  zone upward during
the  peak  of  the last  helium  pulse  on  the  AGB, depends  on  the
occurrence of overshooting in the  He-flash convection zone and on the
number of previous thermal pulses.   Except for the less massive white
dwarfs,  this intershell  region is  eroded by  diffusion by  the time
evolution has  reached the  domain of the  pulsating ZZ Ceti  stars by
$T_{\rm  eff}  \approx$12,000K (Althaus  et  al.   2010a).  Above  the
intershell and below the H  envelope, the He-rich buffer which results
from prior H  burning can be found.  During the AGB,  the mass of this
buffer goes from  almost zero --- at the pulse  peak when the He-flash
convection zone  is driven  close to the  base of  the H layer  --- to
$0.01 \, M_{\odot}$.

The theory of stellar evolution  provides upper limits for the mass of
the  various  intershells.  For  a  typical  white  dwarf of  $0.6  \,
M_{\odot}$,  the  maximum  H  envelope  mass  that  survives  the  hot
pre-white dwarf  stages is about  $10^{-4}\, M_{\odot}$ and  the total
mass of the  He buffer and the He-intershell amounts  to $\sim 0.02 \,
M_{\odot}$.  These values  strongly depend on the stellar  mass of the
white  dwarf  (Renedo et  al.  2010), and  on  the  occurrence of  LTP
episodes, which  may reduce the  mass of the  remnant H and  He buffer
considerably, see Sect. \ref{lowH}.  Asteroseismological inferences on
individual pulsating white dwarfs provide in some cases constraints to
the thickness  of the H and  He envelopes and the  core composition of
white    dwarfs    (Winget   \&    Kepler    2008).   For    instance,
asteroseismological  studies of  DAVs  carried out  by Castanheira  \&
Kepler (2008, 2009) indicate that the mass of the H envelope is in the
range $10^{-4}  \gtrsim M_{\rm H}/M_*  \gtrsim 10^{-10}$, with  a mean
value of  $M_{\rm H}/M_*= 5  \times 10^{-7}$, thus suggesting  that an
important  fraction of  DAs characterized  by  envelopes substantially
thinner than  those predicted by  standard evolution could  exist.  We
emphasize that the initial  chemical stratification of white dwarfs is
not known with sufficient detail  from the stellar evolution theory or
from observations.  This, in turn,  leads to some uncertainties in the
evaluation of the cooling times.

\subsection{Changes in the chemical abundance distribution}
\label{changesx}

There  are  numerous  physical   processes  that  alter  the  chemical
abundance  distribution  with which  a  white  dwarf  is formed.   The
effects  of these  changes  on the  evolutionary  properties of  white
dwarfs may be  important depending on the cooling  stage or luminosity
at which they occur.  The most important of these processes is element
diffusion.   In   particular,  gravitational  settling   and  chemical
diffusion  strongly   influence  the  abundances   produced  by  prior
evolution. Gravitational settling is responsible for the purity of the
outer layers that characterize most white dwarfs.  In fact, because of
the  extremely large  surface gravity  of white  dwarfs, gravitational
settling  rapidly leads to  the formation  of a  pure H  envelope, the
thickness of which gradually  increases as evolution proceeds. In some
cases, the occurrence of accretion processes may lead to the existence
of  cool white  dwarf with  traces of  heavy elements  in  their outer
layers.  At  the chemical interfaces, characterized  by large chemical
gradients,  chemical diffusion strongly  smears the  chemical profile,
see Fig.   \ref{quimi}.  Here, the diffusion time  scale is comparable
to the  white dwarf evolutionary  time scale, so  equations describing
diffusion  have  to  be   solved  simultaneously  with  the  equations
describing white  dwarf evolution (Iben \& Macdonald  1985; Althaus et
al. 2003, 2005a).

Under  certain  circumstances,  diffusion  processes may  trigger  the
occurrence of  thermonuclear flashes  in white dwarfs.  In particular,
note  from  Fig.  \ref{quimi}  that  inside the  He  buffer,  chemical
diffusion has led to  a tail of H extending from the  top layers and a
tail of carbon  from the bottom layers.  If the  white dwarf is formed
with a thin  enough He-rich buffer, a diffusion-induced  H shell flash
may  be initiated,  thus leading  to the  formation of  a self-induced
nova, during  which the white  dwarf increases its luminosity  by many
orders of  magnitude in  a very short  time (Iben \&  MacDonald 1986).
During this  process, the  mass of  the H envelope  is believed  to be
strongly  reduced.   Another  physical  process  that  may  alter  the
chemical composition  of a  white dwarf is  convection.  Specifically,
for  cool white  dwarfs with  H  envelopes thin  enough ($<10^{-6}  \,
M_{\odot}$), convective  mixing will  lead to dilution  of H  with the
underlying He  layer, thus leading  to the formation of  He-rich outer
layers. In  addition, changes in the outer  layer chemical composition
are  expected during  the hot  stages of  white dwarf  evolution  as a
result   of   residual   nuclear   burning.  Finally,   an   efficient
redistribution  of   carbon  and  oxygen   upon  crystallization,  and
sedimentation of minor species such as $^{22}$Ne are expected to occur
in the  white dwarf core, with important  energetics consequences, see
sections \ref{phasesep} and \ref{additional}.

\subsection{Energy sources of white dwarfs}

As  we  saw  in  Sect.  \ref{mestel}, white  dwarf  evolution  can  be
described essentially  as a cooling  process where the source  of star
luminosity  is approximately provided  by the  change in  the internal
energy  stored  in the  ions.  However,  there  are additional  energy
sources and sinks  that may considerably impact the cooling times, and
hence they must be taken into account in detailed calculations:

\subsubsection{Gravitational energy}

Although white  dwarfs evolve at  almost constant radius, the  role of
contraction is by no means  negligible. The radius at the beginning of
the  white  dwarf  phase  can  be up  to  twice  the  zero-temperature
degenerate  radius, and thus  the contribution  of compression  to the
energy output of  the star can be important in  very hot white dwarfs.
In addition, changes in the internal density distribution owing to the
increase in the core mass from H  burning via the CNO cycle lead to an
important release  of gravitational potential energy from  the core of
pre-white  dwarfs   (or  very  young   white  dwarfs).   Gravitational
contraction  is also relevant  in the  final evolutionary  phases when
Debye cooling has  already depleted the thermal energy  content of the
core.   Here,  the residual  contraction  of  the thin  subatmospheric
layers  may  provide up  to  30\% of  the  star  luminosity.  But,  as
mentioned  in   \ref{mestel},  for  most  of   white  dwarf  evolution
compression barely contributes to  the surface luminosity.  This stems
from the fact that the compression work released in the core is almost
completely employed  by degenerate  electrons to increase  their Fermi
energy.
 
\subsubsection{Nuclear energy}

Stable H shell burning via the  CNO cycle and He shell burning are the
main  source of  luminosity during  the evolutionary  stages preceding
white  dwarf  formation.  As  a  result  of  CNO burning,  the  H-rich
envelope  is  consumed. Thus,  below  $\sim  100\, L_{\odot}$  nuclear
burning  becomes a minor  contribution to  surface luminosity  --- the
density and temperature at the  base of the H-rich envelope become too
low once  $M_{\rm H}\sim  10^{-4}\, M_{\odot} $.   Thus, in  a typical
white  dwarf, the  role of  nuclear burning  as a  main  energy source
ceases as soon as the hot  white dwarf cooling branch is reached.  But
it never stops  completely, and depending on the  stellar mass and the
exact  amount  of  H  left   by  prior  evolution,  it  may  become  a
non-negligible  energy  source  for  old DA  white  dwarfs.   Detailed
evolutionary   calculations  show  that,   for  white   dwarfs  having
low-metallicity  progenitors, stable H  burning via  the proton-proton
chains may contribute  by more than 50\% to  the surface luminosity by
the time  cooling has proceeded  down to luminosities ranging  from $L
\sim 10^{-2}\, L_{\odot}$ to  $10^{-3}\, L_{\odot}$ (Iben \& MacDonald
1985).  More  recent  full  calculations  show  that  this  H  burning
contribution  at such  low luminosities  reaches 30\%  in the  case of
white  dwarfs  resulting   from  progenitors  stars  with  metallicity
$Z=0.001$  (Renedo et al.  2010).  However,  predictions of  the exact
value of $M_{\rm  H}$ and hence of the role of  residual H burning are
tied  to the  precise mass  loss history  along the  previous  AGB and
post-AGB phases,  and particularly to  the occurrence of  late thermal
pulses.  In this  sense it is worth mentioning  that a small reduction
in the  H envelope by a  factor of about  2 with respect to  the upper
theoretical limit is sufficient to  strongly inhibit H burning, a fact
that explains  the different role  of H burning obtained  by different
authors.  Hence,  a correct assessment  of nuclear burning  during the
white  dwarf stage  requires the  computation of  the  pre-white dwarf
evolution, something that has not been possible until recently.

During  the early stages  of white  dwarf evolution,  unstable nuclear
burning may also be relevant. As we mentioned in \ref{lowH}, a late He
thermonuclear  flash  may  take   place  after  H  burning  is  almost
completely  extinguished,  thus  eventually  leading to  a  born-again
episode and  to the  consumption of  most of the  H content.   Also, a
diffusion-induced H shell flash may occur at intermediate luminosities
on  the white dwarf  cooling track,  resulting in  the formation  of a
self-induced  nova  (see   Sect.  \ref{changesx}).   Again,  a  proper
assessment of these  nuclear burning episodes and their  impact on the
subsequent white dwarf evolution  requires a detailed knowledge of the
history of the progenitor star.

Finally, stable  H burning may  be dominant in low-mass  white dwarfs,
delaying  their cooling  for significant  periods of  time  (Driebe et
al. 1998; Althaus et al. 2001).   These stars are the result of binary
evolution  and  are characterized  by  thick  H  envelopes.  In  these
low-mass white dwarfs, H burning can also become unstable, thus giving
rise  to  thermonuclear  flashes   at  early  stages  of  white  dwarf
evolution.  This  bears important consequences  for the interpretation
of low-mass white dwarf companions to millisecond pulsars.

\subsubsection{Neutrino losses}

Energy is lost from white dwarfs  not only in the form of photons, but
also through the  emission of neutrinos. Neutrinos are  created in the
very deep interior of white dwarfs  and provide a main energy sink for
hot white dwarfs.  Because of their extremely small cross-sections for
interaction with matter --- at  a central density of $10^6$ gr/cm$^3$,
typical of  white dwarfs,  the mean free  path of neutrinos  is $l_\nu
\approx 3,000  R_{\odot}$ --- once created, neutrinos  leave the white
dwarf core without interactions,  carrying away their energy. In young
white  dwarfs,  neutrinos increase  the  cooling  rate  and produce  a
temperature inversion.  Neutrinos  result from pure leptonic processes
as  a   consequence  of  the  electro-weak   interaction.   Under  the
conditions prevailing in hot  white dwarfs the plasma-neutrino process
is   usually  dominant,   but  for   massive  white   dwarfs  neutrino
bremsstrahlung must also be taken into account (Itoh et al. 1996).

\subsubsection{Additional energy sources}
\label{additional}

As  discussed  in  Sect.  \ref{phasesep}, the  energy  balance  during
crystallization  of  the  white  dwarf  core must  take  into  account
additional energy  sources that  have to be  considered in  a detailed
treatment  of white dwarf  evolution.  This  comprises the  release of
latent heat  and the release  of gravitational energy  associated with
changes  in  the  carbon-oxygen  profile  induced  by  crystallization
(Salaris  et al.  1997).   As previously  explained  in that  section,
partial  separation of  carbon and  oxygen, and  hence changes  of the
initial  distribution of these  elements, is  expected to  occur after
crystallization.   The resulting  release  of gravitational  potential
energy depends on the initial fractions of carbon and oxygen and their
distribution in the  interior. The impact of these  two energy sources
on the  white dwarf cooling times  depends on the stellar  mass of the
white  dwarf, resulting  less relevant  for the  case of  more massive
white dwarfs, due to their larger luminosities at cyrstallization.

Finally, the gravitational settling of  minor species in the core also
make a contribution to the white dwarf luminosity.  In particular, the
slow diffusion  of $^{22}$Ne in  the liquid white dwarf  core releases
enough gravitational  potential energy as to impact  the cooling times
of massive  white dwarfs (Deloye \&  Bildsten 2002; Garc\'{\i}a--Berro
et al. 2008; Althaus et al.  2010b).  In fact, $^{22}$Ne, which is the
product  of He  burning on  $^{14}$N during  prior evolution,  has two
extra  neutrons (relative  to $A=2Z$).  This results  in  an imbalance
between the  gravitational and  electric fields, and  leads to  a slow
gravitational settling of $^{22}$Ne  towards the center.  As predicted
by  Deloye  \&  Bildsten  (2002),  the possible  impact  of  $^{22}$Ne
sedimentation  on white  dwarf cooling  could be  better seen  in old,
metal-rich clusters,  such as NGC 6791, where  the $^{22}$Ne abundance
expected in  the cores of its white  dwarfs could be as  high as $\sim
4\%$ by  mass. As a matter  of fact, the occurrence  of these physical
separation   processes  ---   carbon-oxygen   phase  separation   upon
crystallization and  $^{22}$Ne sedimentation ---  in the core  of cool
white  dwarfs and the  associated slow  down of  the cooling  rate has
recently been demonstrated by Garc\'\i  a--Berro et al. (2010) to be a
fundamental aspect  to reconcile  the longstanding age  discrepancy in
NGC 6791 (Bedin et al. 2008).

\subsection{Energy transport in the outer layers}

Although very  thin, the non-degenerate outer layers  control the rate
at which energy flows from  the interior into space.  Because of this,
a detailed knowledge of the processes responsible for energy transport
in the  outer layers constitutes  a crucial aspect for  an appropriate
assessment  of the cooling  times (D'Antona  \& Mazzitelli  1990). The
treatment  is  difficult  because   it  involves  energy  transfer  by
radiation  and/or  convection  in  a partially  degenerate,  partially
ionized and highly non-ideal gas. This difficulty becomes particularly
relevant  in  cool white  dwarfs,  where  convection  has reached  the
degenerate core.  In this case,  the central temperature, and thus the
cooling  of  white dwarfs,  will  be  influenced  by the  thermal  and
pressure  profiles of  the  subatmospheric layers,  which  have to  be
provided by non-gray model atmospheres. Over the years, detailed model
atmospheres  that include  non-ideal effects  in the  gas  equation of
state  and chemical equilibrium,  and a  complete treatment  of energy
absorption  processes  such as  collision-induced  opacity, have  been
developed (Bergeron et al. 1991;  Saumon \& Jacobson 1999; Rohrmann et
al. 2002).   In particular, the  use of non-gray model  atmospheres to
derive  outer  boundary  conditions  gives  rise  to  shallower  outer
convection zones, as compared with  the standard gray treatment of the
atmosphere.  Finally,  as a result of  collision-induced absorption by
molecular  collisions  at  low  effective temperatures,  the  emergent
spectrum departs significantly from the blackbody emission and becomes
blue as  the effective  temperature is lowered  (Saumon et  al.  1994;
Hansen 1998; Rohrmann 2001),  a prediction corroborated by deep Hubble
Space Telescope exposures in globular clusters (Richer et al. 2006).

\begin{figure}[t]
\centering
\includegraphics[width=0.8\columnwidth,clip]{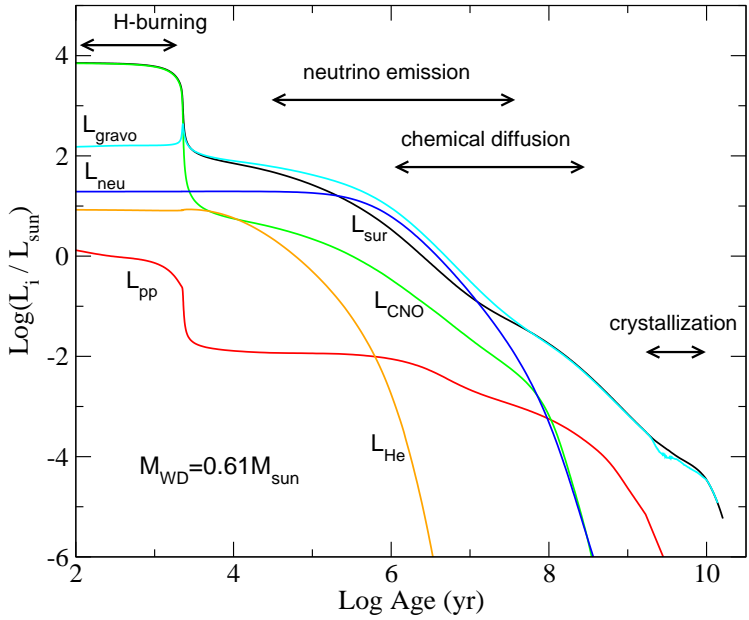}
\caption{Time dependence of the different luminosity contributions for
         a $0.61 \, M_{\odot}$ white dwarf: photon luminosity, $L_{\rm
         sur}$,  luminosity due  to nuclear  reactions (proton-proton,
         $L_{\rm pp}$, CNO bicycle, $L_{\rm CNO}$, He burning, $L_{\rm
         He}$),   neutrino  losses,   $L_{\rm  neu}$,   and   rate  of
         gravothermal  (compressional  plus  thermal)  energy  release
         $L_{\rm gravo}$. The white dwarf progenitor corresponds to an
         initially  $2 \, M_{\odot}$  star with  metallicity $Z=0.01$.
         Calculations include element diffusion, the release of latent
         heat upon  cyrstallization, and outer  boundary conditions as
         given by non-gray  model atmospheres.  The different physical
         processes  of  revelance  during  white dwarf  evolution  are
         indicated.}
\label{lumi}
\end{figure}

\subsection{Results from detailed calculations} 

The time dependence  of the luminosity contributions due  to H burning
($L_{\rm  nuc}$)  via proton  proton  and  CNO  nuclear reactions,  He
burning ($L_{\rm He}$),  neutrino losses ($L_{\rm neu}$), gravothermal
energy ($L_{\rm gravo}$) --- the release of thermal plus gravitational
potential energy --- and  photon emission (surface luminosity, $L_{\rm
  sur}$) for a typical white  dwarf is shown in Fig.  \ref{lumi}.  The
evolution from the  planetary nebulae stage to the  domain of the very
cool white dwarfs  is shown.  Luminosities are in  solar units and the
age is expressed in years from the moment at which the remnant reaches
$\log T_{\rm eff}=4.87$ at high luminosity.

At early times,  that is, for young white  dwarfs, nuclear burning via
CNO is  the first contributor to  the surface luminosity  of the star.
Here, there is a near  balance between $L_{\rm nuc}$ and $L_{\rm sur}$
and between $L_{\rm  gravo}$ and $L_{\rm neu}$ (like  in an AGB star).
During  this  stage,  $L_{\rm  gravo}$  results from  the  release  of
gravitational  potential energy owing  to the  change in  the internal
density caused  by the increase in  the core mass from  CNO H burning.
This a short-lived  phase (a few thousand years)  and, thus, given the
long-lived cooling times of white  dwarfs, it is totally negligible in
terms of age.  Nevertheless, this  phase is important as it configures
the final thickness of the  hydrogen-rich envelope of the white dwarf.
After $\sim 10^{4}$ yr  of evolution, nuclear reactions abruptly cease
and  the star  begins to  descend to  the white  dwarf domain  and the
surface luminosity of the star begins to decline steeply.  Thereafter,
the  evolution is  dictated  by  neutrino losses  and  the release  of
gravothermal energy --- now essentially the release of internal energy
from the  core.  Nuclear  burning (via CNO)  is a  minor contribution.
Note that during  this stage, neutrino losses are  the dominant energy
sink  ($L_{\rm gravo}\approx  L_{\nu}$).  In  fact,  $L_{\nu}\approx 5
L_{\rm  sur}$ and  as a  result the  white dwarf  cooling  is strongly
accelerated.  To  this stage  belong the pulsating  DOV and  DBV white
dwarfs, so  an eventual  measurement of the  rate of period  change in
some  of these  stars  would allow  to  constrain the  plasma-neutrino
production  rates (Winget  et  al.   2004).  The  time  scale for  CNO
burning is larger than the  evolutionary time scale. Burning occurs in
the tail of H distribution and as  a result the mass of the H envelope
is not reduced by burning during this stage.

\begin{figure}[t]
\centering
\includegraphics[width=0.8\columnwidth,clip]{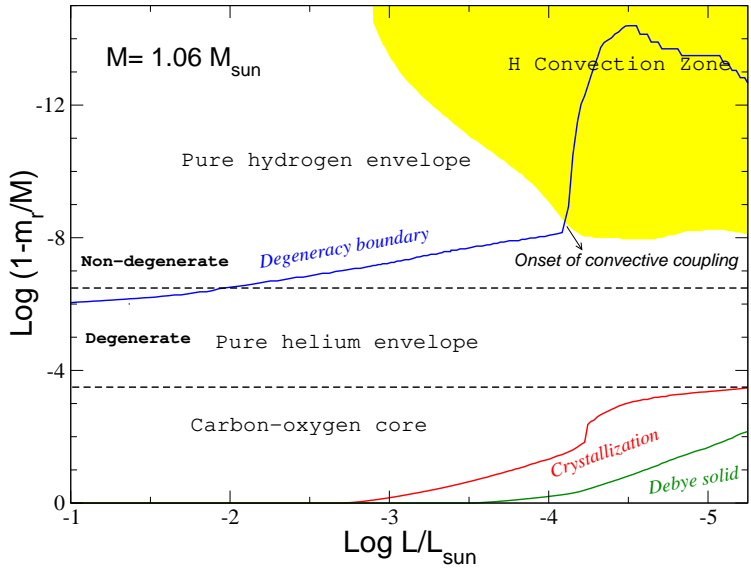}
\caption{The  logarithm of  the  outer mass  fraction  of an  evolving
         massive white dwarf as  a function of the surface luminosity.
         The location of the different chemical regions and the domain
         of  the  physical  processes  relevant for  the  white  dwarf
         evolution are indicated. }
\label{interior}
\end{figure}

When $t_{\rm cool}\sim 10^{7}$ yr, the white dwarf is cool enough that
neutrino is no longer relevant. $L_{\rm gravo}\approx L_{\rm sur}$ and
the white dwarf verifies the  Mestel approximations.  This is the best
understood stage in  the life of a white dwarf. During  it the star is
characterized by a core still in the weakly-coupled Coulomb regime and
a  radiative  envelope.  There  is  no  convection,  nor neutrinos  or
nuclear burning or crystallization, and  the ion thermal energy is the
main source of energy.  Below $\log(L/L_{\odot}) \sim -3$, the picture
changes  appreciably, as  exemplified in  Fig. \ref{interior}  for the
case of a massive white dwarf. At this luminosity, envelope convection
becomes  relevant  and  core  crystallization  sets  in.   As  already
mentioned, crystallization  leads to the  release of both  latent heat
and  gravitational  energy   from  phase  separation,  and  convection
eventually  leads  to  the  onset  of convective  coupling,  with  the
consequent release  of excess thermal  energy, and the  resulting slow
down  in  the  cooling   rate.   In  intermediate-mass  white  dwarfs,
convective  coupling  and  crystallization  take place  more  or  less
simultaneously, thus markedly changing the slope of the cooling curve.
However,  in massive  white dwarfs,  convective coupling  occurs after
most  of the  white dwarf  has crystallized,  as can  be seen  in Fig.
\ref{interior}  (Renedo et  al.  (2010).  In  this  case, both  energy
contributions  influence the  cooling rate  at  different evolutionary
stages.  In addition, during  these stages, appreciable energy release
from $^{22}$Ne  sedimentation in the  white dwarf core takes  place in
the  case of white  dwarfs stemming  from metal-rich  progenitors (see
\ref{additional}), notably  impacting the rate of  cooling (Althaus et
al. 2010b).

The  proton-proton burning  shell  may still  make  a contribution  to
surface luminosity at advanced  stages, thus reducing the cooling rate
of the  white dwarf.   As already discussed,  this is relevant  in the
case of white dwarfs  with low-metallicity progenitors, which are left
with larger  H envelopes  on the cooling  track.  For a  typical white
dwarf resulting from a  solar-metallicity progenitor, as shown in Fig.
\ref{lumi},  residual proton-proton burning  is a  minor contribution.
Eventually,  H  burning  becomes   virtually  extinct  at  the  lowest
luminosities, and  the surface luminosity  is given by the  release of
internal energy from the core.  During these stages the contraction of
the  very  outer layers  may  contribute  to  the surface  luminosity.
Finally,  in the  case of  massive white  dwarfs, the  onset  of Debye
cooling  with the  consequent  rapid decrease  in  the thermal  energy
content of  the white dwarf core,  is expected to  occur at observable
luminosities, as shown in Fig.  \ref{interior}.


\section{H-deficient white dwarfs}
\label{hdeficient}

\subsection{PG 1159 stars and DO white dwarfs}

White dwarf stars with  H-deficient and He-rich atmospheres constitute
about 15\%  of the  white dwarf population,  and are usually  known as
non-DA white  dwarfs, see Sect.   \ref{spectro}.  Most of  these white
dwarfs are  thought to be  the progeny of  PG 1159 stars  (Dreizler \&
Werner  1996; Unglaub \&  Bues 2000;  Althaus et  al. 2005a),  hot and
luminous  stars  with H-deficient  and  He-,  carbon- and  oxygen-rich
surface  layers that  constitute a  transition stage  between post-AGB
stars  and most  of the  H-deficient  white dwarfs  (Werner \&  Herwig
2006).  Typical mass abundances observed  in PG 1159 stars are $X_{\rm
  He}\simeq 0.33$,  $X_{\rm C}\simeq 0.5$ and  $X_{\rm O}\simeq 0.17$,
though notable  variations are  found from star  to star  (Dreizler \&
Heber  1998; Werner  2001).   A large  fraction  of PG  1159 stars  is
believed to  be the  result of  a born-again episode  --- a  VLTP, see
Sect.  \ref{lowH} ----  which causes a hot white  dwarf star to return
to the AGB and then to the  domain of the central stars of a planetary
nebula  as a  hot H-deficient  object and  with a  surface composition
resembling  that of  the  intershell region  chemistry  of AGB  stars.
Eventually, gravitational  settling acting during the  early stages of
white dwarf evolution causes He to float and heavier elements to sink,
giving rise  to a He-dominated surface,  and turning the  PG 1159 star
into a DO white dwarf (Unglaub  \& Bues 2000).  During this stage, the
evolution of the  star is dictated essentially by  neutrino losses and
the release  of gravothermal energy (O'Brien \&  Kawaler 2000; Althaus
et al.  2005a).  Because these  white dwarfs are the hottest ones, the
evolution  during  the  DO  stage  proceeds very  fast,  with  typical
evolutionary time scales $\sim$1 Myr.

\begin{figure}
\centering
\includegraphics[width=0.8\columnwidth,clip]{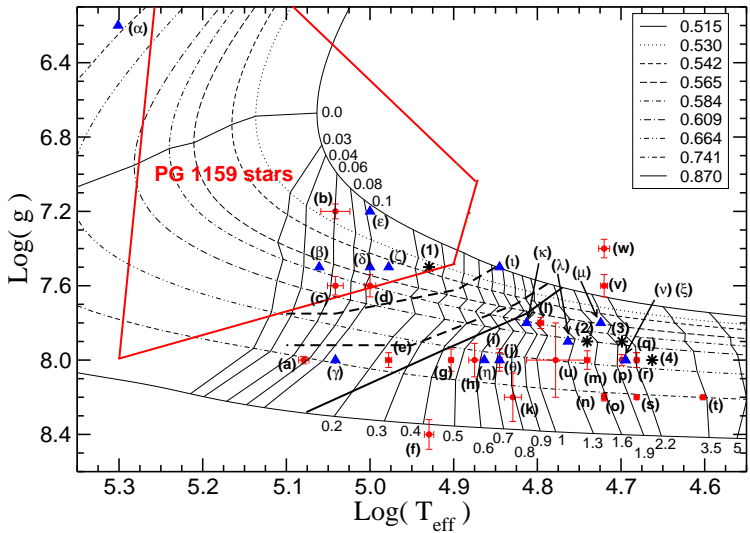}
\caption{Surface gravity, $g$ (in  cm/s$^2$), as a function of $T_{\rm
         eff}$ for the H-deficient white dwarf sequences of Althaus et
         al.   (2009a).   From bottom  to  top,  curves correspond  to
         sequences  with   stellar  masses  from   0.870  to  $0.515\,
         M_{\odot}$.  Also plotted are isochrones ranging from 0.03 to
         5 Myr measured from  the highest effective temperature point.
         The       observed      DO       white       dwarfs      with
         spectroscopically-determined    effective   temperature   and
         gravity as  analyzed by H\"ugelmeyer et  al. (2006), Dreizler
         \& Werner (1996), and Dreizler  et al.  (1997) are also shown
         (filled   circles   and    triangles,   and   star   symbols,
         respectively).  The thick solid  line corresponds to the wind
         limit for  PG 1159 stars  taken from Unglaub \&  Bues (2000),
         and the two thick dashed lines to the wind limits for PG 1159
         stars  with  H/He=  0.1  and  0.01  (upper  and  lower  line,
         respectively).  The region occupied  by PG 1159 stars is also
         shown. From Althaus et al. (2009a).}
\label{dos}
\end{figure}

Several  works have been  devoted to  the identification  and spectral
analysis of H-deficient  objects like PG 1159 and  DO white dwarfs ---
see Dreizler \& Werner (1996),  Dreizler et al.  (1997) and references
therein for  earlier studies  of spectroscopically confirmed  DO white
dwarfs.  With the  advent of large surveys, like  the SDSS, the number
of   spectroscopically-identified  H-deficient  stars   has  increased
considerably ---  see H\"ugelmeyer  et al.  (2006)  and Kepler  et al.
(2007) for DO and DB white dwarfs, respectively.  The determination of
basic parameters of  H-deficient white dwarfs, such as  their mass and
age, requires detailed evolutionary calculations appropriate for these
stars.    Such  evolutionary   calculations  need   to   consider  the
evolutionary  history  that  leads  to the  formation  of  H-deficient
stars. Moreover, if consistent  mass determinations of DO white dwarfs
and PG 1159 stars are to  be done, such calculations need to cover the
entire evolutionary stage  from the domain of luminous  PG 1159 to the
hot white  dwarf stage (Dreizler \& Werner  1996). Evolutionary tracks
and mass-radius relations appropriate for both non-DA white dwarfs and
PG 1159 stars have recently  been computed by Althaus et al.  (2009a).
These sequences have been derived from models which cover the complete
evolutionary history of progenitor stars with different initial masses
evolved through the born-again  episode and an up-to-date constitutive
physics.  The  evolution of these  white dwarf sequences in  the $\log
T_{\rm eff}-\log  g$ plane is  shown in Fig.  \ref{dos}  together with
the effective  temperature and gravity  of all known DO  white dwarfs,
and  the approximate  location of  PG 1159  stars in  this  plane. The
departure of the white  dwarf cooling tracks from the zero-temperature
predictions is noticeable during the early evolutionary stages at high
effective   temperatures,  particularly  in   the  case   of  low-mass
sequences.  Note that  the bulk of observed DO  white dwarfs have ages
between 0.6 and 2.5 Myr, though  the hottest members of the sample are
indeed very young, with ages less than 0.10 Myr.

\begin{figure}
\centering
\includegraphics[width=0.8\columnwidth,clip]{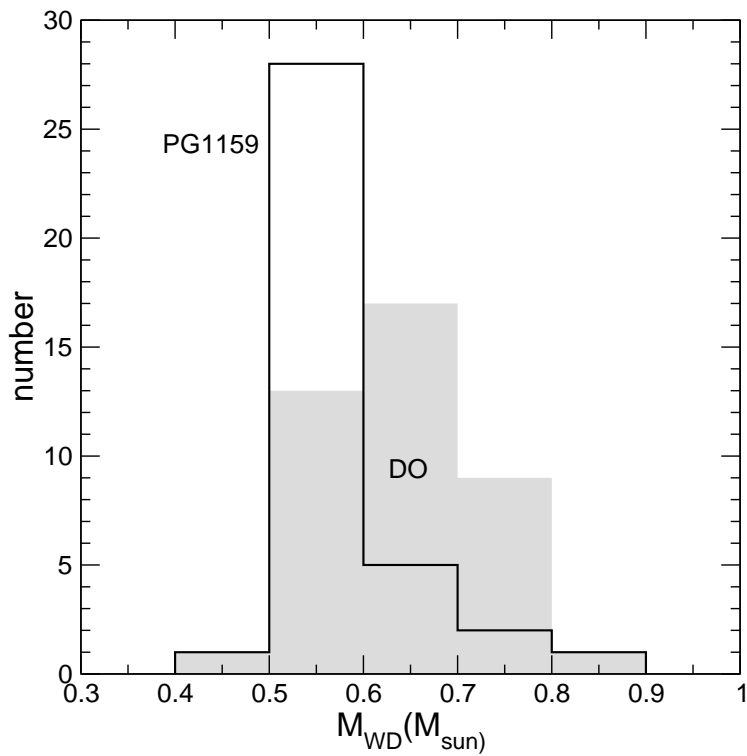}
\caption{The solid  line and the shaded  histogram show, respectively,
         the  mass  distribution for  the  PG  1159  and DO  stars  as
         inferred from  the evolutionary calculations  for these stars
         by Althaus et al.  (2009a).  Clearly, massive DO white dwarfs
         outnumber   massive   PG   1159   stars.  From   Althaus   et
         al. (2009a).}
\label{histo}
\end{figure}

The mass distribution for DO  white dwarfs and PG 1159 stars resulting
from the evolutionary tracks  shown in Fig.  \ref{dos}, is illustrated
in Fig.  \ref{histo}, shaded region and solid line, respectively.  The
size of  the bins  in both distributions  is $0.1\,  M_{\odot}$.  Note
that massive DOs  outnumber massive PG 1159 stars.  A  mean DO mass of
$0.644 \,  M_{\odot}$ is obtained,  considerably higher (by  $0.071 \,
M_{\odot}$)   than  the  mean   mass  of   PG  1159   stars,  $0.573\,
M_{\odot}$. A possible explanation  for this difference is simply that
evolution through  the PG 1159 stage proceeds  considerably faster for
massive stars (Miller Bertolami  \& Althaus 2006), thus decreasing the
possibility of detection.  However,  analyzing the PG 1159+DO group as
a whole, it  can be seen that the difference  in the mass distribution
reflects the  existence of some other  important evolutionary channels
operating within  PG 1159  and DO stars  (Althaus et al.  2009a).  For
instance, it is possible that not all PG 1159 stars evolve to DOs.  In
particular,  those PG  1159 stars  resulting from  a LTP  --- low-mass
remnants are more  prone to experience a LTP  episode --- are expected
to evolve  into DA white dwarfs  (near the winds limits  shown in Fig.
\ref{dos} with  solid dashed  lines), thus avoiding  the DO  stage, as
shown by Unglaub \& Bues (2000).  This is so because, in contrast to a
VLTP event,  in a  LTP, H is  not burned,  but instead diluted  to low
surface abundances (Miller Bertolami \& Althaus 2006).  Alternatively,
some DO white  dwarfs could result from evolutionary  channels that do
not  involve only  PG 1159  stars.  For  instance, they  could  be the
result of post-merger evolution  involving the giant, H-deficient RCrB
stars, via  the evolutionary link RCrB\ $\rightarrow$  EHe (extreme He
stars)\  $\rightarrow$ He-SdO$^+$ $\rightarrow$O(He)  $\rightarrow$ DO
--- see Rauch et  al. (2006) for a connection  between O(He) stars and
RCrB stars.  This  is reinforced by the recent study  by Werner et al.
(2008a,b) of the star KPD0005+5106,  the hottest known DO white dwarf.
These authors  present evidence that KPD0005+5106 is  not a descendant
of PG  1159 stars, but  more probably related  to the O(He)  stars and
RCrB stars.

The  group  of   young  and  hot  DO  white   dwarfs  is  particularly
relevant. Note that they are located considerably above the wind limit
for PG  1159 stars (see Fig.   \ref{dos}).  In fact,  above this line,
mass loss  is large enough  to prevent gravitational  settling. Hence,
the transformation of  a PG 1159 star into a DO  white dwarf should be
expected  approximately  below  this  line.   Although  the  estimated
mass-loss rates  could be overestimated by  more than a  factor of ten
--- in which  case the transition from PG  1159 to DO would  be near a
line with $\log  g\approx 7.5$ (Unglaub \& Bues 2000)  --- it could be
possible that the  less massive and young DOs  are not the descendants
of PG  1159 stars. Also relevant is  the fact that for  ages less than
about 0.3--0.4  Myr, the  DO white dwarf  population is  markedly less
massive than at greater ages.  Indeed the number of detected old, less
massive DOs  is smaller  than the number  of young, less  massive DOs,
despite the  evolutionary timescales  being a factor  about 5  for the
former.   This  could indicate  that  some  of  the less  massive  DOs
experience  a  tranformation in  their  spectral  type after  $\approx
0.3-0.4$ Myr of evolution.  In  particular, traces of H could turn the
white dwarf spectral  type from DO to DA as  a result of gravitational
settling.   An approximate location  where this  transformation should
occur is provided by the  thick dashed lines shown in Fig.  \ref{dos},
which  provides the  wind  limits  for PG  1159  stars with  different
initial H abundances  (Unglaub \& Bues 2000). It  is possible that the
origin of  low-mass, hot DOs can  be connected with the  merger of two
low-mass, He-core white dwarfs (Guerrero et al.  2004; Lor\'en-Aguilar
et al.   2009), which will evolve  to become a  low-gravity extreme He
star and then a hot subdwarf (Saio \& Jeffery 2000). The presence of H
expected in this case, could lead  to a transformation of the DO white
dwarfs into DA ones.

\begin{figure}
\centering
\includegraphics[width=0.9\columnwidth,clip]{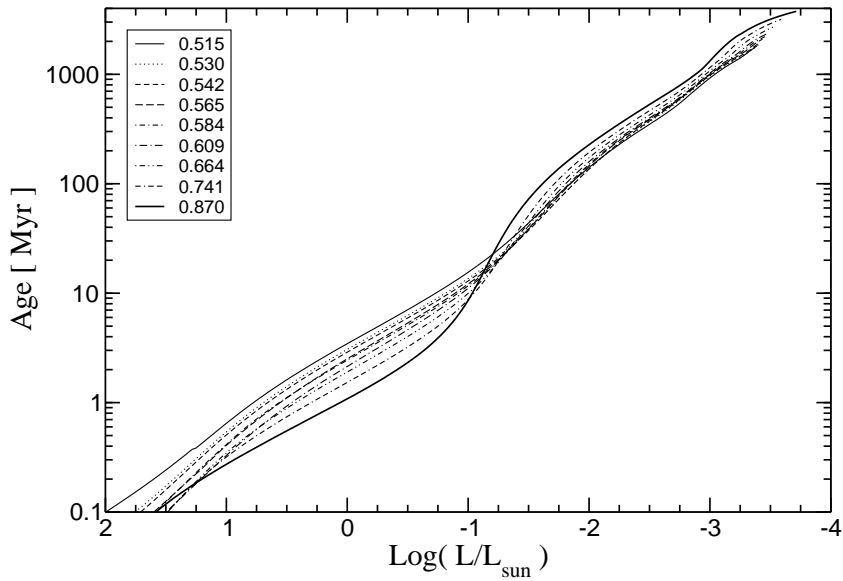}
\caption{Age  (in Myr)  versus  luminosity (in  solar  units) for  the
         H-deficient  white   dwarf  sequences  from   Althaus  et  al
         (2009a).}
\label{edad_he}
\end{figure}

Finally, in  Fig.  \ref{edad_he}  it is  shown the age  (in Myr)  as a
function  of  the luminosity  (in  solar  units)  for the  H-deficient
sequences  of Althaus  et al.   (2009a).  At  very  high luminosities,
residual He  shell-burning constitutes an appreciable  fraction of the
surface  luminosity.  Except  for these  very high  luminosity stages,
evolution during the  early stages is driven by  neutrino emission and
the release  of gravothermal  energy.  In particular,  neutrino losses
exceed photon luminosity during most of the hot white dwarf stages. As
a matter of  fact, except for the hot and  low-gravity DO sample shown
in Fig.  \ref{dos}, it is  during this ``neutrino epoch'' that most DO
white  dwarfs are observed.   Note from  Fig.  \ref{edad_he}  that the
imprints of neutrino emission on the cooling curve are more noticeable
as the  mass increases.   Less massive sequences  are older  than more
massive sequences for $\log(L/L_{\odot})  \gtrsim -1$.  At this stage,
neutrino emission  arrives at  its end, thus  causing a change  in the
slope  in the  cooling  curve around  $\log(L/L_{\odot})\sim -1$,  and
later for the less  massive sequences.  At lower luminosities, Coulomb
interactions  become  more important,  increasing  the specific  heat,
until crystallization  and the associated  release of energy  start at
the  center.  At this  point, cooling  can be  well understood  on the
basis  of the  simple cooling  theory of  Mestel, which  predicts less
massive white dwarfs to cool faster (see Sect.\ref{mestel}).

\subsection{Hot DQ white dwarfs}
\label{hotdq}

The recent discovery of a new lukewarm population of white dwarfs with
carbon-rich  atmospheres ---  known as  hot DQs  (Dufour et  al. 2007;
2008a)  --- has  sparked the  attention  of researchers  since it  has
raised the possibility of the  existence of a new evolutionary channel
of formation of H-deficient white dwarfs.  Dufour et al.  (2008a) have
reported  that nine  hot white  dwarfs identified  in the  Fourth Data
Release  of the  SDSS are  characterized by  atmospheres  dominated by
carbon.  The existence of these new white dwarfs --- all of them found
in a  narrow effective temperature  strip (between $\approx$  18,000 K
and 24,000 K)  --- poses a challenge to  the stellar evolution theory,
which cannot adequately explain their origin. As proposed by Dufour et
al.  (2008a)  the hot DQ population  could be related to  the very hot
($\approx 200,000$ K) and massive  ($0.83 \, M_{\odot}$) member of the
PG  1159  family,  the  enigmatic  star H1504+65,  for  which  a  post
born-again  origin   is  not   discarded  (Althaus  et   al.   2009c).
Interestingly enough, H1504+65  is the only known star  with no traces
of either  H or  He until the  discovery of  the hot DQ  white dwarfs.
Dufour et al.  (2008a) have outlined an evolutionary scenario in which
undetected traces of He remaining in the carbon- and oxygen-rich outer
layers of  H1504+65 would diffuse  upwards leading to a  He-rich white
dwarf.  In  this picture,  a carbon-rich atmosphere  should eventually
emerge  as  the  result  of  convective mixing  at  smaller  effective
temperatures. The  first quantitative assessment  of such evolutionary
scenario  was  presented in  Althaus  et  al.  (2009b). These  authors
demonstrated that mixing between the outer He convection zone with the
underlying  convective carbon  intershell  gives rise  to white  dwarf
structures  with the  appropriate effective  temperatures  and surface
compositions inferred from the observation  of hot DQs.  With the help
of detailed born-again simulations  for massive remnants, the study of
Althaus et al.  (2009b) provides the first theoretical evidence for an
evolutionary  link  between  the  H-  and He-deficient  PG  1159  star
H1504+65    with    the    high-gravity    DQ   white    dwarf    SDSS
J142625.70+575218.4, a  connection that can  be traced back  to strong
mass-loss episodes  after the  VLTP.  For the  intermediate-gravity DQ
white dwarfs, the scenario  predicts the formation of carbon-dominated
envelopes  for residual  He  contents  of the  order  of $M_{\rm  He}=
10^{-8}\,  M_{*}$  or smaller.   The  existence  of H-deficient  white
dwarfs with such small He content could be indicative of the existence
of a new  evolutionary scenario for the formation  of white dwarfs, as
suggested by  Dufour et al.   (2008a).  Alternatively, in view  of the
evolutionary  connection between  H1504+65 and  the  high-gravity SDSS
J142625.70+575218.4,  it cannot  be discarded  that the  hot  DQ white
dwarf population  of intermediate-gravity could be  the descendants of
some PG  1159 stars  with small He  contents. Finally, we  recall that
some of the possible  evolutionary paths that H-deficient white dwarfs
may follow are outlined in Fig. \ref{spectral}.


\section{Stellar pulsations and variable white dwarfs}
\label{pulsations}

Until about a couple of decades ago, almost all what we knew about the
properties  of stars was  derived from  observations of  their surface
layers.   Fortunately,  the  handsome  circumstance  that  many  stars
pulsate opened a new opportunity  to probe the interior of stars.  The
study of  stellar pulsations, also called {\sl  stellar seismology} or
{\sl  asteroseismology},  is  a  relatively  young  field  of  stellar
astrophysics that, by means  of the confrontation between the observed
frequencies   (periods)  of  pulsating   stars  and   the  appropriate
theoretical  models, is  able  to infer  details  about their  origin,
internal structure and evolution. The larger the number of frequencies
detected in a given  star, the more powerful asteroseismology becomes.
Asteroseismological inferences  are not limited  to global quantities,
such as gravity, effective  temperature or stellar mass. They provide,
in  addition, information  about  the internal  rotation profile,  the
chemical composition, the  run of the local sound  speed, the presence
and strength of magnetic fields, the extent of convective regions, and
several other  interesting properties.  It  is important to  note that
the main observable  of asteroseismology, the oscillation frequencies,
are the quantities that most accurately one can measure for a star.

Stellar  pulsations are  the eigenmodes  of stars,  that  is, standing
waves characterized  by specific discrete values of  frequency --- the
eigenfrequencies  of the  star ---  and the  associated time-dependent
perturbations   in  displacement,  pressure,   density,  gravitational
field,\ldots  are  called eigenfunctions.   Each  star  has an  unique
spectrum of  eigenfrequencies, which  is fixed by  the details  of its
structure.  Pulsation  modes that maintain the  spherical symmetry are
called radial modes.   They are the simplest oscillations  that a star
is able  to sustain, and are  the type of pulsations  that undergo the
classical  variable  stars  such   as  Cepheids  and  RR  Lyra.   When
pulsations  depart  from  spherical  symmetry, the  modes  are  called
non-radial modes.   Radial pulsations can  be considered as  a special
case of non-radial  modes.  In recent years it  has been realized that
different  kinds of  stellar  objects at  different  locations in  the
Hertzsprung-Russell   diagram   and   evolutionary   stages,   undergo
non-radial pulsations:  solar-like, $\gamma$ Doradus,  $\delta$ Scuti,
roAp, SPB,  $\beta$ Cephei, R CrB,  WR/LBV, V361 Hya,  V1093 Her, and,
relevant for  this review, variable  white dwarf stars.   For specific
details, we refer  the reader to the excellent  reviews of Cox (1980),
Unno et al.   (1989), Brown \& Gilliland (1994),  and Gaustchy \& Saio
(1995, 1996).

\begin{figure}
\centering
\includegraphics[width=0.9\columnwidth,clip]{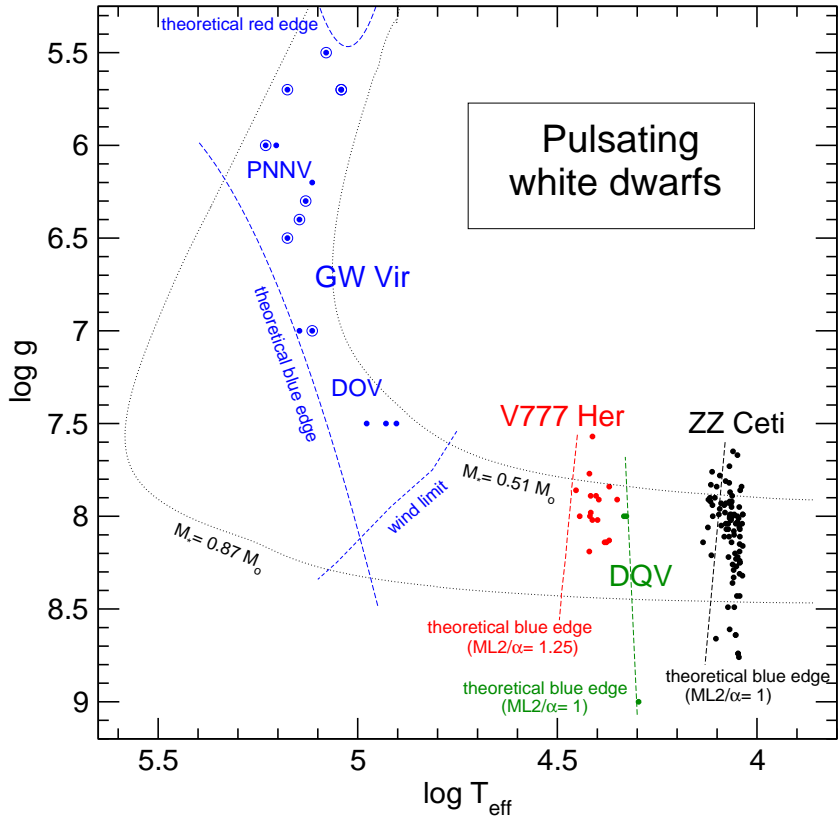}
\caption{The location of the  several classes of pulsating white dwarf
         stars in the $\log  T_{\rm eff}-\log g$ plane.  Two post-VLTP
         (H-deficient)    evolutionary   tracks   are    plotted   for
         reference.  Also shown is  the theoretical  blue edge  of the
         instability  strip  for  the   GW  Vir  stars  (C\'orsico  et
         al. 2006), the V777 Her  stars (C\'orsico et al.  2009a), the
         DQV stars  (C\'orsico et  al. 2009d), and  the ZZ  Ceti stars
         (Fontaine \& Brassard 2008).}
\label{teff-g-plane}
\end{figure}
 
In the course  of their lives, white dwarfs  cross several instability
strips  where   they  become  non-radial   pulsating  variable  stars.
Pulsations manifest  themselves as  periodic bright variations  in the
optical and also  in the FUV regions of  the electromagnetic spectrum.
Typical light curves show peak-to-peak amplitudes between 0.4 mmag and
0.3 mag. Pulsating white dwarfs are found at four regimes of effective
temperature and  gravity (see Fig.   \ref{teff-g-plane}).  The hottest
known class  ($80,000$ K  $\leq T_{\rm eff}  \leq 180,000$ K  and $5.5
\leq \log g \leq 7.5$)  is constituted by variable H-deficient, C-, O-
and He-rich atmosphere pre-white  dwarf stars, named pulsating PG 1159
stars.  This family includes some objects that are still surrounded by
a nebula, they  are the variable planetary nebula  nuclei, designed as
PNNVs. Pulsating  PG 1159  stars that lack  a nebula are  called DOVs.
Both  groups (DOVs and  PNNVs) are  frequently referred  to as  GW Vir
variable stars after  the prototype of the class,  \pp\ (McGraw et al.
1979).  Going to substantially lower effective temperatures and higher
gravities ($22,000$ K  $\leq T_{\rm eff} \leq 29,000$  K and $7.6 \leq
\log  g \leq  8.2$), we  find the  pulsating DB  (He-rich atmospheres)
white  dwarfs,  also called  DBVs  or  V777  Her variable  stars,  the
existence  of  which was  theoretically  predicted  by  Winget et  al.
(1982b)  before  their  discovery.    The  next  class  of  degenerate
pulsators are  the cool ($10,500$ K  $\leq T_{\rm eff}  \leq 12,500$ K
and  $7.8 \leq  \log g  \leq 8.8$)  pulsating DA  (H-rich atmospheres)
white dwarfs, also called DAVs or  ZZ Ceti variable stars.  It was the
first class of  pulsating white dwarfs to be  observed (Landolt 1968).
Between the DBV  and the DAV instability strips,  we find the recently
discovered class of hot DQ  variable white dwarfs, or DQVs ($19,000$ K
$\leq T_{\rm eff} \leq 22,000$ K).   They are white dwarfs with C- and
He-rich  atmospheres.  The  analysis of  the prototype  of  this class
(SDSS  J142625.71+575218.3)  carried  out  by  Green  et  al.   (2009)
appeared  to   confirm  the  pulsating  nature  of   this  star,  thus
unequivocally ruling  out the  interacting binary hypothesis  that was
initially put forward  by Montgomery et al.  (2008a)  in the discovery
paper.   Table \ref{tabla-properties},  adapted  from O'Brien  (2003),
presents  a summary  of  the  main characteristics  of  each class  of
pulsating white dwarf stars. The  first column corresponds to the name
of each  class and the  year of discovery  of the first member  of the
group, the second  column shows the number of members,  columns 3 to 6
list  the  visual  magnitudes,  effective  temperatures,  periods  and
amplitudes,  respectively,  and  the  last column  shows  the  driving
mechanism supposed to act in each case.

\begin{table*}
\scriptsize
\caption{Properties of pulsating white dwarfs.}
\begin{tabular}{ccccccc}
\hline
\hline
\noalign{\smallskip}
Class & Known     & Visual     &  $T_{\rm eff}$ & Periods & Amplitudes &  Driving \\
      & pulsators & magnitudes &  [kK]          & [min]   & [mag]      & mechanism \\
\noalign{\smallskip}
\hline
\noalign{\smallskip}
PNNV     & 10  & $11.8-16.6$        &  $110-170$  & $7-100$    & $0.01-0.15$   & $\kappa-\gamma$ (C/O)       \\
(1984)   &     &                    &             &            &               &                             \\
         &     &                    &             &            &               &                             \\
DOV      &  6  & $14.8-16.7$        &  $80-160$   & $5-43$     & $0.02-0.1$    & $\kappa-\gamma$ (C/O)       \\
(1979)   &     &                    &             &            &               &                             \\
         &     &                    &             &            &               &                             \\
V777 Her & 20  & $13.6-16.7$        & $22.4-28.4$ & $2-18$     & $0.05-0.3$    & $\kappa-\gamma$ (He{\sc ii})\\
(1982)   &     &                    &             &            &               & (convection?)               \\
         &     &                    &             &            &               &                             \\
DQV      &  3  & $17.7-19.6^{\dag}$ &  $19-22$    & $4-18$     & $0.005-0.015$ & $\kappa-\gamma$ (C)         \\
(2008)   &     &                    &             &            &               & (convection?)               \\
         &     &                    &             &            &               &                             \\
ZZ Ceti  & 145 & $12.2-16.6$        & $10.4-12.4$ & $1.6-23.3$ & $0.01-0.3$    & $\kappa-\gamma$ (H{\sc i})  \\
(1968)   &     &                    &             &            &               & convection                  \\
\noalign{\smallskip}
\hline
\hline
\end{tabular}
\\
{\footnotesize $^{\dag}g$ magnitude of the SDSS $ugriz$ system.}
\label{tabla-properties}
\end{table*}

Among the several  classes of pulsating stars, the  study of pulsating
white dwarfs is probably the  area that has experienced the most rapid
progress.  According  to the classical  review of Winget  (1988), this
has been  the result  of several fortunate  facts: (i)  the relatively
simple  physical   structure  of  white  dwarfs,   combined  with  the
apparently slow  rotation rate and weak magnetic  field, which ensures
simplicity  also in  their  pulsation properties;  (ii) the  pulsation
amplitudes are small  enough as to be treated in  the framework of the
linear theory, but large enough  still to be readily detectable; (iii)
the multiperiodic nature of white dwarf pulsations, that provides many
independent clues to their underlying stellar structures; and (iv) the
periods are typically short  enough, thus enabling observers to detect
many  cycles in  a  single  run. For  recent  excellent reviews  about
pulsating  white dwarf  stars we  refer the  reader to  the  papers of
Winget \& Kepler (2008) and Fontaine \& Brassard (2008).

\subsection{Brief history of discovery}

A full  account of the history  of discovery of  pulsating white dwarf
stars can  be found  in the work  of Vuille  (1998) and in  the review
articles of Winget (1988) and  Winget \& Kepler (2008).  The notion of
pulsations  in white dwarfs  goes back  to the  mid-twentieth century,
when  Sauvenier-Goffin (1949)  and Ledoux  \&  Sauvenier-Goffin (1950)
explored for  the first time  the possibility that white  dwarfs could
undergo  stellar  pulsations.   They  found  that  rapid  variability,
associated to radial modes driven by H burning, should be exhibited by
white  dwarfs. The absence  of high  frequency variability  typical of
radial  modes in these  stars led  to the  conclusion that  the energy
source of white dwarfs could not be nuclear reactions.  This, in turn,
clarified the role  of release of thermal energy as  the main agent of
white  dwarf evolution,  paving  the  way to  the  development of  the
Mestel's (1952) model of white dwarf cooling (Sect.  \ref{mestel}).

The  first  variable  white  dwarf,  HL Tau  76,  was  serendipitously
discovered by A. Landolt (Landolt 1968), who did not even suspect that
the  finding would  result later  in a  new and  fascinating  field of
stellar research. In fact, in  the abstract of his paper, Landolt just
announced:  ``Photoelectric  data  which  point to  a  $12^{\rm  m}.5$
variation   in  the   brightness  of   a  white-dwarf-like   star  are
discussed''.  The luminosity variations detected  in HL Tau 76, with a
period of about 740 s, were too long to be attributed to radial modes,
according to the theoretical  studies available at that time (Faulkner
\& Gribbin 1968; Ostriker \&  Tassoul 1968).  Two other variable white
dwarfs  were discovered  in the  meantime by  Lasker \&  Hesser (1969,
1971): G  44$-$32 (with periods  in the range  600$-$800 s) and  R 548
(with  periods  in  the  range  200$-$300 s).   The  newly  discovered
variabilities in  these compact objects were  attributed to non-radial
gravity  modes, rather than  purely radial  modes (Warner  \& Robinson
1972;  Chanmugam  1972),  in  agreement  with the  at  that  available
theoretical estimations  of Baglin \&  Schatzman (1969) and  Harper \&
Rose (1970).

Further confirmation that  the variability of white dwarfs  was due to
non-radial  gravity  modes  came  from  the  detection  of  rotational
splitting in R  548 (Robinson et al. 1976),  where rotation breaks the
mode  degeneracy  and  each  frequency  is split  into  multiplets  of
frequencies.  McGraw (1979) carried out a very important observational
work that provided a strong confirmation of the gravity-mode nature of
variable  white  dwarfs.    The  multicolour  Str\"{o}mgen  high-speed
photometric  data of that  work suggested  that the  bright variations
were  entirely  due to  surface  temperature  variations,  and not  to
changes  in radius,  in  line with  the  fact that  gravity modes  are
characterized  by a predominantly  tangential displacement  (see Sect.
\ref{overview}).  The  very important work  by Robinson et  al. (1982)
provided  unequivocal  theoretical  support   to  the  idea  that  the
luminosity  variations   are  entirely  due   to  surface  temperature
variations, and that the variations  of radius are as small as $\delta
R_*/R_*  \sim 10^{-4}$.   Note that  the lineal  theory  of non-radial
pulsations does not provide any clue about the value of the fractional
change in  radius, since the  governing equations are  homogeneous and
the normalization of the eigenfunctions is arbitrary (Cox 1980).

A  breakthrough  in  the  field  of white  dwarf  pulsations  was  the
foundation  of the  Whole Earth  Telescope consortium  (Nather  et al.
1990),  most  commonly known  as  WET,  the  first of  several  global
networks  for asteroseismology  that enabled  to  obtain uninterrupted
light curves spanning one to two weeks, thus avoiding the regular gaps
in the data  caused by the inevitable regular  appearance at each site
of  the  Sun.   Details  about  WET and  its  first  applications  are
described in Nather et al.  (1990) and Winget et al.  (1991, 1994).

The study of pulsating  white dwarf stars through asteroseismology has
in the recent years enabled  astronomers to access a wealth of details
about the  internal structure of white  dwarfs, otherwise inaccessible
by means  of traditional techniques.  Before going  on the pulsational
properties  of  white dwarfs  and  the  asteroseismological tools  and
applications,  we   briefly  review   below  the  theory   of  stellar
pulsations, with emphasis in non-radial gravity-mode pulsations.

\section{Overview of non-radial pulsations}
\label{overview}

In  what  follows,  we   briefly  describe  the  basic  properties  of
non-radial modes.  This overview will  be necessarily succinct,  so we
refer the  reader to the excellent  monographs by Unno  et al.  (1989)
and  Cox  (1980)  for  further  details about  the  theory  of  linear
non-radial stellar pulsations.  Non-radial  modes are the most general
kind  of   stellar  oscillations.   There  exist   two  subclasses  of
non-radial  pulsations,  namely,  {\sl  spheroidal} modes,  for  which
$\left(\nabla \times  \vec{\xi}\right)_r= 0$ and $\sigma  \neq 0$, and
{\sl toroidal} modes,  for which $\left(\nabla \cdot \vec{\xi}\right)=
0$ and $\sigma=  0$, where $\vec{\xi}$ is the  displacement vector and
$\sigma$ the pulsation  frequency. Of interest in this  review are the
spheroidal  modes, which are  further classified  into $g$-,  $f$- and
$p$-modes  according  to  the  main  restoring  force  acting  on  the
oscillations, being gravity (buoyancy)  for the $g$- and $f$-modes and
pressure gradients for the $p$-modes.

For a  spherically symmetric star, a linear  non-radial pulsation mode
can be represented as a three-dimensional standing wave of the form:

\begin{equation}
\psi^{\prime}_{k \ell m}(r,\theta,\phi,t)= \psi^{\prime}_{k \ell m}(r)\;
Y^m_{\ell}(\theta,\phi)\; e^{i \sigma_{k  \ell m} t},
\end{equation}

\noindent where the prime indicates a small Eulerian perturbation of a
given    quantity   $\psi$    (like   the    pressure,   gravitational
potential,\ldots) and  $Y^m_{\ell}(\theta,\phi)$ are the corresponding
spherical  harmonics.   Geometrically,  $\ell=  0,1,2,\ldots$  is  the
number of nodal lines in the stellar surface and $m= 0, \pm 1, \ldots,
\pm \ell$ is  the number of such nodal lines  in longitude.  Note that
radial pulsations  are recovered  when $\ell= 0$.   In absence  of any
physical agent able to remove spherical symmetry (like magnetic fields
or rotation),  the eigenfrequencies $\sigma_{k \ell  m}$ are dependent
upon  $\ell$ but are  $(2\ell+1)$ times  degenerate in  $m$.  Finally,
$\psi^\prime_{k \ell m}(r)$ is  the radial part of the eigenfunctions,
which for  realistic models  must be necessarily  computed numerically
together with $\sigma_{k \ell m}$.  The eigenfunctions and eigenvalues
are obtained  as solution of a  eigenvalue problem of  fourth order in
space, if  we are considering  the adiabatic approximation  ($dS= 0$),
which  has  analytical solution  only  for  a  unrealistic star  model
consisting of  a homogeneous compressible sphere  (Pekeris 1938).  The
index  $k=  0,1,\ldots$,  known  as  the radial  order  of  the  mode,
represents (in the frame of  simple stellar models like those of white
dwarf  stars) the  number  of nodes  in  the radial  component of  the
eigenfunction.   They constitute  concentric spherical  surfaces where
radial displacement is  null.  For $g$-modes, the larger  the value of
$k$,  the lower  the oscillation  frequency, while  for  $p$-modes the
opposite holds.

White dwarf and pre-white  dwarf stars pulsate in non-radial $g$-modes
with periods between about 100 s  and 1,200 s, although PNNVs are able
to pulsate with much longer  periods, up to about 6,000 s.  Curiously,
the $g$-mode periods of white dwarfs  are of the order of magnitude of
$p$-mode periods for non-degenerate pulsating stars. It is interesting
to note that  radial modes and $p$-modes with  periods between 0.1 and
10  s have  been found  to be  pulsationally unstable  in a  number of
theoretical stability analysis (Saio  et al.  1983; Kawaler 1993) but,
however, these short  period modes have not been  ever observed in any
white dwarf star, although the interest in the topic has been recently
renewed (Silvotti et  al. 2007).  Finally, we mention  that until now,
only modes with $\ell= 1$ and $\ell= 2$ have been identified with high
confidence  in pulsating white  dwarfs.  Modes  with higher  values of
$\ell$ are  thought to be  excited, but due to  geometric cancellation
effects (Dziembowski  1977), they should  be very difficult  to detect
(however, see Sect. \ref{identifi}).

\subsection{Classes of spheroidal modes and critical frequencies} 

Generally  speaking, $g$-modes  are characterized  by  low oscillation
frequencies (long  periods) and by displacements of  the stellar fluid
essentially  in  the  horizontal  direction.   The  structure  of  the
$g$-mode  period  spectrum  is  governed  by  the  Brunt-V\"ais\"al\"a
frequency, given by:

\begin{equation}
N^2= g \left( \frac{1}{\Gamma_1} \frac{d\ln p}{dr} -\frac{d\ln \rho}{dr}\right)
\label{brunt-vaisala}
\end{equation}

\noindent  where  $g$ is  the  local  gravitational acceleration.   If
$N^2>0$ then $N$ is the (real) frequency of oscillation of a parcel of
stellar  fluid around  its equilibrium  level under  gravity  (Unno et
al. 1989). On  the other hand, $p$-modes have  high frequencies (short
periods) and are characterized  by essentially radial displacements of
the stellar  fluid. The critical  frequency for $p$-modes is  the Lamb
frequency, defined as:

\begin{equation}
L_{\ell}^2= \ell \left(\ell+1\right) \frac{c_{\rm s}^2}{r^2}
\end {equation}

\noindent $c_{\rm  s}$ being the local adiabatic  sound speed, defined
as   $c_{\rm   s}^2=   \Gamma_1   p/\rho$,   with   $\Gamma_1=   (d\ln
p/d\ln\rho)_{\rm ad}$.  A sound wave travels a distance $\approx 2 \pi
r / \ell$  horizontally in a time interval  of $\approx 2\pi/L_{\ell}$
(Unno et al. 1989).

Finally,  there is a  single $f$-mode  for a  given $\ell$  ($\geq 2$)
value.   Usually, this  mode  does not  have  any node  in the  radial
direction  ($k= 0$)  and possesses  an intermediate  character between
$g$- and $p$-modes.  In fact,  its eigenfrequency lies between that of
the low order $g$- and  $p$-modes, and generally slowly increases when
$\ell$ increases.

\subsection{The asymptotic behavior for high radial order}
\label{asymptotic}

When the radial order is high enough ($k \gg 1$) and for a small value
of  $\ell$,  the frequency  of  $p$-modes  is  approximately given  by
(Tassoul 1980):

\begin{equation} 
\sigma_{k \ell} \approx \frac{\pi}{2} \left(2 k + \ell + n
 + \frac{1}{2} \right)
\left[ \int_0^{R_*} \frac{1}{c_{\rm s}(r)} dr\right]^{-1},
\label{asin-p}
\end{equation}

\noindent  Here  $n$  is   the  politropic  index  characterizing  the
structure of the surface layers of a stellar model. On the other hand,
for  completely  radiative  or  completely convective  and  chemically
homogeneus stellar models, the frequency of $g$-modes for $k \gg 1$ is
given by:

\begin{equation} 
\frac{1}{\sigma_{k \ell}} \approx \frac{\pi}{2} \left(2 k + \ell + n + 
\frac{1}{2} \right) \frac{1}{\sqrt{\ell(\ell+1)}}  \left[ \int_0^{R_*} 
\frac{N(r)}{r} dr\right]^{-1},
\label{asin-g}
\end{equation}

\noindent From  Eq.  (\ref{asin-p}) we have, for  $p$-modes with fixed
$\ell$:

\begin{equation} 
\Delta \sigma^{\rm a}= \sigma_{k+1 \ell} - \sigma_{k \ell} =
\pi \left[ \int_0^{R_*} \frac{1}{c_{\rm s}(r)} dr\right]^{-1}= 
\mbox{constant},  
\label{asin-p-delta}
\end{equation}

\noindent that  is, the asymptotic  frequency spacing of  $p$-modes is
constant (and not  dependent from $\ell$) in the  limit of high radial
order. The  value of the constant  depends only on the  profile of the
adiabatic sound  speed, $c_{\rm s}(r)$,  in the interior of  the star.
Similarly, from  Eq. (\ref{asin-g}) we have, for  $g$-modes with fixed
$\ell$,

\begin{equation} 
\Delta \Pi_{\ell}^{\rm a}=  \Pi_{k+1 \ell} - \Pi_{k \ell}=
\frac{2 \pi^2}{\sqrt{\ell(\ell+1)}} \left[ \int_0^{R_*} \frac{N(r)}{r} 
dr\right]^{-1}= \mbox{constant},  
\label{asin-g-delta}
\end{equation}

\noindent  where  $\Pi_{k \ell}=  2  \pi  /  \sigma_{k \ell}$  is  the
pulsation period. Thus, the  asymptotic period spacing of $g$-modes is
a  constant value (which  dependens on  $\ell$) in  the limit  of high
radial order.  This constant value is fixed by the Brunt-V\"ais\"al\"a
frequency.

\subsection{Local analysis and propagation diagrams}

Important insights on the properties of non-radial pulsation modes can
be derived  by means  of a ``local  analysis''.  If we  consider modes
with very short  wavelengths (large wave number $k_r$),  that is, high
radial  order modes  ($k \gg  1$),  and in  the frame  of the  Cowling
approximation  --- $\Phi^{\prime}=0$,  $\Phi$ being  the gravitational
potential  (Cowling  1941)  ---  the  coefficients  in  the  pulsation
equations vary slowly in  comparison with the eigenfunctions, and then
the solutions  are proportional to $e^{i  k_r r}$. The  result of this
local analysis is the following dispersion relation:

\begin{equation}
k_r^2= \frac{1}{\sigma^2 c_s^2} \left( \sigma^2 - L_{\ell}^2\right) 
\left(\sigma^2 - N^2  \right)
\label{rel-disp}
\end{equation}

If $\sigma^2 >  N^2, L_{\ell}^2$ or $\sigma^2 <  N^2, L_{\ell}^2$ then
$k_r^2 >  0$ and $k_r \in$  Re, being the solutions  able to propagate
radially.   These  inequalities  define  two  propagation  regions  or
``resonant  cavities''  in the  interior  of  the  star: the  p-region
(corresponding to propagating  $p$-modes) and the g-region (associated
with propagating $g$-modes).   If, on the contrary, $N^2  > \sigma^2 >
L^2_{\ell}$ or $L^2_{\ell} > \sigma^2 > N^2$ then $k_r^2 < 0$ and $k_r
\in$ Im and thus the  solutions are evanescent, that is, the amplitude
of oscillation increases or decreases exponentially with $r$.

\begin{figure}
\centering
\includegraphics[width=0.9\columnwidth,clip]{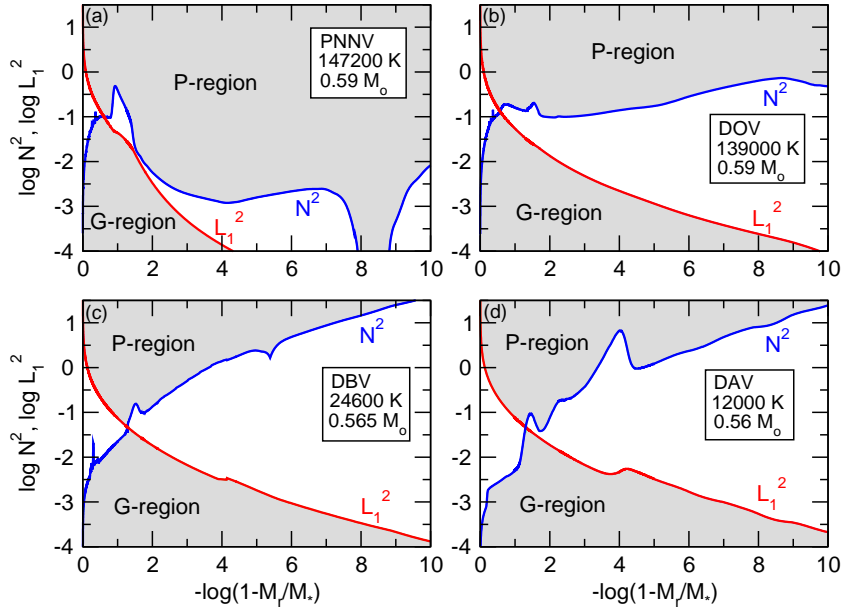}
\caption{Propagation diagrams ($\log  N^2,L_{\ell}^2$ as a function of
         $-\log(1-M_r/M_*)$) corresponding to selected pre-white dwarf
         and  white  dwarf  stellar  models.  Panel  (a)  depicts  the
         situation  of a  PNNV star,  panel (b)  corresponds to  a DOV
         star, panel  (c) is representative  of a DBV star,  and panel
         (d) depicts the case of  a DAV star. Gray zones correspond to
         wave propagation regions.}
\label{propa}
\end{figure}

To  illustrate the behavior  described by  the dispersion  relation of
Eq. (\ref{rel-disp}) in the specific  case of white dwarfs, we display
in  Fig. \ref{propa}  the propagation  diagrams corresponding  to four
selected pre-white  dwarf and white  dwarf stellar models.   Panel (a)
corresponds to a PG 1159  model at stages before the evolutionary knee
(see Fig. \ref{teff-g-plane}), which is representative of a PNNV star;
panel (b)  shows the situation  of a PG  1159 model located  after the
evolutionary  knee,  representative  of  a  DOV  star;  panel  (c)  is
representative of a DBV star;  and finally, panel (d) corresponds to a
DAV  star. In  each diagram,  the profile  of  the Brunt-V\"ais\"al\"a
frequency and the Lamb frequency (for $\ell= 1$) are plotted.

Two  outstanding  features  shown  in  this figure  are  worth  to  be
commented.  To begin with, while  the Lamb frequency exhibit a similar
qualitative behavior for all  the four models, the Brunt-V\"ais\"al\"a
frequency shows very important  changes, basically produced by changes
in the strength of electronic degeneracy at the different evolutionary
stages considered. This is  particularly notorious when we compare the
case of the DOV  model with the DBV model, and, in  turn, with the DAV
model.   The  changes in  the  Brunt-V\"ais\"al\"a  frequency lead  to
important  changes in the  extension and  location of  the propagation
cavities.   For instance,  in the  case of  the PNNV  and  DOV models,
$g$-modes are allowed  to propagate mainly at the  deep regions of the
core due to the large values of $N^2$ there.  This is because electron
degeneracy is  not important at  these stages.  The $p$-modes,  on the
other  hand, are  able to  propagate in  more external  regions.  This
behavior is  shared by  all the ``normal''  (non-degenerate) pulsating
stars. Note that  in the case of  the PNNV model, there is  a range of
frequencies  for which  the  modes  are allowed  to  propagate in  the
p-region as well  as in the g-region. They correspond  to modes with a
mixed ($p$ and $g$) character.  In the case of the DAV model, the most
degenerate  configuration, the  Brunt-V\"ais\"al\"a frequency  is very
small at the core region, forcing $g$-modes to propagate mainly in the
stellar  envelope.   On  the  contrary, $p$-modes  are  restricted  to
propagate in the inner zones of the star.  In summary, while $g$-modes
probe mainly  the central regions in  the case of PNNV  and DOV stars,
they are strongly sensitive to the outer layers in the case of DBV and
DAV stars. Note,  however, that some $g$-modes with  low-order $k$ are
still able to probe the deep  interior of DAVs and DBVs, thus allowing
to make some inferences on the structure of the carbon-oxygen core.

\begin{figure}
\centering
\includegraphics[width=0.9\columnwidth,clip]{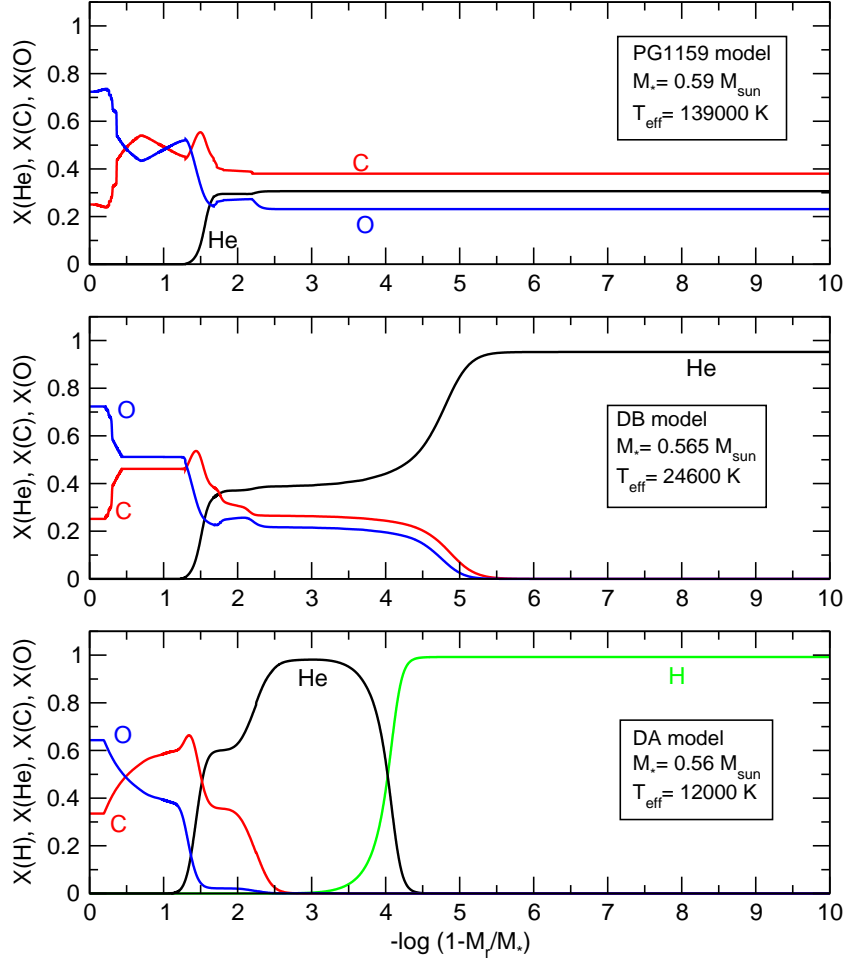}
\caption{The  internal  chemical  structure  of the  white  dwarf  and
         pre-white dwarf models described in Fig. \ref{propa}.}
\label{quimip}
\end{figure}

The second important feature  exhibited by the propagation diagrams of
Fig.  \ref{propa}  is the presence of  several bumps and  peaks in the
run  of  the  Brunt-V\"ais\"al\"a  frequency.   They are  due  to  the
presence of  steep density gradients  in the interior of  white dwarfs
models,   induced  by   several  chemical   transition   regions  (see
\ref{chemical}).   If white dwarfs  were chemically  homogeneous, then
the  Brunt-V\"ais\"al\"a  frequency   should  exhibit  a  very  smooth
behavior.   The  chemical  structure  of  realistic  white  dwarf  and
pre-white  dwarf   models  (with  different   evolutionary  histories)
described in Fig.  \ref{propa},  is shown in Fig.  \ref{quimip}.  This
figure shows  the fractional abundances of the  main chemical species.
For  white  dwarfs   and  pre-white  dwarfs,  the  Brunt-V\"ais\"al\"a
frequency  must  be numerically  assessed  by  means  of the  relation
(Tassoul et al. 1990; Brassard et al. 1991):

\begin{equation} \label{eq2}
N^2   =   \frac{g^2\   \rho}{P}\   \frac{\chi_{_T}}{\chi_{\rho}}\
\left(\nabla_{\rm ad} - \nabla + B \right),
\label{brunt-vaisala-ledoux}
\end{equation}

\noindent where $\chi_{_T} = \left({\partial \ln{P}}/{\partial \ln{T}}
\right)_{\rho}$  and  $\chi_{\rho}= \left({\partial  \ln{P}}/{\partial
\ln{\rho}} \right)_{T}$, $\nabla$ and $\nabla_{\rm ad}$ are the actual
and adiabatic temperature gradients, respectively, and $B$, the Ledoux
term, is given by

\begin{equation} \label{eq3}
B=-\frac{1}{\chi_{_{T}}} \sum^{n-1}_{{i}=1} \chi_{_{X_{i}}}
\frac{d\ln {X}_{i}}{d\ln P}.
\end{equation}

\noindent Here $X_{i}$ is the abundance by mass of species $i$, $n$ is
the total number of considered species and

\begin{equation} \label{eq4}
\chi_{_{X_{i}}}= \left( \frac{\partial \ln{P}}
{\partial \ln{X_{i}}} \right)_{\rho,T,\{X_{ j \neq i}\} }.
\end{equation}

This formulation (the ``modified Ledoux'' treatment) has the advantage
of   avoiding   the   numerical   problems  that   appear   when   Eq.
(\ref{brunt-vaisala})  is  employed to  compute  $N$  in the  strongly
degenerate matter  typical of white  dwarf interiors (Brassard  et al.
1991),  and  at   the  same  time  it  explicitly   accounts  for  the
contribution to $N$ from any  change in composition in the interior of
model (the zones  of chemical transition) by means  of the Ledoux term
$B$   of   Eq.    (\ref{eq3}).    The   correct   treatment   of   the
Brunt-V\"ais\"al\"a   frequency   in   the  interfaces   of   chemical
composition in stratified white dwarfs is very important, particularly
in  connection with  the resonance  effect  of modes  known as  ``mode
trapping'' (see Sect.  \ref{mode-trapping} below).

\subsection{Effects of slow rotation and weak magnetic fields}
\label{rota-mag}

We have seen that, in absence of rotation and/or magnetic fields, each
eigenfrequency is ($2\ell+1$)-fold degenerated.  This degeneracy takes
place  due to  the  absence of  a  preferential axis  of symmetry  for
oscillations. In  the presence of slow  rotation, as it  appears to be
the case of most white dwarfs, we have $\Omega \ll \sigma_{k \ell m}$,
where $\Omega$  is the angular  frequency of rotation.  In  this case,
the theory of perturbations can be  applied to a first order, and then
each frequency is split in ($2\ell+1$) equally-spaced components, thus
completely lifting the degeneracy  in frequency. Each component of the
multiplet is given by:

\begin{equation} 
\sigma_{k \ell m}(\Omega) = \sigma_{k \ell}(\Omega=0) + 
\delta \sigma_{k \ell m}. 
\label{rota}
\end{equation}

\noindent If  rotation is rigid ($\Omega=$ constant), the correction to
the eigenfrequencies is expressed as:

\begin{equation}
\delta \sigma_{k \ell m}= -m\ \Omega \left( 1-C_{k \ell} \right),
\label{coefi1}
\end{equation}

\noindent with $m= 0, \pm 1, \ldots, \pm \ell$ and $C_{k  \ell}$
being coefficients  that  depend on  the
details of  the stellar structure  and the eigenfunctions  obtained in
the  non-rotating case.  Such  coefficients are  given by  (Cowling \&
Newing 1949; Ledoux 1951):

\begin{equation}
C_{k \ell}= \frac
{\int_0^{R_*} \rho r^2 \left[ 2 \xi_r \xi_t + \xi_t^2 \right] dr}
{\int_0^{R_*} \rho r^2 \left[ \xi_r^2 + \ell(\ell+1) \xi_t^2 \right] dr}
\label{coefi}
\end{equation}

\noindent where $\xi_r$ and $\xi_t$  are the radial and the tangential
eigenfunctions.  We  note  that  in  the case  of  $p$-modes,  if  $k$
increases then $\xi_r \gg  \xi_t$ and thus $C_{k \ell} \rightarrow 0$.
In the  case of $g$-modes, on the  other hand, when $k$  is large then
$\xi_r  \ll \xi_t$, in  such a  way that  $C_{k \ell} \rightarrow  1 /
\ell(\ell+1)$.

If  the  condition  of  rigid   rotation  is  relaxed  and  we  assume
spherically symmetric rotation, $\Omega= \Omega(r)$, the correction to
the eigenfrequencies is given by (Hansen et al. 1977):

\begin{equation}
\delta \sigma_{k \ell m}= -m\ \int_0^{R_*} \Omega(r) K_{k\ell}(r) dr,
\label{coefi2}
\end{equation}

\noindent  where  $K_{k\ell}(r)$ is  the  first-order rotation  kernel
computed  from the  rotationally unperturbed  eigenfunctions according
to:

\begin{equation}
K_{k \ell}(r)= \frac
{\rho r^2 \left\{\xi_r^2 - 2 \xi_r \xi_t - 
\xi_t^2 \left[1- \ell(\ell+1) \right] \right \}}
{\int_0^{R_*} \rho r^2 \left[ \xi_r^2 + \ell(\ell+1) \xi_t^2 \right] dr}
\label{rot-kernel}
\end{equation}

The presence of a weak magnetic  field in a pulsating white dwarf also
destroys the  frequency degeneracy, but  at variance with the  case of
rotation, the degeneracy is only partially lifted. In this case it can
be shown that, after applying the perturbative approach and assuming a
simple  configuration  for  the  magnetic  field  of  the  star,  each
eigenfrequency is split into $(\ell+1)$ components:

\begin{equation} 
\sigma_{k \ell m}(B) = \sigma_{k \ell}(B=0) + \sigma_{k \ell}^{\prime}(B) 
\label{magne}
\end{equation}

\noindent  where  the  correction  $\sigma_{k  \ell}^{\prime}$,  which
involves   complex   mathematical   expressions,  depends   on   $B^2=
|\vec{B}|^2$, and  also on $m^2$ instead  $m$ (Jones et  al. 1989). We
note  that in the  presence of  a magnetic  field, even  the component
$m=0$  can  be  shifted.   This  is  at  variance  with  the  case  of
rotation. Also, we mention  that generally $g$-modes are more affected
than $p$-modes by the presence of a magnetic field.

\subsection{Mode identification}
\label{identifi}

Mode identification consists in  assigning the indices $k$, $\ell$ and
$m$ to a  given observed pulsation mode. This is the  first step in an
asteroseismological  analysis  because   it  connects  the  models  to
observations.  It is generally a very difficult task, in particular in
the case  of the  radial order  $k$, because it  is not  an observable
quantity.  Traditionally,  identification of  $\ell$ and $m$  in white
dwarfs has  been performed through rotational mode  splitting.  Due to
the  presence  of rotation,  each  frequency  appears  as a  multiplet
containing $(2\ell+1)$ components (see Sect. \ref{rota-mag}).  Thus, a
triplet of modes corresponds to dipole ($\ell= 1$) modes, a quintuplet
corresponds to quadrupole ($\ell=2$) modes, and so on.  In some cases,
it  is possible to  assign $m$  to each  component of  the multiplets.
Another way  to obtain the  value of $\ell$  of a series  of pulsation
modes  is  using the  period  spacing.  The  idea  is  to compare  the
observed mean period spacing between adjacent modes ($\overline{\Delta
  \Pi_k}=  \overline{\Pi_{k+1} -  \Pi_k}$) and  the  asymptotic period
spacing  ($\Delta  \Pi_{\ell}^{\rm  a}$),  which,  for  $g$-modes,  is
dependent  on  $\ell$  ---  see  Eq.   (\ref{asin-g-delta})  in  Sect.
\ref{asymptotic}.  Thus, by matching  the observed period spacing with
the theoretically  predicted value of $\Delta  \Pi_{\ell}^{\rm a}$ the
value of $\ell$ can be inferred.  The observation of $\overline{\Delta
  \Pi_k}$ also is  employed to infer the stellar  mass of white dwarfs
(see Sect. \ref{mass} below).

The two above mentioned methods  have been applied successfully to the
DOV stars \pp\  (Winget et al.  1991) and \pr\  (Kawaler et al. 1995),
and to  the DBV white  dwarf GD 358  (Winget et al.  1994).   This has
been possible thanks to the long and continuous time series photometry
obtained  with the  WET  array, that  allowed  the rotation  multiplet
structure  to  be  resolved  and   a  large  number  of  modes  to  be
detected. This has  allowed to derive the period  spacing. However, in
spite of the employment of the WET, it has not been possible to obtain
the same success for any DAV star.  This is due to the small number of
periods  detected and/or  the complex  and changing  structure  of the
pulsation spectrum characterizing DAVs. Robinson et al. (1995) devised
a method  for mode  identification that applies  even in the  cases in
which there is  a very reduced number of  periods detected. The method
is based on the effects of limb-darkening on the pulsation amplitudes.
A   non-radial  $g$-mode  in   a  white   dwarf  manifest   itself  as
perturbations  in the temperature  which, in  turn, create  bright and
dark regions on the stellar  surface.  Non-radial modes of each $\ell$
value experience geometrical cancellation due to the presence of these
regions   with   opposing   phases.    At   short   wavelenghts,   the
limb-darkening is stronger, and reduces the effects of cancellation on
modes  with high values  of $\ell$.   As a  result, the  amplitudes of
pulsations are sensitive  to $\ell$ (and are insensitive  to $m$), and
then it is possible to  derive the harmonic degree.  Using this method
(called ``time-resolved  ultra-violet spectroscopy'') allowed Robinson
et al.  (1995)  to infer that the largest amplitude mode  of the DAV G
117$-$B15A had $\ell= 1$.  Kepler  et al.  (2000) employed this method
and found  that the pulsations in the  DAVs G 226-29 and  G 185-32 are
consistent with $\ell= 1$.  Nitta  et al. (1998) also used this method
for the  DBV white dwarf  GD 358  and were only  able to rule  out the
presence of modes with $\ell > 2$.

A  variant  of the  above  mentioned  method  to identify  $\ell$  was
proposed by van Kerkwijk et al.  (2000) and Clemens et al.  (2000) and
applied  to  the  DAV  white   dwarf  G  29-38.   The  method,  called
``time-resolved optical spectroscopy'', is  based on the dependence of
the amplitudes  of non-radial pulsation modes within  the Balmer lines
at  visual wavelengths  only. They  found that  for this  star several
modes  are   consistent  with  $\ell=   1$,  and  that  one   mode  is
characterized by $\ell= 2$.  Kotak  et al.  (2002a) studied the DAV HS
0507+0434B and  Kotak et al.  (2003)  studied the DBV  GD358.  In both
cases they found that $\ell= 1$ for most of the modes present in these
stars.   Kotak  et  al.  (2004)  also studied  G  117$-$B15A,  and  by
comparing  the  fractional  wavelength  dependence  of  the  pulsation
amplitudes  (chromatic  amplitudes)  with  models,  they  confirmed  a
previous report that the strongest mode, at 215 s, has $\ell= 1$.  The
chromatic amplitude  for the 272  s mode, on  the other hand,  is very
puzzling, showing an increase  in fractional amplitude with wavelength
that  cannot be reproduced  by the  models for  any $\ell$  at optical
wavelengths.  Recently, Thompson et al. (2008) revisited the case of G
29-38, and  confirmed several $\ell$ identifications and  add four new
values, including an additional $\ell = 2$ and a possible $\ell = 4$.


\section{Asteroseismological tools}
\label{tools}

When a rich spectrum of non-radial $g$-modes of a variable white dwarf
is  available,  and the  eigenmodes  have  been correctly  identified,
asteroseismology can be applied to infer the fundamental parameters of
the  star.  Obviously, the  larger the  number of  independent periods
detected, the larger  the amount of information that  can be inferred.
Below, we describe  the asteroseismological methods currently employed
to probe  the interiors  of pulsating white  dwarfs. They  can provide
information   about   the   stellar   mass,  the   internal   chemical
stratification,  the  rate  of  rotation, the  existence  of  magnetic
fields,  the cooling  timescale and  core composition,  and  the outer
convection zone.

\subsection{Period spacing: determination of the stellar mass}
\label{mass}

\begin{figure}
\centering
\includegraphics[width=0.9\columnwidth,clip]{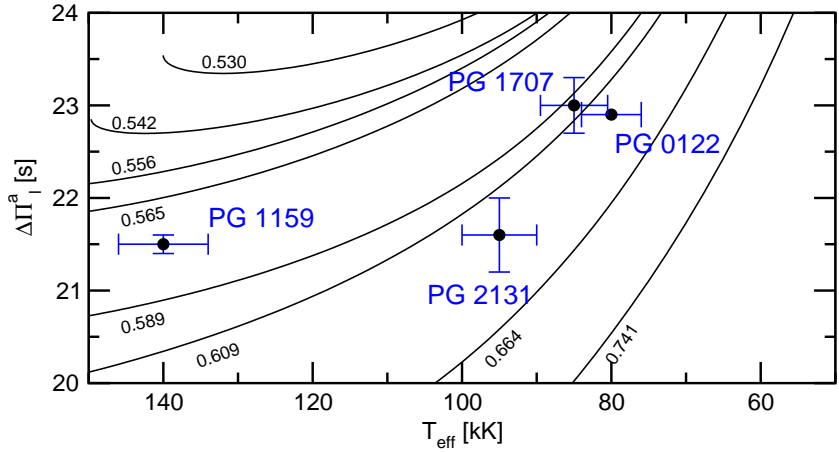}
\caption{The  dipole ($\ell=  1$) asymptotic  period  spacing ($\Delta
         \Pi_{\ell}^{\rm a}$) for different stellar masses in terms of
         the effective  temperature.  Numbers along  each curve denote
         the stellar mass (in solar units) of the PG 1159 evolutionary
         sequences.   The   plot  also  shows  the   location  of  the
         high-gravity, low-luminosity GW Vir  stars \pg, \pp, \pr, and
         \pt.}
\label{asint-obs}
\end{figure}

The stellar mass  of a pulsating white dwarf can  be obtained from the
observed period spacing once the harmonic degree $\ell$ is identified.
>From Eq.  (\ref{brunt-vaisala}) or Eq. (\ref{brunt-vaisala-ledoux}) it
is   clear  that   the  Brunt-V\"ais\"al\"a   frequency   is  directly
proportional to  the gravity.  Thus,  the period spacing  is inversely
proportional to the gravity  through Eq.  (\ref{asin-g-delta}), and to
the stellar mass. Therefore, by measuring the period spacing in a real
star, the total  mass can be inferred.  This  is particularly true for
DOVs and  PNNVs stars  because in these  cases the periods  and period
spacings depend almost exclusively on  the stellar mass, and to a much
smaller extent,  on the  luminosity and the  thickness of  the He-rich
outer  envelope. This  was first  recognized by  Kawaler  (1987).  The
method, however,  is not directly  appliable to DBVs and  DAVs because
for those stars  the periods and period spacings  are sensitive to the
thickness of  the He  and/or H envelopes,  in addition to  the stellar
mass.   The  approach  consists  in computing  the  asymptotic  period
spacing for  white dwarf  models of different  effective temperatures,
corresponding to  evolutionary sequences of  different stellar masses.
By adopting  the effective temperature  of the target star  as derived
from spectroscopy, the  observed period spacing can be  matched to the
asymptotic  period  spacing.  We  illustrate  this  procedure in  Fig.
\ref{asint-obs}  for the  case of  DOV stars.   Specifically,  in this
figure we show the  asymptotic period spacing ($\Delta \Pi_{\ell}^{\rm
a}$) for $\ell= 1$ corresponding  to PG 1159 evolutionary sequences of
different  stellar  masses.   In  addition,  selected  low-luminosity,
high-gravity pulsating PG 1159 stars have been included by using their
observed  $T_{\rm eff}$  and $\Delta  \Pi$ values.   The value  of the
stellar  mass  for  each  star   can  be  obtained  by  simple  linear
interpolation.

The method described above is computationally inexpensive and has been
widely  used in  the  past  because it  does  not involve  pulsational
calculations.  However,  we must emphasize that the  derivation of the
stellar mass using  the asymptotic period spacing may  not be entirely
reliable in  pulsating white dwarfs  that have modes  characterized by
low   and   intermediate  radial   orders.    This   is  because   Eq.
(\ref{asin-g-delta}), which  describes the asymptotic  behavior of the
period  spacing, is  strictly valid  only in  the limit  of  very high
radial  order (long  periods) and  for chemically  homogeneous stellar
models,  while  white  dwarf  stars  are  expected  to  be  chemically
stratified  and characterized  by strong  chemical gradients  built up
during   the  progenitor  star   life  (see   Sect.   \ref{chemical}).
Generally, this  method can  lead to an  over-estimate of  the stellar
mass, except for stars that pulsate with very high radial orders, such
as the PNNV stars (Althaus et al. 2008a).

\begin{figure}
\centering
\includegraphics[clip,width=0.9\columnwidth]{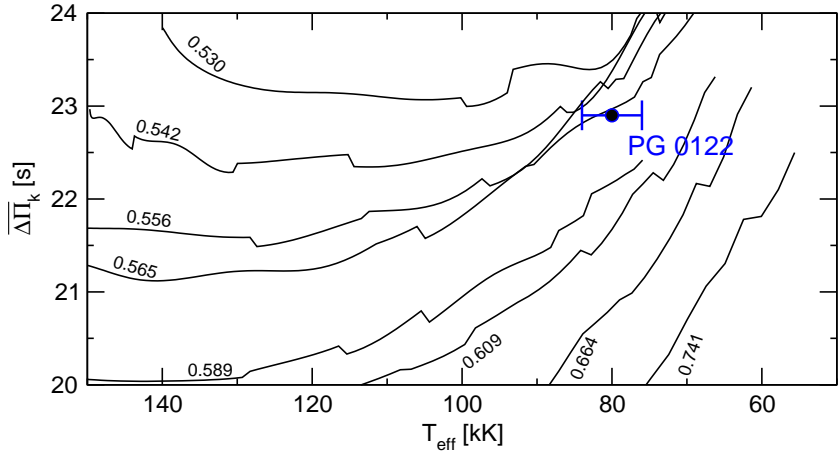}
\caption{Same  as Fig.  \ref{asint-obs}, but  for the  average  of the
         computed period spacings ($\overline{\Delta \Pi_{\ell}}$).}
\label{psp}
\end{figure}

A more realistic  approach to infer the stellar mass  of PG 1159 stars
is  to compare the  observed period  spacing with  the average  of the
computed period spacings.  The average of the computed period spacings
is  computed  as  $\overline{\Delta \Pi_{\ell}}=  (n-1)^{-1}  \sum_k^n
\Delta  \Pi_k  $,  where   $\Delta  \Pi_k=  \Pi_{k+1}-\Pi_k$  are  the
calculated period spacings  and $n$ is the number  of computed periods
laying  in  the  range  of  the  observed  periods.   Fig.   \ref{psp}
illustrates this method for the case of \pg.  In this case, the curves
(notably  jagged   and  irregular)  correspond   to  $\overline{\Delta
\Pi_{\ell}}$  for each  PG 1159  evolutionary sequence  with different
stellar masses.   Again, the location of  the target star  is given by
the observed  $T_{\rm eff}$ and  $\Delta \Pi$ values, and  the stellar
mass of the target star can be inferred by simple interpolation.  This
method is  more reliable  for estimating the  stellar mass of  PG 1159
stars than that described above  because, provided that the average of
the computed period spacings is  evaluated at the appropriate range of
periods,  the   approach  is  adequate  for  the   regimes  of  short,
intermediate and long  periods (i.e., for all $k$).   When the average
of  the computed  period spacings  is taken  over a  range  of periods
characterized by high $k$ values,  then the predictions of the present
method  become  closer  to  those  of the  asymptotic  period  spacing
approach.   On the other  hand, the  present method  requires detailed
period  computations, at  variance  with the  method described  above.
Note that in the case of  \pg, the employment of the asymptotic period
spacing overestimates the stellar mass, as compared with the inference
from  the  average of  the  computed  period  spacing (0.62  vs  0.565
$M_{\odot}$)  Finally, we  note  that the  methods  described in  this
section  for assessing  the  stellar mass  rely  on the  spectroscopic
effective temperature, and the results are unavoidably affected by its
associated uncertainty.

\subsection{Mode trapping: hints about the chemical stratification}
\label{mode-trapping}

The  period  spectrum  of  chemically homogeneous  stellar  models  is
characterized by  a constant period separation,  given very accurately
by   Eq.    (\ref{asin-g-delta}).    However,   current   evolutionary
calculations  predict  that  white   dwarfs  must  have  one  or  more
composition gradients in  their interiors (Sect. \ref{chemical}).  The
presence of one or  more abrupt chemical transitions strongly modifies
the character of  the resonant cavity in which  modes should propagate
as  standing  waves.    Specifically,  chemical  interfaces  act  like
reflecting boundaries that partially  trap certain modes, forcing them
to  oscillate  with larger  amplitudes  in  specific regions  (bounded
either by two interfaces or by one interface and the stellar center or
surface)  and  with smaller  amplitudes  outside  those regions.   The
requirement for  a mode to  be trapped is  that the wavelength  of its
radial  eigenfunction  matches  the  spatial  separation  between  two
interfaces or between one interface and the stellar center or surface.
This  mechanical resonance, known  as {\sl  mode trapping},  was first
studied by  Winget et al. (1981)  and has been the  subject of intense
study in  the context of  stratified DA and  DB white dwarfs  --- see,
e.g., Brassard et  al.  (1992a) and C\'orsico et  al.  (2002a). In the
field of PG 1159 stars, mode trapping has been extensively explored by
Kawaler \& Bradley (1994) and C\'orsico \& Althaus (2006).

\begin{figure}
\centering
{\includegraphics[clip,width=0.9\columnwidth]{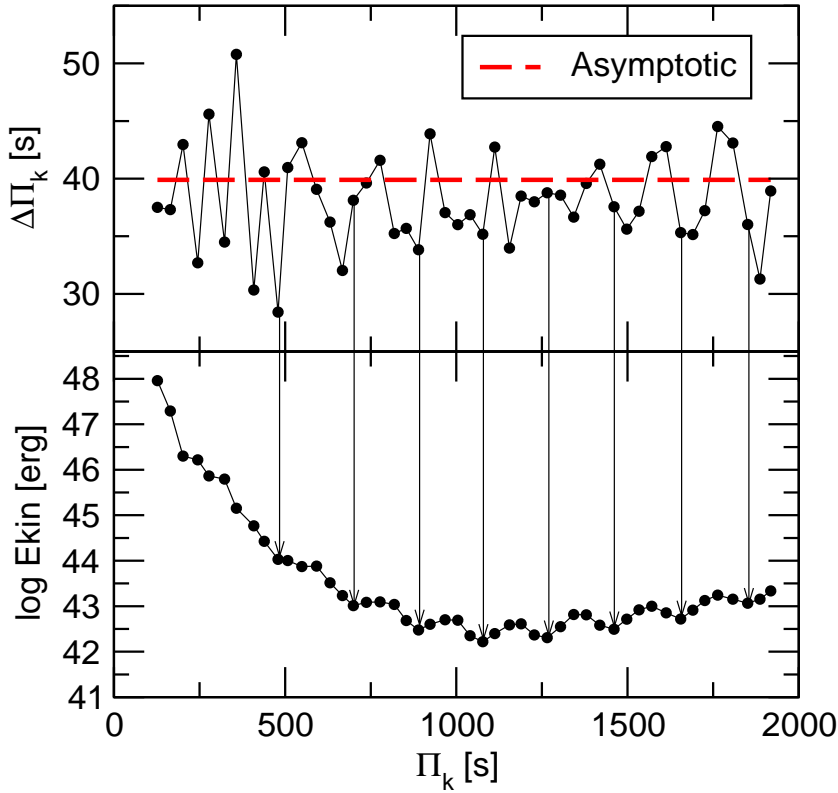}}
\caption{The forward  period spacing $\Delta \Pi_k$  (upper panel) and
         the pulsation  kinetic energy (lower  panel) in terms  of the
         pulsation periods  for a typical DB white  dwarf model. Modes
         trapped in the outer He-rich envelope correspond to minima in
         $E_{\rm kin}$.}
\label{delp}
\end{figure}

There are  (at least) two  ways to identify  trapped modes in  a given
stellar model.  First, we can consider the oscillation kinetic energy.
Because the  amplitude of eigenfunctions is  arbitrarily normalized at
the surface of the model, the  values of the kinetic energy are useful
only  in  a  relative   sense.   The  oscillation  kinetic  energy  is
proportional   to   the  integral   of   the   squared  amplitude   of
eigenfunctions, weighted  by the density.  Thus,  modes propagating in
the deep interior  of the white dwarf, where  densities are very high,
will exhibit  larger values  than modes which  are oscillating  in the
low-density, external regions.  When  only a single chemical interface
is  present, modes can  be classified  as modes  trapped in  the outer
layers, modes confined in the core regions, or simply ``normal modes''
--- which oscillate  everywhere in the star ---  characterized by low,
high,  and   intermediate  values  of   kinetic  energy,  respectively
(Brassard et al. 1992a).

A  second (and  more important  from an  observational point  of view)
consequence  of  mode trapping  is  that  the  period spacing  $\Delta
\Pi_{k}$,  when plotted in  terms of  the pulsation  period $\Pi_{k}$,
exhibits  strong departures  from uniformity.   The  period difference
between an observed mode and its adjacent modes ($\Delta k= \pm 1$) is
an  observational diagnostic of  mode trapping,  at variance  with the
kinetic energy,  whose value is  very difficult to estimate  only from
observation.  For  stellar models  characterized by a  single chemical
interface,  local  minima   in  $\Delta  \Pi_{k}$  usually  correspond
directly to modes trapped in the outer layers, whereas local maxima in
$\Delta \Pi_{k}$ are associated to modes confined in the core region.

The   diagrams  $\Delta  \Pi_{k}$   versus  $\Pi_{k}$   have  valuable
information about the structure of  the outer layers and the thickness
of the outer envelope of  a white dwarf. In particular, the difference
of  period between  successive trapped  modes (successive  minima), or
``cycle of  trapping'', is inversely proportional to  the thickness of
the outer  envelope where the modes  are trapped.  On  the other hand,
the amplitude  of the variation  of $\Delta \Pi_k$, or  ``amplitude of
trapping'', is a  measure of the thickness of  the chemical transition
region.  The potential use of mode trapping to measure the thicknesses
of the outer envelopes in DA and DB white dwarfs led to several groups
of  researchers to  derive  a  relation between  the  period of  modes
trapped  in the  outer envelope  and  the location  of the  transition
region  responsible  for such  a  trapping,  the  so called  ``trapped
mode-period'' relation --- see Kawaler \& Weiss (1990) and Brassard et
al.  (1992a) for DAVs, Bradley et al.  (1993) for DBVs, and Kawaler \&
Bradley (1994) for DOVs.

This method  for estimating  the thickness of  the outer  envelopes in
white   dwarfs,  very   appealing   and  extremely   useful  from   an
asteroseismological  point of  view, relies  however on  two important
assumptions: (i) the existence  of a {\sl unique} relevant composition
gradient in the  interior of the star, that is,  the most external one
(He/H in the  DAVs, C/He in the DBVs),  thus neglecting other possible
chemical  interfaces  located  at  the  core  regions;  and  (ii)  the
assumption of diffusive equilibrium in the trace element approximation
(Tassoul  et al.   1990)  to model  the  chemical transition  regions.
C\'orsico et al.  (2001a, 2002a, 2002b) published a serie of pulsation
studies  based   on  detailed  evolutionary  DA   white  dwarf  models
characterized   by   outer   chemical   interfaces   modeled   through
time-dependent element  diffusion, and  the chemical structure  at the
deep  regions of  the core  resulting from  the  progenitor evolution.
They  showed that  the mode-trapping  features produced  by  the outer
chemical  transition  region  is  actually much  less  important  than
thought  before, due  to the  fact  that chemical  profiles shaped  by
chemical diffusion are very smooth as compared with the predictions of
the   trace  element  approximation.    In  addition,   these  studies
demonstrated that the chemical structure  of the core regions can be a
relevant source of mode  trapping/confinement, in particular for modes
with low radial order.

Thus, the rather simple picture  of mode trapping in simplified models
characterized by only one  chemical interface becomes markedly complex
when  the   stellar  model   is  characterized  by   several  chemical
composition gradients, in particular for the regime of low order modes
(short  periods), where  it  is very  difficult  to disentangle  which
specific chemical interface is responsible for a given feature of mode
trapping.   In  addition, Montgomery  et  al.   (2003)  have found  an
intrinsic degeneracy in  the way that pulsation modes  sample the core
and the  envelope of  white dwarfs, which  can potentially lead  to an
ambiguity in the  asteroseismologically inferred locations of internal
chemical structures.  Fig. \ref{delp} shows the period spacing and the
kinetic energy for a realistic  evolutionary DB white dwarf model with
$M_*= 0.565 \, M_{\odot}$ and $T_{\rm eff}= 28\,400$ K.  In this case,
the strong  mode-trapping features are found even  at high-order modes
(long periods). Note that modes trapped in the outer He-rich envelope,
which are characterized by minima of kinetic energy, do not correspond
to  minima in  period spacing.   The  dashed line  corresponds to  the
asymptotic period  spacing.  The chemical  structure of this  model is
plotted in  the central  panel of Fig.   \ref{quimip}.  The  model has
three  different chemical transition  regions, which  combined produce
the complex pattern of mode trapping displayed in Fig. \ref{delp}.

In  summary, when realistic  evolutionary models  of white  dwarfs are
considered in  pulsational studies, the potential of  mode trapping to
infer the  location of the  chemical interfaces (and the  thickness of
the  envelope)  is  strongly  weakened  and  becomes  almost  useless.
Instead, all we can infer is  that white dwarfs are in fact chemically
stratified  stars,  thus confirming  the  predictions  of the  stellar
evolution theory.

\subsection{Frequency splitting: constraining the rotation rate and/or 
            the magnetic field}

We have seen in Sect. \ref{rota-mag} that slow and solid-body rotation
of  a  star produces  a  set  of  equally-spaced frequencies,  with  a
separation between  each component of  the multiplet given  by $\delta
\sigma_{k  \ell  m}= m\  (1-C_{  k  \ell})  \Omega$.  This  rotational
splitting of frequencies is found in a large number of pulsating white
dwarf  stars. This  fact can  be  exploited to  estimate the  rotation
period of the star ($P_{\rm R}= 2\pi /\Omega$), as well as to identify
the  harmonic  degree  $\ell$  and  the azimuthal  order  $m$  of  the
modes. In practice, generally there  are one o more components missing
in the observed multiplets,  and frequently the present components can
show very  different amplitudes. The mechanism  of selection operating
here  is  at   present  unknown  (Winget  \&  Kepler   2008).   As  an
illustration of the inference of  the rotation period in white dwarfs,
we consider  the prototypical  star \pp.  For  this star,  20 triplets
($\ell= 1, m= -1,  0, +1$) and 8 quintuplets ($\ell= 2,  m= -2, -1, 0,
+1, +2$) have been identified (Winget et al.  1991).  By computing the
average of the frequency spacings between the components of multiplets
($\overline{\delta \sigma_{k \ell m}}$),  a rotation period of $P_{\rm
R}= 1.38 \pm 0.01$ days has been inferred.

Employing  a more  self-consistent approach,  Charpinet et  al. (2009)
have recently  revised the issue of  the stellar rotation  of \pp.  On
the basis of the asteroseismological model of C\'orsico et al.  (2008)
for that  star, Charpinet et al.   (2009) have carried  out a detailed
analysis of the individual  multiplets by assuming solid-body and also
differential internal rotation profiles,  and have arrived at the firm
conclusion that the  star is rotating as a solid  body with a rotation
period of $P_{\rm  R}= 33.61 \pm 0.59$ h, in  very good agreement with
the previous result by Winget et al. (1991).

Finally,  pulsating  white   dwarfs  can  exhibit  magnetic  frequency
splitting.  In this case, we  have seen in Sect. \ref{rota-mag} that a
magnetic field can  lead to a mode characterized  by a harmonic degree
$\ell$ to split into $\ell+1$ components, instead $2\ell+1$ components
due to  rotational splitting.  Since  the frequency separation  of the
multiplets depends  on $B^2$,  by measuring the  frequency separation,
the magnitude of  the magnetic field can be,  in principle, estimated.
Historically, the prototypical  pulsating white dwarf showing magnetic
splitting (doublets  of frequency)  is the DAV  star R  548.  However,
solid  arguments lead  to the  conclusion that  the  supposed magnetic
doublets of  this star  can be actually  rotational triplets  with the
central  component $m= 0$  having very  small amplitudes  (Fontaine \&
Brassard 2008).

\subsection{The  rate of  period  changes: clues  to the core chemical 
            composition}

The rates of period change  of pulsation $g$-modes in white dwarfs are
a very important observable  quantity that can yield information about
the core chemical composition. As  a variable white dwarf evolves, its
oscillation periods  vary in response  to evolutionary changes  in the
mechanical structure of the star.  Specifically, as the temperature in
the core of  a white dwarf decreases, the  plasma increases its degree
of  degeneracy  so  the  Brunt-V\"ais\"al\"a frequency  ---  the  most
important physical quantity in  $g$-mode pulsations --- decreases, and
the pulsational spectrum of the star shifts to longer periods.  On the
other hand,  residual gravitational  contraction (if present)  acts in
the  opposite  direction,   thus  shortening  the  pulsation  periods.
Competition  between  the   increasing  degeneracy  and  gravitational
contraction  gives rise  to a  detectable temporal  rate of  change of
periods ($\dot{\Pi}\equiv  d\Pi/dt$).  Roughly, the rate  of change of
the  pulsation  period  is related  to  the  rates  of change  of  the
temperature at the  region of the period formation,  $\dot{T}$, and of
the stellar radius, $\dot{R_*}$ (Winget et al. 1983):

\begin{equation}
\frac{\dot{\Pi}}{\Pi} \approx - a \frac{\dot{T}}{T} + 
b \frac{\dot{R_*}}{R_*},
\label{rchp}
\end{equation}

\noindent where $a$  and $b$ are constants whose  values depend on the
details of the white dwarf  modeling (however, both $a$ and $b \approx
1$). The  first term in Eq.   (\ref{rchp}) corresponds to  the rate of
change in period induced by the cooling of the white dwarf and it is a
positive contribution,  while the second  term represents the  rate of
change  due  to  gravitational   contraction  and  it  is  a  negative
contribution.  In principle,  the rate of change of  the period can be
measured  by observing  a  pulsating  white dwarf  over  a large  time
interval when one or more very stable pulsation periods are present in
their light curves.  In the case of pulsating DA white dwarfs, cooling
dominates  over gravitational  contraction,  in such  a  way that  the
second  term in  Eq.  (\ref{rchp})  is negligible,  and  only positive
values of  the rate of  change of period  are expected.  For  the high
effective temperatures characterizing  the DOV/PNNV instability strip,
gravitational contraction is still  significant, to such a degree that
its influence on $\dot{\Pi}$ can  overcome the effects of cooling.  In
this  case the  second term  in  Eq.  (\ref{rchp})  is not  negligible
anymore  and,  consequently, either  positive  or  negative values  of
$\dot{\Pi}$  are  possible.   For  these hot  pre-white  dwarfs,  what
determines that  $\dot{\Pi}$ is positive or negative  is the character
of the mode: if the pulsation period corresponds to a mode confined to
the deep regions of the star,  then a positive value of $\dot{\Pi}$ is
expected, while  if the  mode is  a mode trapped  mostly in  the outer
layers, we  would expect a  negative value of $\dot{\Pi}$  (Kawaler \&
Bradley 1994).   As for the DBV  stars, in which the  influence of the
gravitational contraction  on $\dot{\Pi}$  is negligible, it  is found
from theoretical  evolutionary calculations that the  expected rate of
period change should be positive.

\begin{figure}
\centering
\includegraphics[width=1.3\columnwidth,clip]{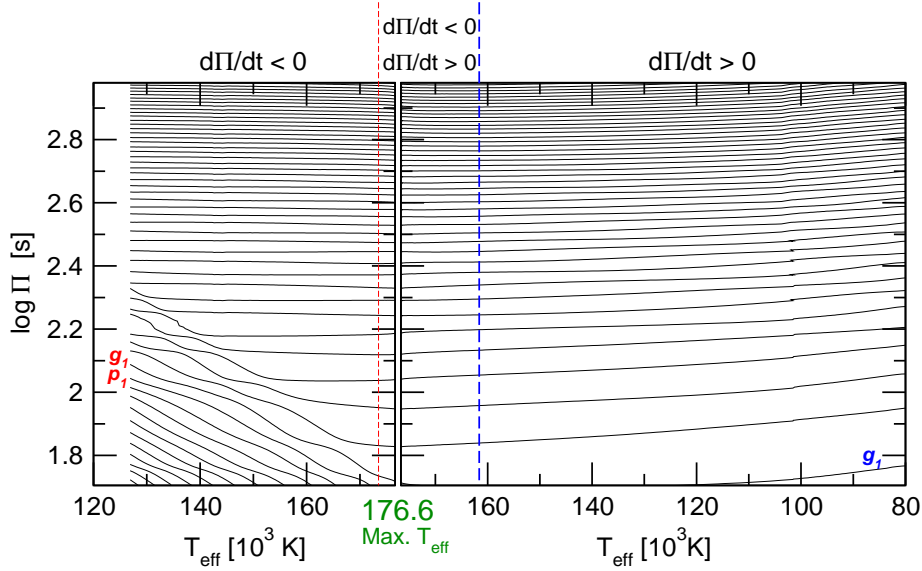}
\caption{The  evolution of the  $\ell= 1$  pulsation periods  $\Pi$ in
         terms  of  the effective  temperature,  corresponding to  the
         sequence   of   PG  1159   models   with   $M_*=  0.5895   \,
         M_{\odot}$. The  left panel depicts the  situation before the
         models reach the maximum  value of effective temperature, and
         right panel shows the  situation when the models have already
         passed through  this stage. Note that the  orientation of the
         $T_{\rm  eff}$ axis  in the  left panel  is opposite  that of
         right panel.}
\label{per-teff}
\end{figure}

As an illustrative example of the temporal drift of periods in a white
dwarf, we show  in Fig.  \ref{per-teff} the evolution  of periods with
the  effective temperature  corresponding to  a PG  1159  model before
(left panel)  and after  (right panel) the  model reaches  the turning
point    in   the   Hertzsprung-Russell    diagram   ---    see   Fig.
\ref{teff-g-plane}.  Pulsation periods decrease in stages in which the
model is still approaching  its maximum effective temperature ($T_{\rm
eff} \approx  176\,600$ K).  When  the model has already  settled into
the cooling  phase, the  periods increase. Note  in the left  panel of
Fig.   \ref{per-teff}  that  when  $T_{\rm  eff} <  176\,600$  K,  the
low-order  periods  ($\Pi_k  \lesssim   200$  s)  exhibit  a  behavior
reminiscent of  the well-known ``avoided  crossing''.  When a  pair of
modes experiences avoided crossing, the modes exchange their intrinsic
properties  (Aizenman et  al.  1977).   These are  modes with  a mixed
character, that is, modes that behave as $g$-modes in certain zones of
the star and as $p$-modes in other regions.  Finally, we have included
in  Fig. \ref{per-teff}  two thin  dashed vertical  lines  that define
three  different regimes for  the theoretical  rate of  period change:
models  exhibit  negative  (positive)   rates  of  period  change  for
effective temperatures to  the left (right) of that  of the red (blue)
line, and positive and negative rates between both lines (C\'orsico et
al.  2008).  Thus,  we expect a pulsating PG 1159  star located in the
region  of the  evolutionary knee  to exhibit  a mix  of  positive and
negative values of $\dot{\Pi}$.  A  very recent study (Costa \& Kepler
2008) suggests that the prototype star \pp\ could be located precisely
at the evolutionary knee in the Hertzsprung-Russell diagram, making it
an excellent target to test the evolutionary models of pulsating white
dwarfs.

What  can be  learned from  the rate  of period  change? Observational
measurements  of $\dot{\Pi}$  can  provide a  sensitive  probe of  the
structure and evolution  of white dwarf stars.  As  shown by Winget et
al. (1983), a measurement of the  rate of period change of a pulsating
white  dwarf constitutes, particularly  for DAV  and DBV  white dwarfs
(which evolve at almost constant  radius), a direct measurement of the
cooling rate of the star. This, in turn, provides valuable information
about the  core chemical composition.   For a given mass  and internal
temperature distribution,  the rate of period change  increases if the
mean atomic weight of the core is increased. The mean atomic weight of
the core of a white dwarf can be inferred, in principle, comparing the
observed   value   of  $\dot{\Pi}$   with   theoretical  values   from
evolutionary sequences of white dwarfs with  cores made of C and O (or
heavier species such as  Ne, Mg,\ldots) with different mass fractions.
A simple relation can be obtained by using the Mestel cooling law (see
Sect.  \ref{mestel}).  The result is $\dot{\Pi}= (3-4) \times 10^{-15}
(A/14)$ [s/s] (Kepler et al. 2005).
 
The measured  $\dot{\Pi}$ can also be  used to detect  the presence of
planets  orbiting  around pulsating  white  dwarfs.   If  a planet  is
orbiting  around  a  star,  its   distance  to  the  Sun  will  change
periodically  as the  star orbits  around the  center of  mass  of the
system. If the star is a stable pulsator, then the orbital motion will
change the pulse arrival times as  seen from Earth, as compared with a
constant period. The variation of the pulse arrival times is given by:

\begin{equation}
\tau= \frac{a_{\rm p } m_{\rm p} \sin i}{M_* c}
\end{equation}

\noindent where $a_{\rm p}$ and  $m_{\rm p}$ are the orbital semi-axis
and the mass of the planet, $M_*$  is the total mass of the host star,
$c$ is the speed light, and $i$ is the inclination of the orbit to the
line  of sight.   Recently, Mullally  et al.   (2008) have  studied 15
DAVs.  For the  star GD 66 ($T_{\rm eff}= 11\,980$  K, $\log g= 8.05$)
they  found evidence  $(d\Pi/dt=  1.35 \times  10^{-12}$  s/s) of  the
presence  of a planet  with a  mass of  $m_{\rm p}=  2 M_{\rm  J}$, an
orbital semi-axis  of $a_{\rm p}= 2.35$  AU, and an  orbital period of
$4.5$  yr.   In comparison,  the  rate  of  period change  by  cooling
($d\Pi/dt  \sim   1  \times  10^{-15}$  s/s)  and   by  proper  motion
($d\Pi/dt\sim  1  \times 10^{-16}$  s/s)  are  much  smaller.  If  the
existence  of a planet  orbiting around  GD 66  is confirmed  it would
indicate that a planet can survive the expansion in the giant phase of
its host star.

In addition,  the measurement of $\dot{\Pi}$ in  variable white dwarfs
can be  employed to  set constraints on  particle physics.   McGraw et
al. (1979) were  the first to suggest that  hot pulsating white dwarfs
could be  employed to  determine the effect  of neutrino cooling  as a
star becomes  a white dwarf ---  see also Winget et  al.  (1983).  The
influence  of neutrino  energy loss  on $\dot{\Pi}$  was  discussed in
detail by  Kawaler et  al. (1986) for  the case  of DBV and  DOV white
dwarfs,  see also  Winget et  al (2004).   In addition,  Isern  et al.
(1992), C\'orsico et al. (2001b)  and Bischoff-Kim et al. (2008b) have
explored the  effect on the  rate of period  change of DAV stars  of a
hypothetical  axion emissivity.   Finally, O'Brien  \&  Kawaler (2000)
have   discussed  the   possibility   of  inferring   limits  on   the
theoretically determined plasmon  neutrino emission rates by employing
DOV white dwarfs.

\subsection{Period-to-period fits: the asteroseismological models}
\label{period-to-period}

In  this  approach  we  seek  pulsation models  that  best  match  the
individual pulsation periods of  the target pulsating white dwarf.  To
measure the  goodness of the  match between the  theoretical pulsation
periods  ($\Pi_k^{\rm   T}$)  and  the   observed  individual  periods
($\Pi_i^{\rm O}$), a ``quality function'' is used:

\begin{equation}
\chi^2(M_*,T_{\rm eff})= \frac{1}{n}\sum_{i=1}^{n}   
\min\left[(\Pi_i^{\rm O}-\Pi_k^{\rm T})^2\right]
\end{equation}

\noindent where $n$ is the  number of observed periods.  For each star
of  interest, the model  that shows  the lowest  value of  $\chi^2$ is
adopted as the ``best-fit model''.

\begin{figure}
\centering
\includegraphics[width=0.9\columnwidth,clip]{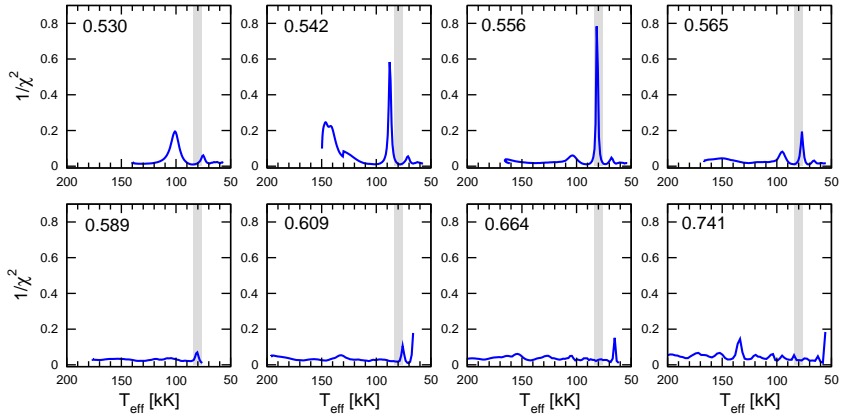}
\caption{The  inverse of  the quality  function of  the period  fit in
         terms of  the effective temperature.   Each panel corresponds
         to  a given  value of  the stellar  mass.  The  vertical grey
         strip indicates the spectroscopic  value of $T_{\rm eff}$ and
         its uncertainties.  Note the strong maximum of the inverse of
         the quality  function for the case $M_*=  0.556 \, M_{\odot}$
         and $T_{\rm  eff}= 81\,500$ K, corresponding  to the best-fit
         model  for the  pulsating  PG 1159  star  PG 0122+200.   From
         C\'orsico et al (2007b).}
\label{chi2} 
\end{figure}

We illustrate  this procedure in  Fig. \ref{chi2}, where  the quantity
$(\chi^2)^{-1}$ in  terms of  the effective temperature  for different
stellar  masses is  displayed  for  the pulsating  PG  1159 star  \pg,
together with  the corresponding spectroscopic  effective temperature.
The quantity $(\chi^2)^{-1}$ is  preferred instead of $\chi^2$ because
it emphasizes those  models that provide good a  agreement between the
observed and the theoretical  periods.  Obviously, the lower the value
of $\chi^2$, the better the  period match.  Note the strong maximum of
the inverse of  the quality function in Fig.   \ref{chi2} for the case
$M_*=  0.556  \,  M_{\odot}$   and  $T_{\rm  eff}=  81\,500$  K.   The
associated PG  1159 model  is adopted as  the best-fit model  for \pg.
The quality  of the period fit  can be quantitatively  measured by the
average of the  absolute period differences, $\overline{\delta \Pi_i}=
(\sum_{i=1}^n |\delta  \Pi_i|)/n$, where $\delta  \Pi_i= \Pi_i^{\rm O}
-\Pi_k$,  and  by  the  root-mean-square  residual,  $\sigma_{_{\delta
\Pi_i}}=   \sqrt{(\sum  |\delta  \Pi_i|^2)/n}   $.   In   the  example
considered  here  for  \pg,  $\overline{\delta  \Pi_i}=  0.88$  s  and
$\sigma_{_{\delta  \Pi_i}}=   1.27$  s  was   obtained  (C\'orsico  et
al. 2007b).  It is worth mentioning  that in some cases it is possible
to get a stellar model (the best-fit model) that nicely reproduces the
period  spectrum  observed  in  pulsating stars  without  artificially
tuning  the value of  structural parameters  which, instead,  are kept
fixed at the values  predicted by the evolutionary calculations.  This
is  the case  of  the  asteroseismological studies  on  PG 1159  stars
carried out by C\'orsico et al. (2007ab, 2008, 2009b).

We close this section by noting that the period-fit approach described
here yields an asteroseismological model  from which one can infer, in
addition to  $M_*$, the luminosity,  radius, gravity, and  distance of
the target star. In addition, the period-fit approach does not require
--- in principle  --- external  constraints such as  the spectroscopic
value  of $T_{\rm  eff}$,  i.e.,  the method  works  ``by letting  the
pulsation modes speak for themselves''  --- see Metcalfe (2005) for an
interesting discussion about this issue.

\subsection{Non-linear light-curve fits: constraining convection?}
\label{nonlinear}

Light curves of pulsating white dwarfs exhibit a variety of shapes, in
some cases being very simple and nearly sinusoidal, and in other cases
exhibiting complex,  strongly non-linear features  that frequently are
variable with  time --- see  Figs.  17 to  19 of Fontaine  \& Brassard
(2008).   In  the  last  case,  the Fourier  transform  contains  many
frequencies  that  are not  genuine  eigenfrequencies  (that is,  they
cannot be  labelled with numbers $\ell,  m, k$), but  instead they are
just linear  combinations and harmonics of  real eigenfrequencies.  As
such, they  do not carry information  about the interior  of the star.
However, they can  be exploited to extract information  of the surface
layers.  As such, the study of non-linearities in the light curve of a
pulsating white  dwarf can be considered  as other asteroseismological
(but non-conventional) tool.

Assuming that non-linearities  in the light curves appear  as a result
of the  non-linear response of the convection  zone, Montgomery (2005)
proposed  a new  technique for  fitting observed  non-sinusoidal light
curves  to study  convection in  white dwarfs,  allowing to  infer the
depth  of the  outer convection  zone, and  how the  thickness  of the
convection zone changes as  a function of effective temperature. Also,
this technique allows to place constraints on mode identifications for
the pulsation modes.  Montgomery (2005) has been able to reproduce the
pulse shape of a large-amplitude mode  in the DAV G 29$-$38 and in the
DBV star  PG 1351+489. More recently, Montgomery  (2008b) has improved
and extended his technique and has applied it to the multiperiodic DBV
star GD  358 and has  also re-examined the  case of PG  1351+489.  His
results indicate that GD 358  must have a thicker convection zone than
PG 1351+489.

We stress  that the  technique of Montgomery  (2005) assumes  that the
non-linearities in the light curves  of pulsating white dwarfs are due
exclusively to the  non-linear response of the convection  zone to the
sinusoidal temperature perturbation entering  at its basis. Adopting a
different point of view, Brassard  et al. (1995) have assumed that the
non-linearities in the light curve  are due to the non-linear response
of   the  emergent   radiative   flux  to   a   perturbation  of   the
temperature. Brassard  et al. (1993)  and Fontaine \&  Brassard (1994)
have  been able  to explain  the  presence of  eight non-linear  peaks
(harmonic and cross-frequencies of  the three eigenmodes) in the light
curve  of the  DAV star  G 117$-$B15A,  without the  necessity  of the
existence of  a convection  zone, a  fact that is  not crucial  in the
theory developed by Brassard et al.  (1995). This does not mean in any
way that convection  does not affect the non-linearities  in the light
curve of  a star, but instead,  that there must  be non-linear effects
due exclusively  to the non-linear response of  the emergent radiative
flux,  irrespective of  the  presence  or not  of  a outer  convective
region.


\section{The various families of white dwarf pulsators}
\label{families}

Below, we  review our knowledge  of asteroseismology of  the different
classes of  pulsating white dwarfs,  and we summarize the  most recent
results.

\subsection{ZZ Ceti stars}
\label{ZZceti}

Pulsating white dwarfs  with H-rich atmospheres, the ZZ  Ceti stars or
simply DAVs, are the most  numerous among the degenerate pulsators. ZZ
Ceti  stars are  found  within  a very  narrow  interval of  effective
temperatures ($10,500$  K $\lesssim  T_{\rm eff} \lesssim  12,500$ K).
Their  brightness  variations,  which  reach  up to  $0.30$  mag,  are
interpreted as being caused by spheroidal, non-radial $g$-modes of low
degree ($\ell  \leq 2$) and  low and intermediate overtones  $k$, with
periods  between  100 and  1,200  s.   Radial  modes ($\ell=  0$)  and
non-radial  $p$-modes,  although  found  overstables in  a  number  of
theoretical  studies  of pulsating  DA  white  dwarfs  --- see,  e.g.,
Vauclair (1971) and Saio et al.  (1983) --- have been discarded as the
cause of variability  in such stars, because the  periods involved are
shorter than $10$ s.  Observationally, these high-frequency signatures
have not been  detected thus far (Silvotti et  al. 2007).  With regard
to the mechanism that drives pulsations, the $\kappa-\gamma$ mechanism
is the traditionally accepted one (Dolez \& Vauclair 1981; Dziembowski
\& Koester 1981; Winget et al.  1982a).  Nonetheless, Brickhill (1991)
proposed the ``convective driving'' mechanism as being responsible for
the overstability  of $g$-modes in DAVs  --- see also  Goldreich \& Wu
(1999).  Although  both mechanisms  predict roughly the  observed blue
edge of the instability strip, none  of them is capable to explain the
observed red edge, where pulsations of DA white dwarfs seemingly cease
(Kanaan 1996; Kotak et al. 2002b; Mukadam et al.  2006).

\subsubsection{Observations of DAVs: the purity of the instability strip}

A fundamental observational  problem related to the DAVs  is the quest
for purity of the ZZ Ceti instability strip.  That is, the question is
if all DAs located within the instability strip are variable, or there
exist stars in this region  which do not exhibit flux variations. This
problem is of particular relevance because if the instability strip is
pure  all  DA  white  dwarfs  must experience  a  phase  of  pulsation
instability.   This implies  that any  information about  the internal
structure of  DAVs inferred using  asteroseismology is also  valid for
all  the  DAs  in  general.   The  study of  the  purity  of  the  DAV
instability strip  is sensitive to  two factors: (i) the  accuracy and
sensitivity of the photometric  observations, and (ii) the accuracy of
the determination  of $T_{\rm  eff}$ and $\log  g$.  As for  the first
point, in  searching for photometric  variations, a given star  can be
pulsating with amplitudes below  the detection limit of the telescopes
employed.      This    can    render     a    pulsating     star    as
non-variable. Concerning the second point, if a non-pulsating star has
its values  of $\log  g$ and $T_{\rm  eff}$ poorly determined,  it may
appear  within  the  instability  strip  very  close  to  the  genuine
variables.  Employing high  signal-to-noise spectra,  Bergeron  et al.
(1995,  2004), and Gianninas  et al.  (2005, 2006)  have found  a pure
instability strip  for the brightest  DAVs.  On the other  hand, there
are about 20 DA stars  located within the instability strip, for which
pulsations are not detected (Mukadam et al. 2004).  These stars suffer
from a  poor determination of  their surface parameters. On  the other
hand, Castanheira et al.   (2007) have found low amplitude variability
for two stars previously  reported as non-variables. At present, there
is solid  evidence suggesting that  the DAV instability strip  is pure
(Winget \& Kepler 2008).

It is worth  noting as well that  a large number of DAVs  are known, a
fact that makes them as a very attractive subject of study to the eyes
of asteroseismology.  Indeed, to date  (late 2009) there are about 143
DAVs known, of which 83 were discovered in the SDSS since 2004 (Winget
\& Kepler 2008).  This number  is adequate to attempt a detailed study
of the group  properties of ZZ Ceti stars (Mukadam  et al.  2006). The
first steps  in this direction have  been done by  Clemens (1993). The
instability  strip is  populated by  two categories  of  objects.  The
first group is composed by hot ZZ Ceti stars (or hDAVs) that are close
to  the blue edge.  These stars  are characterized  by a  very reduced
number of periods, low amplitude of pulsations, and light curves which
are almost  sinusoidal and  very stable in  time. The second  group is
made of cool ZZ Ceti stars (or cDAVs) which are close to the red edge,
exhibit  many periods, large  amplitudes of  pulsation, non-sinusoidal
light  curves  which  are  very  unstable  on  short  timescales,  and
populated  by  many  linear  combinations  and  harmonics  of  genuine
eigenfrequencies.  Mukadam  et al. (2006) suggest  introducing a third
class of  ZZ Ceti stars, to  be called the  intermediate DAVs (iDAVs).
It  would be  the evolutionary  subclass adjoining  the hDAVs  and the
cDAVs, showing a large range of pulsation periods. Generally, there is
a  clear correlation  between  effective temperature  and period:  the
cooler (hotter) the star, the longer (shorter) the periods (Mukadam et
al.   2006).  In  addition, there  is a  (weaker)  correlation between
pulsation period and gravity (Fontaine \& Brassard 2008).

\subsubsection{DA white dwarf models and asteroseismological 
studies of DAVs}

The  first published  complete set  of DA  and DB  white  dwarf models
suitable  for asteroseismology was  that presented  by Tassoul  et al.
(1990).  A  large parameter  space was explored  in such  a monumental
study, and  for a  long time  (since the early  eighties) this  set of
models  represented the  state-of-the-art in  the area.  The pulsation
properties of  these models  were thoroughly explored  in a  series of
important  papers  by  Brassard  et  al.   (1991,  1992a,  1992b).  As
important as these models were at that time, they suffer from a number
of  shortcomings:  the core  of  the  models is  made  of  pure C,  in
opposition to the results  of standard evolutionary calculations, that
indicate that the cores of typical  white dwarfs are made of a mixture
of C and  O; the C/He and He/H chemical interfaces  are modeled on the
basis  of the  assumption of  the diffusive  equilibrium in  the trace
element approximation, an approach that involves a quasi-discontinuity
in the  chemical profiles at  the chemical interfaces which,  in turn,
leads  to peaked  features in  the Brunt-V\"ais\"al\"a  frequency that
produces  excessive mode-trapping  effects (C\'orsico  et  al.  2002a;
2002b).  These models were employed for asteroseismological inferences
of the DAVs G 226$-$29 (Fontaine et al.  1992) and GD 154 (Pfeiffer et
al. 1996).

The  models  presented by  Bradley  (1996)  constituted a  substantial
improvement in  the field.  These  models have carbon-oxygen  cores in
varying proportions,  and the CO/He  and He/H chemical  interfaces are
more realistic. Maybe  the most severe shortcoming of  these models is
the (unrealistic)  ramp-like shape of the  core carbon-oxygen chemical
profiles.   These DA  models  were  the basis  of  the very  important
asteroseismological studies on the DAVs G 29$-$38 (Bradley \& Kleinman
1997), G  117$-$B15A and  R 548  (Bradley 1998), GD  165 and  L 19$-$2
(Bradley 2001), and G 185$-$32 (Bradley 2006).

The next step in improving the modeling of DAVs was given by C\'orsico
et al. (2002b) and Benvenuto  et al. (2002), who employed evolutionary
models  characterized by  He/H  chemical interfaces  resulting from  a
time-dependent  element  diffusion  treatment  (Althaus  \&  Benvenuto
2000), and the core chemical structure extracted from the evolutionary
calculations of Salaris et al.  (1997).  The employment of very smooth
chemical interfaces,  as shaped  by chemical diffusion,  revealed that
the  use  of  the  trace  element approximation  is  inappropriate  in
pulsational  studies.   This  grid   of  models  was  employed  in  an
asteroseismological study of G 117$-$B15A (Benvenuto et al. 2002).  It
is worth  noting that the starting configurations  for these sequences
were obtained through an artificial  procedure, and not as a result of
evolutionary calculations of the progenitor stars.

Recently,  Castanheira \&  Kepler  (2008, 2009)  have  carried out  an
extensive asteroseismological  study of DAVs employing  DA white dwarf
models  similar to  those of  Bradley  (1996), but  with a  simplified
treatment of  the core  chemical structure. Specificallym,  they fixed
the  central abundances  to 50  \% of  O  and 50  \% of  C.  The  He/H
chemical interfaces adopted for  these models are a parametrization of
chemical  profiles  resulting  from time-dependent  element  diffusion
(Althaus et  al.  2003).  This  study includes the  ``classical'' DAVs
and  also the recently  discovered SDSS  DAVs.  In  total, 83  ZZ Ceti
stars were analyzed.  A very  relevant result of these studies is that
the thickness of the H  envelopes inferred from asteroseismology is in
the  range $10^{-4} \gtrsim  M_{\rm H}/M_*  \gtrsim 10^{-10}$,  with a
mean  value of  $M_{\rm  H}/M_*= 5  \times  10^{-7}$.  This  important
result  suggests that  DA  white dwarfs  with envelopes  substantially
thinner  than predicted by  the standard  theory of  stellar evolution
could  exist.  However,  these  results do  not  include the  possible
effects  of  realistic  carbon-oxygen  profiles on  the  fits.  Almost
simultaneously  to the study  of Castanheira  \& Kepler  (2008, 2009),
Bischoff-Kim et al.  (2008a)  have performed a new asteroseismological
study  on G  117$-$B15A  and R  548  employing DA  white dwarf  models
similar to those used by  Castanheira \& Kepler (2008, 2009), but with
core chemical profiles similar to those of Salaris et al. (1997).  The
results of this  work are in agreement with  previuos studies of these
ZZ  Ceti  stars.  Recently,  the  models  and the  asteroseismological
approach of Bischoff-Kim  et al.  (2008a) have been  employed to infer
the internal structure and seismological  distance of the ZZ Ceti star
KUV  02464$+$3239  (Bogn\'ar  et  al.  2009).   Finally,  Bischoff-Kim
(2009) has presented the results of an asteroseismological analysis of
two rich DAVs, G38$-$29 and R 808.

A summary of the  asteroseismological inferences on individual ZZ Ceti
stars performed so far is presented in Table 7 of Fontaine \& Brassard
(2008).  As  evident from  what we said  in the  preceding paragraphs,
there is no  asteroseismological study of DAVs to  date based on fully
evolutionary   models  (generated  from   the  ZAMS)   that  considers
physically sound  treatments to model  the chemical interfaces  at the
core  {\sl  and}  the envelope  of  white  dwarfs.  This would  be  an
important  point for  future  asteroseismological studies  of ZZ  Ceti
stars.  Realistic  stellar models are crucial to  correctly assess the
adiabatic period  spectrum and mode-trapping properties of  the DAs, a
very relevant aspect of  white dwarf asteroseismology (Brassard et al.
1991; Brassard et al. 1992a,b; Bradley 1996; C\'orsico et al. 2002).

\subsubsection{The paradigmatic ZZ Ceti star G 117$-$B15A}

Now, we  briefly revise  some outstanding results  in the field  of DA
white dwarf asteroseismology.  In particular,  we focus here on a very
well studied  ZZ Ceti star: G  117$-$B15A.  This star  is an otherwise
typical DA  white dwarf,  the variability of  which was  discovered by
McGraw  \& Robinson  (1976) and,  since  then, it  has been  monitored
continuously.   The mass and  the effective  temperature of  this star
have been the subject of numerous spectroscopic re-determinations.  In
particular,   values  of   $0.59   \,  M_{\odot}$   and  $11,620$   K,
respectively, have  been derived by Bergeron et  al.  (1995).  Koester
\& Allard  (2000) have reported a  somewhat lower value  for the mass,
$M_*= 0.53\,  M_{\odot}$ and  a higher effective  temperature, $T_{\rm
eff}=  11,900 \pm  140$ K.   G 117$-$B15A  has oscillation  periods of
215.2, 271 and 304.4 s (Kepler et al.  1982).  As mentioned, this star
has  been one of  the DAVs  analyzed by  Bradley (1998).   This author
obtains two  different structures according  the $k$-identification of
the modes exhibited by the star.   If the periods at 215, 271, and 304
s are associated with $k= 1$,  $k= 2$, and $k= 3$, this author obtains
an  asteroseismological  model with  $T_{\rm  eff}=  12,160$ K,  $M_*=
0.55\,  M_{\odot}$, $M_{\rm  H}/M_*=  3 \times  10^{-7}$, and  $M_{\rm
He}/M_*= 10^{-2}$. If,  instead, the periods have $k=  2$, $k= 3$, and
$k=  4$, the  asteroseismological  model is  characterized by  $T_{\rm
eff}= 12,530$  K, $M_*= 0.55\, M_{\odot}$, $M_{\rm  H}/M_*= 1.5 \times
10^{-4}$, and  $M_{\rm He}/M_*= 10^{-2}$.   Note that there  are three
orders  of magnitude  of  difference in  the  mass of  the H  envelope
between  the  two  possible  (and equally  valid)  asteroseismological
solutions.    This  degeneracy  of   seismological  solutions   for  G
117$-$B15A  has also been  found by  Benvenuto et  al.  (2002)  on the
basis  of  independent   stellar  and  pulsation  modeling.   Finally,
Bischoff-Kim et al.   (2008a) also find two classes  of solutions, one
characterized  by  ``thin''  H  envelopes and  other  associated  with
``thick''  H envelopes,  although their  ``thick''  envelope solutions
($M_{\rm  H}/M_*= 6  \times  10^{-7}$) are  considerably thinner  than
those of the previous works.

G  117$-$B15A has  a very  remarkable property:  its 215  s  period is
extremely stable, a fact that has enabled to detect its rate of period
change,  after more  than 30  years of  observations.   In particular,
Kepler et  al. (2005)  have reported a  value of $\dot{\Pi}=  3.57 \pm
0.82 \times 10^{-15}$ s/s. This implies  a variation of 1 s in $\sim 9
\times  10^{6}$ yr,  making this  star the  most stable  known optical
clock. Theoretical models indicate that this mode is varying its 215 s
period  by virtue  of  cooling of  the  core.  The  inferred value  of
$\dot{\Pi}$ is  compatible with  a core made  of carbon and  oxygen in
(yet)  unknown  proportions.   It   is  expected  that  the  value  of
$\dot{\Pi}$ will be refined in the  next years so that we will be able
to obtain the fractions of carbon and oxygen in the core and, thus, to
place   constraints   on   the   cross   section   of   the   reaction
$^{12}$C$(\alpha,   \gamma)^{16}$O.   The   most   recent  value   for
$\dot{\Pi}\  (215\ {\rm  s})$ reported  by Kepler  (2009)  is somewhat
larger, of $4.77 \pm 0.59 \times 10^{-15}$ s/s.

As  a further  application of  the  measurement of  $\dot{\Pi}$ for  G
117$-$B15A, we briefly describe the attempts to put constraints on the
mass of  axion, a weakly  interacting particle whose  existence (still
not demonstrated) has been postulated to solve the CP symmetry problem
in the  standard model. Axions are very  interesting particles because
they are  plausible candidates for  dark matter. The theory  of axions
does not provide any clue about  the value of the axion mass, although
it  is known  that  the more  massive  the axion,  the stronger  their
interaction with matter, and in turn, the larger the emissivity. Isern
et al. (1992) devised the following ingenious method to infer the mass
of the axion.  They considered  the evolution of white dwarfs with and
without  axion  emissivity, and  compared  the  theoretical values  of
$\dot{\Pi}$ for increasing masses of  the axion with the observed rate
of  period change of  G 117$-$B15A.   Isern et  al. (1992)  obtained a
value  of  8.7 meV  (assuming  $\cos^2\beta= 1$)  on  the  basis of  a
semi-analytical treatment.   Later, C\'orsico et al.  (2001b) found an
axion mass of  4 meV using a detailed  asteroseismological model for G
117$-$B15A.   Recently, Bischoff-Kim  et al.  (2008b) have  obtained a
upper limit  of $13.5-26.5$ meV using  an improved asteroseismological
model for  the star and  a better treatment  of the error bars  on the
calculated $\dot{\Pi}$  values. Since both  the value of  the measured
$\dot{\Pi}$, as  well as  the modeling of  DA white dwarfs,  have been
changing over the  years, it is expected that  these estimates for the
axion mass could change in the near future (Isern et al. 2010).

\subsection{DQVs}

Hot DQ white dwarfs characterized by carbon-dominated atmospheres were
recently  discovered  at   effective  temperatures  of  between  $\sim
18,000$~K and $\sim  24,000$~K by Dufour et al.   (2007). This finding
could be  indicative of  the existence a  new evolutionary  channel of
white-dwarf  formation (see  Sect.  \ref{hotdq}).   Shortly  after the
discovery of hot  DQs, Montgomery et al. (2008a)  reported the finding
of the  first variable hot  DQ star, SDSS  J142625.70$+$575218.4 (with
$\log g  \sim 9$ and  $T_{\rm eff} \sim  19,800$ K), with  a confirmed
period  $\Pi \approx  418$ s.   Shortly later,  Barlow et  al.  (2008)
reported the discovery  of two additional variable hot  DQ stars, SDSS
J220029.08$-$074121.5 ($\log  g \sim 8$, $T_{\rm eff}  \sim 21,240$ K)
and  SDSS J234843.30$-$094245.3 ($\log  g \sim  8$, $T_{\rm  eff} \sim
21,550$ K), with periods $\Pi \approx 656$ s and $\Pi \approx 1052$ s,
respectively.    The  measured   periodicities  were   interpreted  as
non-radial $g$-mode pulsations, similar to the well-studied pulsations
of the GW Vir, V777 Her, and ZZ Ceti classes of white-dwarf variables.
The pulsation hypothesis, however,  was defied by the possibility that
these  stars  could be  AM  CVn  systems,  because of  the  similarity
exhibited in  the pulse shape of  the light curves  (Montgomery et al.
2008a).   On  the  other  hand,  a  compelling  argument  against  the
interacting binary hypothesis is that  it does not explain why all hot
DQ white dwarfs are grouped within the same range of temperatures, and
none at higher or lower  effective temperatures (Dufour et al. 2009a).
Recently, the pulsating nature of the variable hot DQ white dwarfs has
been  conclusively   confirmed  at  least   for  one  of   them  (SDSS
J142625.70$+$575218.4) by Green et  al. (2009), who have discovered an
additional period $\Pi \sim 319.7$  s (apart from the already known at
$418$ s).  On the other hand,  Dufour et al.  (2009b) have carried out
additional  observations   of  SDSS  J220029.08$-$074121.5   and  SDSS
J234843.30$-$094245.3. They have found  evidence of two new periods at
254.7 s and 577.6 s for the former star, and also a probable period at
417 s for the latter star. These authors also found evidence that SDSS
J220029.08$-$074121.5 could have  a magnetic field, as is  the case of
SDSS   J142625.70$+$575218.4  (Dufour   et  al.    2008b),   but  SDSS
J234843.30$-$094245.3 not. Table  \ref{dqvs} summarizes the properties
of  the known  DQVs. The  fifth  column gives  the spectroscopic  mass
derived from the evolutionary tracks of Althaus et al. (2009a).

\begin{table*}
\scriptsize
\begin{center}
\caption{Stellar  parameters  and pulsation  properties  of the  three
         known DQV stars.}
\begin{tabular}{lccccccc}
\hline
\hline
 Star            & Mag.    & $T_{\rm eff}$ & $\log g$ & $M_*$         & $\Pi$          & Magnetic? &Ref.\\
                 & ($g$)   & [K]           & [cgs]    & $[M_{\odot}]$ & [s]            &           &    \\
\hline
SDSS J142625.70$+$575218.4 & 19.16 & $19\,800$ & 9 & $\sim 1.2$ & $320-418$  & yes & 1\\
SDSS J220029.08$-$074121.5 & 17.70 & $21\,240$ & 8 & $\sim 0.6$ & $255-578$  & yes & 2\\   
SDSS J234843.30$-$094245.3 & 19.00 & $21\,550$ & 8 & $\sim 0.6$ & $417-1044$ & no  & 2\\
\hline
\hline
\end{tabular}
\end{center}
{\footnotesize References:
$^{1}$Montgomery et al. (2008a) and Green et al. (2009),
$^{2}$Barlow et al. (2008) and Dufour et al. (2009a).} 
\label{dqvs}
\end{table*}  

The  origin  of  variability  in  hot DQ  stars  was  first  addressed
theoretically  by Fontaine  et al.   (2008), following  the hypothesis
that the  variability could be  caused by pulsations. They  found that
$g$-modes can  be excited in the  range of temperature  where real DQs
are  found  (below  $\sim 21,500$  K),  but  only  if the  models  are
characterized by substantial amounts of He ($X_{\rm He} \gtrsim 0.25$)
in their carbon-rich envelopes.   Dufour et al.  (2008a) estimated the
$T_{\rm eff}$, $\log g$, and C/He ratio of the nine known hot DQ stars
and constructed  a dedicated stellar  model for each object  using the
same modeling  as in Fontaine  et al.  (2008).  Using  a non-adiabatic
approach, Dufour et al.  (2008a) predicted that only SDSS J1426$+$5752
should exhibit  pulsations, and failed to predict  variability in SDSS
J220029.08$-$074121.5  and  SDSS  J234843.30$-$094245.3.  However,  it
appears  that the  pulsation models  of  Fontaine et  al.  (2008)  and
Dufour et al.  (2008a) are not entirely consistent with their proposed
evolutionary  picture  for the  formation  of  hot  DQs.  Indeed,  the
background models  they assumed  for their stability  calculations are
characterized by a He content  several orders of magnitude higher than
the  content of  He required  by their  evolutionary scenario  to work
(Dufour  et al.   2008a).  In  fact,  the bottom  of the  He-dominated
envelope in their stellar models is located at a fractional mass depth
of  $\log  q_{\rm  env}   \equiv  \log(1-M_r/M_*)=  -3$  (Fontaine  et
al. 2008).

\begin{figure}
\centering
\includegraphics[width=0.9\columnwidth,clip]{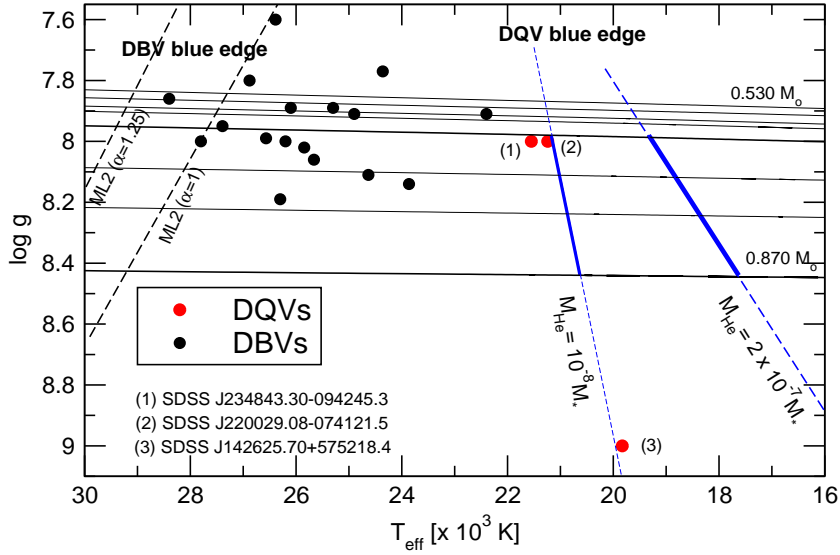}
\caption{A $\log  g-T_{\rm eff}$ diagram  showing the location  of the
         three known variable hot DQs (the DQVs) with red circles, and
         the  known  seventeen  DBVs  (black  circles).   Evolutionary
         tracks corresponding  to DB  white dwarf models  with stellar
         masses of  0.530, 0.542,  0.556, 0.565, 0.589,  0.664, 0.742,
         and $0.87\,  M_{\odot}$ (from top to bottom)  are shown using
         solid  curves.    The  blue  edge  of   the  theoretical  DBV
         instability strip  (taken from  C\'orsico et al.   2009a), is
         displayed with black dashed lines.   The blue edge of the DQV
         instability domain  is drawn  with thick blue  solid segments
         for two  values of $M_{\rm He}$.   For illustrative purposes,
         these  lines are  extended to  high and  low  gravities (blue
         dashed lines).}
\label{logg-logteff-observed}
\end{figure}

Recently,  C\'orsico et al.   (2009d) have  presented a  new pulsation
stability analysis  of carbon-rich hot  DQ white dwarf stars  with the
aim to  test the convective-mixing picture  for the origin  of hot DQs
(Dufour et  al.  2007; Althaus et  al.  2009b). This  is done studying
their pulsational properties.  C\'orsico et al.  (2009d) have employed
the  full evolutionary  models of  hot  DQ white  dwarfs developed  by
Althaus et al.  (2009b), which consistently cover the entire evolution
from the born-again  stage to the DQ white  dwarf stage.  According to
this  study, overstable  $g$-modes in  hot DQ  white dwarf  models are
primarily  driven  by  the  $\kappa$-mechanism caused  by  the  strong
destabilizing effect of the opacity bump due to the partial ionization
of C, the role of the partial ionization of He{\sc ii} being much less
relevant. Also, the  theoretical blue edge of DQVs  is hotter for less
massive     models    than    for     more    massive     ones    (see
Fig. \ref{logg-logteff-observed}),  which is at odds  with the results
of  Fontaine  et al.   (2008)  and  Montgomery  et al.   (2008a).   In
addition, the blue edge for DQV stars is hotter for smaller amounts of
He in the envelopes.  In summary, the calculations of C\'orsico et al.
(2009d)  support  the  diffusive/convective  mixing  picture  for  the
formation  of  hot  DQs,  and  in  particular,  demonstrate  that  the
diffusive/convective mixing  scenario is not only able  to explain the
origin of hot  DQ white dwarfs, but also  accounts for the variability
of these stars.   We stress that the conclusions  reached in C\'orsico
et al.  (2009d),  and also the results of Fontaine  et al.  (2008) and
Dufour et  al.  (2008a), especially  those concerning the  location of
the  blue  edge  of  the  DQV  instability  strip,  could  be  altered
substantially  if  a fully  consistent  treatment  of the  interaction
between  convection  and  pulsation  were included  in  the  stability
analysis.

\subsection{V777 Her stars}

V777  Her (or  DBV)  stars  are $g$-mode  variable  white dwarfs  with
He-rich  atmospheres  (DB)  and  intermediate  effective  temperatures
($T_{\rm eff} \sim  25,000$ K), that pulsate with  periods between 100
and 1,100 s.  The pulsating  nature of DBVs was predicted almost three
decades  ago on  theoretical grounds  by  Winget et  al.  (1982a)  and
shortly  after confirmed  observationally by  Winget et  al.  (1982b).
Since then, considerable effort has been devoted to study these stars.
In  particular, the multiperiodic  star GD  358, the  most extensively
studied  member of the  DBV class,  has been  the subject  of numerous
investigations  devoted  to  disentangle  its internal  structure  and
evolution,  initially  by means  of  ``hands on''  asteroseismological
procedures (Bradley  and Winget  1994b) and later  employing objective
fitting techniques --- see, e.g.,  Metcalfe et al. (2000, 2001, 2002).
In  particular,   Metcalfe  et   al.   (2001)  have   applied  genetic
algorithm-based    procedures   to    place    constraints   on    the
$^{12}$C$(\alpha,\gamma)^{16}$O reaction rate  from inferences for the
abundance of central oxygen in GD 358.

\subsubsection{Observations of DBVs and the presence of H}

There are currently 20 known DBVs,  9 of which have been discovered by
Nitta  et al.   (2009) in  the SDSS  and 2  have recently  reported by
Kilkenny  et  al.  (2009).   In  Table \ref{DBVs}  we  list  the  main
properties of  all known  DBVs. Although the  number of DBVs  has been
considerably enlarged,  it still  is not sufficient  as to  assess the
group characteristics.

\begin{table*}
\scriptsize
\begin{center}
\caption{Stellar parameters and pulsation properties of all known V777
         Her (or DBV) stars.}
\label{DBVs}
\begin{tabular}{lccccccc}
\hline
\hline
 Star            & Mag.    & $T_{\rm eff}$ & $\log g$ &  $\Pi$            & Ref.\\
                 &         & [K]           & [cgs]    &  [s]              &     \\
\hline
KUV 0513+2605              & 16.70 ($V$) & $26\,300$  & 8.19 & $350-900$  & 1 \\       
CBS 114                    & 17.2  ($B$) & $26\,200$  & 8.00 & $230-670$  & 1 \\
PG 1115+158                & 16.1  ($B$) & $25\,300$  & 7.89 & $831-1072$ & 1 \\
PG 1351+489                & 16.4  ($B$) & $26\,100$  & 7.89 & $333-490$  & 1 \\
PG 1456+103                & 15.9  ($B$) & $22\,400$  & 7.91 & $423-854$  & 1 \\
GD 358                     & 13.65 ($V$) & $24\,900$  & 7.91 & $423-810$  & 1 \\    
PG 1654+160                & 16.2  ($B$) & $27\,800$  & 8.00 & $149-851$  & 1 \\
PG 2246+120                & 16.73 ($V$) & $27\,200$  & 7.89 & $256-329$  & 1 \\
EC 20058$-$5234            & 15.0  ($B$) & $28\,400$  & 7.86 & $195-540$  & 1 \\
SDSS J034153.03$-$054905.8 & 18.25 ($g$) & $25\,087$  & 8.02 &  $942$     & 2 \\  
SDSS J085202.44+213036.5   & 18.50 ($g$) & $25\,846$  & 8.02 &  $951$     & 2 \\ 
SDSS J094749.40+015501.8   & 19.95 ($g$) & $23\,453$  & 8.13 &  $255$     & 2 \\ 
SDSS J104318.45+415412.5   & 18.95 ($g$) & $26\,291$  & 7.77 &  $420$     & 2 \\  
SDSS J122314.25+435009.1   & 18.98 ($g$) & $23\,442$  & 7.84 &  $544-687$ & 2 \\
SDSS J125759.03$-$021313.3 & 19.16 ($g$) & $25\,820$  & 7.57 &  $532-729$ & 2 \\
SDSS J130516.51+405640.8   & 17.46 ($g$) & $24\,080$  & 8.14 &  $658-912$ & 2 \\
SDSS J130742.43+622956.8   & 18.83 ($g$) & $23\,841$  & 8.14 &  $890$     & 2 \\
SDSS J140814.63+003838.9   & 19.19 ($g$) & $26\,073$  & 7.98 &  $258-335$ & 2 \\
EC 04207$-$4748            & 15.3 ($V$)  & $27\,300$  & 7.8  &  $223-448$ & 3 \\
EC 05221$-$4725            & 16.6 ($V$)  & $-$        & $-$  &  $704-976$ & 3 \\ 
\hline
\hline
\end{tabular}
\end{center}
{\footnotesize References: 
$^{1}$Beauchamp et al. (1999) and Fontaine \& Brassard (2008),
$^{2}$Nitta et al. (2009),
$^{3}$Kilkenny et al. (2009).} 
\end{table*}  

The  issue of the  purity of  the instability  strip also  matters for
DBVs, as in the case of DAVs.  However, in addition to the sensitivity
of the photometric observations  and the accuracy of the determination
of $T_{\rm eff}$ and $\log g$, for the DBVs the picture is complicated
further  by the  possible presence  of small  quantities of  H  in the
He-rich atmospheres of DBs.  The presence of H is attributed mainly to
accretion from the  interstellar medium. It is expected  that a modest
accretion  rate of about  $10^{-19}-10^{-21}\, M_{\odot}/$yr  could be
enough  to  explain the  presence  of H  in  the  DBs.  The  effective
temperatures of DB stars derived from model atmospheres that contain H
are considerably lower  than those obtained in the case  in which H is
neglected (Beauchamp  et al. 1999;  Castanheira et al.  2006;  Voss et
al. 2007).   Thus, the  presence of (unknown)  quantities of H  in the
He-rich atmospheres of DB stars has strong impact on the determination
of the  effective temperature and  gravity.  Concerning the  purity of
the DBV instability strip, before arriving at a final conclusion it is
crucial to address first the three above mentioned factors.

\subsubsection{The theoretical blue edge}

\begin{figure}
\centering
\includegraphics[width=0.9\columnwidth,clip]{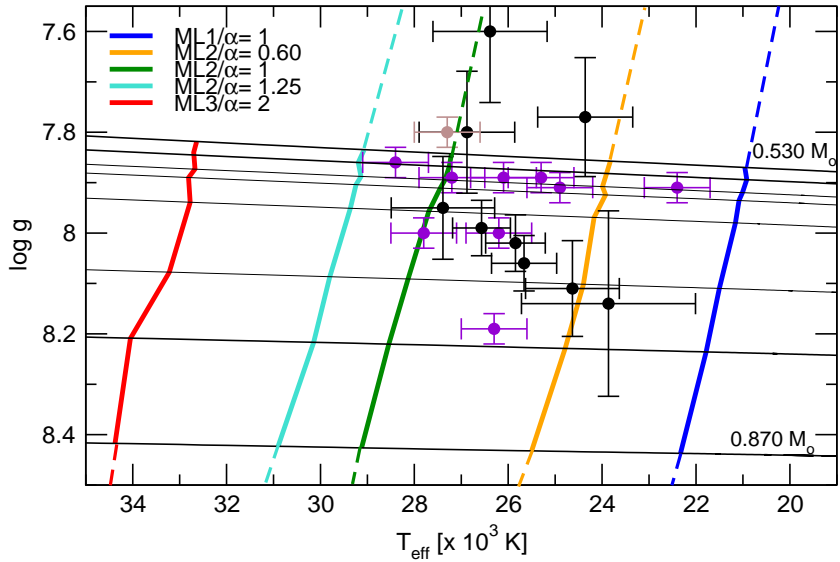}
\caption{The  blue edge  of the  DBV instability  strip  for different
         prescriptions  of the  MLT. The  location of  most  known DBV
         stars is  included --- black  dots: Beauchamp et  al. (1999),
         violet dots: Nitta et al.  (2009), brown dot: Kilkenny et al.
         (2009).  Only the ML2$/\alpha= 1.25$ prescription proposed by
         Beauchamp et al.  (1999) is  able to account for the location
         of all known DBVs.}
\label{banda-dbvs}
\end{figure}

The  pulsations  in DBV  stars  are thought  to  be  triggered by  the
$\kappa-\gamma$ mechanism  acting on the  partial ionization of  He at
the base of the outer convection zone.  The observed instability strip
of the  DBV stars is located  between $T_{\rm eff}  \approx 28\,400$ K
and  $T_{\rm eff}  \approx 22\,500$  K (Winget  \& Kepler  2008).  All
published  stability analysis  of  DB models  ---  see, for  instance,
Bradley \&  Winget (1994a)  and Beauchamp et  al.  (1999)  --- clearly
indicate a strong dependence of location of the theoretical blue (hot)
edge  of the  DBV  instability strip  with  the convective  efficiency
adopted in the  envelope of the stellar models.   At present, there is
no general consensus between different authors about what is the right
convective efficiency in  DB models in order to  fit the observed blue
edge. C\'orsico  et al. (2009a)  have recently explored the  impact of
different  convective  efficiencies and  also  the  presence of  small
amounts of H in the atmospheres of DB stars on the precise location of
the theoretical blue edge of the DBV instability strip.  They employed
a new  set of  fully evolutionary DB  white dwarf models  that descend
from the post-born  again PG 1159 models (Miller  Bertolami \& Althaus
2006) recently presented in Althaus  et al. (2009a).  Details of these
DB models are available in that paper. C\'orsico et al. (2009a) employ
the  following  prescriptions  of  the  MLT, in  order  of  increasing
efficiency:  ML1$/\alpha=  1$,  ML2$/\alpha= 0.60$,  ML2$/\alpha=  1$,
ML2$/\alpha=   1.25$,  and   ML3$/\alpha=  2$,   $\alpha$   being  the
characteristic  mixing length  (Tassoul et  al. 1990).   The resulting
theoretical blue edges of the  DBV instability strip are shown in Fig.
\ref{banda-dbvs},  where the  case  of atmospheres  devoided  of H  is
considered.  Fig.  \ref{banda-dbvs} also  shows the complete set of DB
evolutionary  sequences (solid  lines)  on the  $T_{\rm eff}-\log  g$
diagram, along  with the location of  the known DBV  stars.  Note that
there is a clear dependence of the blue edge with the stellar mass. It
is apparent that  only the ML2$/\alpha= 1.25$ prescription  is able to
account  for  the  location of  all  known  DBVs.   We warn  that  the
non-adiabatic  computations of  C\'orsico et  al.  (2009a)  assume the
``frozen convection''  approximation.  Thus,  the results for  the DBV
theoretical blue edge could somewhat change if the perturbation to the
convective flux were taken into account in the stability analysis.

\begin{figure}
\centering
\includegraphics[width=0.9\columnwidth,clip]{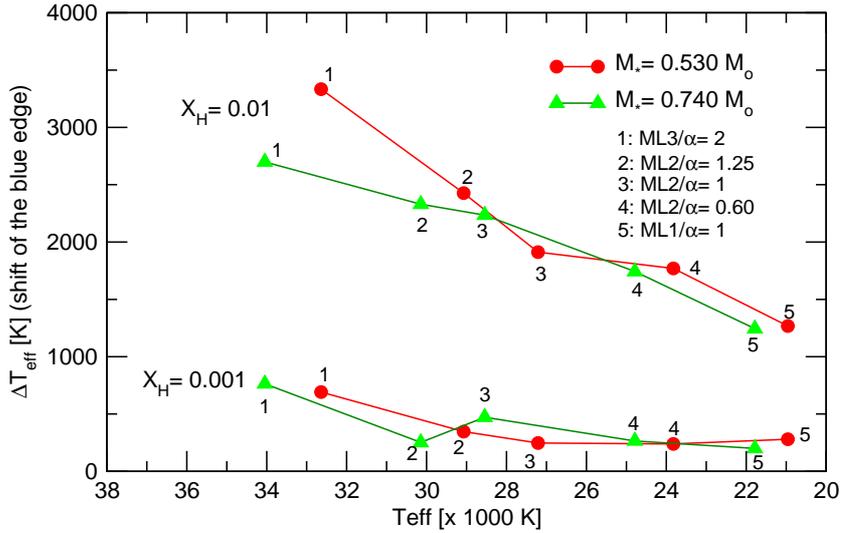}
\caption{The shift (in $T_{\rm eff}$) of  the blue edge as a result of
         the  presence of  H in  the He-rich  atmospheres of  DBVs.  A
         fractional abundance  of H of  0.01 (0.001) is able  to shift
         the  blue edge  towards lower  $T_{\rm eff}$s  in  about 3000
         (800) K for  the case of  ML3.  The shifts  are substantially
         smaller (but non-negligible) for less efficient convection.}
\label{shift}
\end{figure}

The shift  of the effective  temperature of DBVs towards  lower values
when some impurities of H  are considered in the model atmospheres has
a counterpart in the location of the theoretical blue edge of the V777
Her instability strip.  C\'orsico et al.  (2009a) have performed fully
non-adiabatic  pulsation computations  in DB  models  characterized by
some  presence  of  H  in  the almost  He-pure  envelopes.  They  have
restricted themselves to the cases  in which $M_*= 0.530 \, M_{\odot}$
and $M_*=  0.741 \, M_{\odot}$.  For  these two values  of the stellar
mass, they have explored the cases in which $X_{\rm H}= 0.0001, 0.001$
and  $0.01$, corresponding  to  $\log(n_{\rm H}/n_{\rm  He})= \log  (4
X_{\rm  H}/X_{\rm He})=  -3.4,  -2.4$ and  $-1.44$, respectively.  The
results are  depicted in Fig.  \ref{shift}.  When $X_{\rm  H}= 0.0001$
there is  no appreciable effect  on the location  of the blue  edge of
instability.  But  if $X_{\rm H}= 0.001$  the blue edge  is shifted to
lower effective temperatures by $\sim 800$ K (for ML3$/\alpha= 2$) and
$\sim 300$ K (for ML1$/\alpha= 1$). Finally, for $X_{\rm H}= 0.01$ the
shift is of $\sim 3000$ K (for ML3$/\alpha= 2$) and $\sim 1200$ K (for
ML1$/\alpha= 1$).

\subsection{GW Vir stars}

Pulsating PG 1159 stars or  GW Vir variable stars (after the prototype
of the spectral class and the  variable type, \pp\ or GW Vir) are very
hot H-deficient post-AGB  stars with surface layers rich  in He ($\sim
30-85  \%$), carbon  ($\sim 15-60  \%$)  and oxygen  ($\sim 2-20  \%$)
(Werner \& Herwig 2006)  that exhibit multiperiodic, low degree ($\ell
\leq  2$), high  radial  order ($k  \gtrsim  18$) $g$-mode  luminosity
variations with periods  in the range from about  300 to 3000 seconds.
As  mentioned earlier,  some  GW Vir  stars  are still  embedded in  a
planetary nebula  and they  are commonly called  PNNVs.  GW  Vir stars
without  nebulae  are  usually  called  DOV  stars.   PNNV  stars  are
characterized by much higher luminosity  than DOV stars.  GW Vir stars
are  particularly  important  to  infer fundamental  properties  about
pre-white dwarfs in general, such  as the stellar mass and the surface
compositional  stratification (Kawaler \&  Bradley 1994;  C\'orsico \&
Althaus 2006).  In  addition, pulsating PG 1159 stars  have been shown
by C\'orsico \&  Althaus (2005) to be valuable  tools to constrain the
occurrence  of extra mixing  episodes in  their progenitor  stars.  PG
1159  stars are  believed to  be the  evolutionary  connection between
post-AGB stars  and most H-deficient white dwarfs.   For details about
the   origin   and   evolution   of   PG   1159   stars,   see   Sect.
\ref{hdeficient}.

\subsubsection{Observations}

There   exist  numerous   studies   in  the   literature  focused   on
observational aspects  of PG  1159 stars. In  particular, we  draw the
attention of the  reader to the excellent review  article by Werner \&
Herwig  (2006), which  summarizes numerous  spectroscopic  studies and
also theoretical aspects about the  formation of these stars. In Table
\ref{gwvirginis} we  show the main  properties of GW Vir  stars. There
are 16 members of the class known up to now, of which 10 are PPNVs and
6 are  DOVs. The  table also includes  stars that share  the pulsation
properties of GW  Vir stars but do not formally belong  to the PG 1159
spectroscopic class.  They are 5 Wolf-Rayet Central Stars of Planetary
Nebula ([WCE]) type.   Note that NGC 2371$-$2 is  a transition object.
We  have included  PG 1159  stars and  [WCE] stars  in the  same table
because the  physics of mode excitation in  both spectroscopic classes
is the same.  At  least 5 GW Vir stars (\rxj, \pp,  \pt, \pr, and \pg)
have been  intensively observed by  the WET global instrument,  and so
they   are   particularly   interesting  for   asteroseismology   (see
Sect. \ref{astro-infer}).

\begin{table*}
\scriptsize
\begin{center}
\caption{Stellar  parameters  and pulsation  properties  of all  known
         pulsating PG 1159 stars and pulsating [WC] stars.}
\label{gwvirginis}
\begin{tabular}{lccccccc}
\hline
\hline
\noalign{\smallskip}
 Star & $T_{\rm eff}^{\rm a}$ & $\log g^{\rm a}$ & $M_*^{\rm b}$          &  Class & PN? & $\Pi$ & $\Delta \Pi$ \\
      & [kK]                 & [cgs]           & [$M_{\odot}$]        &         &       & [s]   & [s]          \\
\noalign{\smallskip}
\hline
\noalign{\smallskip}
Longmore 4      & 120 & 5.5 & 0.63 & PG 1159      & yes & $831-2325\ ^{\rm c}$   & ---                        \\
Abell 43        & 110 & 5.7 & 0.53 & PG 1159, H   & yes & $2380-6075\ ^{\rm d}$  & ---                        \\
NGC 7094        & 110 & 5.7 & 0.53 & PG 1159, H   & yes & $2000-5000\ ^{\rm e}$  & ---                        \\ 
NGC 246         & 150 & 5.7 & 0.75 & PG 1159      & yes & $1460-4348\ ^{\rm f}$  & ---                        \\
NGC 5189        & 135 & 6.0 & $-$  & [WCE]        & yes & $690\ ^{\rm f}$        & ---                        \\
Sk 3            & 140 & 6.0 & $-$  & [WCE]        & yes & $929-2183\ ^{\rm g}$   & ---                        \\
NGC 2867        & 141 & 6.0 & $-$  & [WCE]        & yes & $769\ ^{\rm f}$        & ---                        \\
NGC 6905        & 141 & 6.0 & $-$  & [WCE]        & yes & $710-912\ ^{\rm f}$    & ---                        \\
NGC 1501        & 134 & 6.0 & $-$  & [WCE]        & yes & $1154-5235 \ ^{\rm h}$ & $22.3\pm 0.3 \ ^{\rm h}$   \\
HE 1429-1209    & 160 & 6.0 & 0.66 & PG 1159      & no  & $919\ ^{\rm i}$        & ---                        \\
RX J2117.1+3412 & 170 & 6.0 & 0.72 & PG 1159      & yes & $694-1530\ ^{\rm j}$   & $21.48\pm 0.04 \ ^{\rm j}$ \\
HS 2324+3944    & 130 & 6.2 & 0.53 & PG 1159, H   & no  & $2005-2569\ ^{\rm k}$  & ---                        \\
NGC 2371-2      & 135 & 6.3 & $-$  & [WC]-PG 1159 & yes & $923-1825\ ^{\rm f}$   & ---                        \\ 
K 1-16          & 140 & 6.4 & 0.54 & PG 1159      & yes & $1500-1700\ ^{\rm l}$  & ---                        \\
Jn 1            & 150 & 6.5 & 0.55 & PG 1159      & yes & $454\ ^{\rm m}$        & ---                        \\
VV 47           & 130 & 7.0 & 0.53 & PG 1159      & yes & $261-4310\ ^{\rm m}$   & $24.21  ^{\rm r}$          \\
NGC 6852        & $-$ & $-$ & $-$  & PG 1159      & yes & $1096-5128\ ^{\rm m}$  & ---                        \\
PG 1159$-$035   & 140 & 7.0 & 0.54 & PG 1159      & no  & $339-982\ ^{\rm n}$    & $21.5\pm 0.1 \ ^{\rm n}$   \\
PG 2131+066     &  95 & 7.5 & 0.55 & PG 1159      & no  & $339-598\ ^{\rm o}$    & $21.6\pm 0.4 \ ^{\rm s}$   \\
PG 1707+427     &  85 & 7.5 & 0.53 & PG 1159      & no  & $335-909\ ^{\rm p}$    & $23.0\pm 0.3 \ ^{\rm p}$   \\
PG 0122+200     &  80 & 7.5 & 0.53 & PG 1159      & no  & $336-612\ ^{\rm q}$    & $22.9\pm 0.0 \ ^{\rm q}$   \\
\noalign{\smallskip}
\hline
\hline
\end{tabular}
\end{center}
{\footnotesize References: 
$^{\rm a}$Werner \& Herwig (2006),
$^{\rm b}$Miller Bertolami \& Althaus (2006),
$^{\rm c}$Bond \& Meakes (1990),
$^{\rm d}$Vauclair et al. (2005), 
$^{\rm e}$Solheim et al. (2007),
$^{\rm f}$Ciardullo \& Bond (1996), 
$^{\rm g}$Bond \& Ciardullo (1993),
$^{\rm h}$Bond et al. (1996),
$^{\rm i}$Nagel \& Werner (2004), 
$^{\rm j}$Vauclair et al. (2002),
$^{\rm k}$Silvotti et al. (1999), 
$^{\rm l}$Grauer et al. (1987),
$^{\rm m}$Gonz\'alez P\'erez et al. (2006),
$^{\rm n}$Costa et al. (2008),
$^{\rm o}$Kawaler et al. (1995),
$^{\rm p}$Kawaler et al. (2004),
$^{\rm q}$Fu et al. (2007),
$^{\rm r}$C\'orsico et al. (2009c),
$^{\rm s}$Reed et al. (2000).} 
\end{table*}  

\subsubsection{Evolutionary models  for PG 1159 stars}

A large variety  of H-deficient models have been  employed in the past
for pulsation studies of PG  1159 stars. We can roughly classify these
pulsation models  into three  main categories: static  envelope models
(Starrfield et al. 1983, 1984,  1985; Bradley \& Dziembowski 1996; Cox
2003;  Quirion et  al.   2004, 2007),  simplified evolutionary  models
(Kawaler et al. 1986; Kawaler 1988; Stanghellini et al.  1991; Kawaler
\&  Bradley 1994;  Saio 1996;  Gautschy 1997),  and  full evolutionary
models  (Gautschy et  al.  2005;  C\'orsico et  al.  2006).   Here, we
restrict ourselves to  briefly describe the set of  models employed in
the study of C\'orsico et  al.  (2006) --- see Sect.  \ref{hdeficient}
for  details.  These  models have  been  generated by  Althaus et  al.
(2005a) and Miller Bertolami \&  Althaus (2006) by taking into account
the complete evolutionary history of the progenitor stars. This is the
unique set of full evolutionary  sequences that covers a wide range of
stellar   masses,   rendering    such   models   very   suitable   for
asteroseismology.   Althaus et  al.  (2005a)  and Miller  Bertolami \&
Althaus (2006) computed the complete evolution of model sequences with
a stellar mass in the ZAMS  within $1$ and $3.75\, M_{\odot}$. All the
sequences were  computed employing the {\tt  LPCODE} evolutionary code
--- see Althaus et al.  (2005a) for details --- and were followed from
the ZAMS  through the  thermally pulsing and  mass-loss phases  on the
AGB.   After  experiencing  several  thermal pulses,  the  progenitors
departed from the AGB and evolved towards high effective temperatures.
Mass loss during  the departure from the AGB  was arbitrarily fixed to
obtain a  final He  shell flash during  the early white  dwarf cooling
phase.   After  the  born-again  episode, the  H-deficient,  quiescent
He-burning remnants evolved at constant luminosity to the domain of PG
1159 stars with a surface  chemical composition rich in He, carbon and
oxygen.   The masses  of the  remnants span  the  range $0.530-0.741\,
M_{\odot}$.  A very  important result of the work  of Miller Bertolami
\& Althaus (2006) is a  new determination of the spectroscopic mass of
PG 1159  stars.  The mean mass  obtained by these  authors is $0.573\,
M_{\odot}$,  $0.044  M_{\odot}$  lower  than the  previously  accepted
value.

\subsubsection{Excitation of pulsations}

A longstanding problem associated with  GW Vir stars is related to the
excitation mechanism, although at  present, this problem appears to be
solved.  The early work by Starrfield et al.  (1983) was successful in
finding the correct destabilizing agent, namely the $\kappa$-mechanism
associated  with the partial  ionization of  the K-shell  electrons of
carbon and/or oxygen in the  envelope of models. However, their models
required a  driving region very poor in  He in order to  be capable to
excite  pulsations;  even very  low  amounts  of  He could  weaken  or
completely extinguish  the destabilizing  effect of carbon  and oxygen
(the so  called ``He poisoning'' effect).  The  latter requirement led
to  the conjecture  that a  composition gradient  would exist  to make
compatible the He-devoid driving  regions and the He-rich photospheric
composition.  Even  modern detailed  calculations still point  out the
necessity  of a  compositional  gradient in  the  envelopes of  models
(Bradley \& Dziembowski  1996; Cox 2003).  The presence  of a chemical
composition gradient is difficult to  explain in view of the fact that
PG 1159  stars are still experiencing  mass loss ---  for instance for
\pp\  the  mass-loss  rate  is $\dot{M}  \sim  10^{-8.1}\,  M_{\odot}$
yr$^{-1}$ (Koesterke et al.  1998) --- a fact that prevents the action
of gravitational settling of carbon  and oxygen, and instead, tends to
homogenize the envelope of hot white dwarfs (Unglaub \& Bues 2000).

Clearly at odds  with the hypothesis of a  composition gradient in the
PG 1159  envelopes, calculations by Saio (1996),  Gautschy (1997), and
Quirion et al.  (2004, 2007) --- based on modern opacity OPAL data ---
demonstrated  that  $g$-mode  pulsations  in  the  correct  ranges  of
effective temperatures and periods could  be easily excited in PG 1159
models having an uniform  envelope composition. In particular, Quirion
et al.  (2004)  found that the presence of  non-variable PG 1159 stars
among GW  Vir stars could be  readily explained by the  presence of an
excessively large abundance of He. Hence, in contrast with the case of
DAVs  and   DBVs,  the  GW   Vir  instability  strip  appears   to  be
intrinsically impure.

As  important  as  they are,  the  vast  majority  of the  studies  of
pulsation  driving in  PG 1159  stars performed  in the  past  rely on
simplified stellar models.  Indeed, the earliest works employed static
envelope  models  and  old  opacity  data.  Even  more  modern  works,
although based on  updated opacity data (OPAL), still  use a series of
static  envelope models  that  do not  represent  a real  evolutionary
sequence,   or   evolutionary   calculations   based   on   simplified
descriptions  of  the  evolution   of  their  progenitors.   The  only
exception is the very important  work of Gautschy et al. (2005), which
employs equilibrium  PG 1159 models  that evolved through the  AGB and
born-again stages, beginning  from a $2.7 \, M_{\odot}$  zero age main
sequence  model star.   Gautschy et  al.  (2005)  analyzed  four model
sequences, with  $0.530, 0.55, 0.589$  and $0.64 \,  M_{\odot}$, being
the  $0.589   \,  M_{\odot}$   sequence  derived  directly   from  the
evolutionary computations  of Althaus  et al.  (2005a).   However, the
remainder sequences  were created from the $0.589\,  M_{\odot}$ one by
artificially changing  the stellar mass  shortly after the end  of the
born-again episode.

\begin{figure}
\centering
\includegraphics[width=0.9\columnwidth,clip]{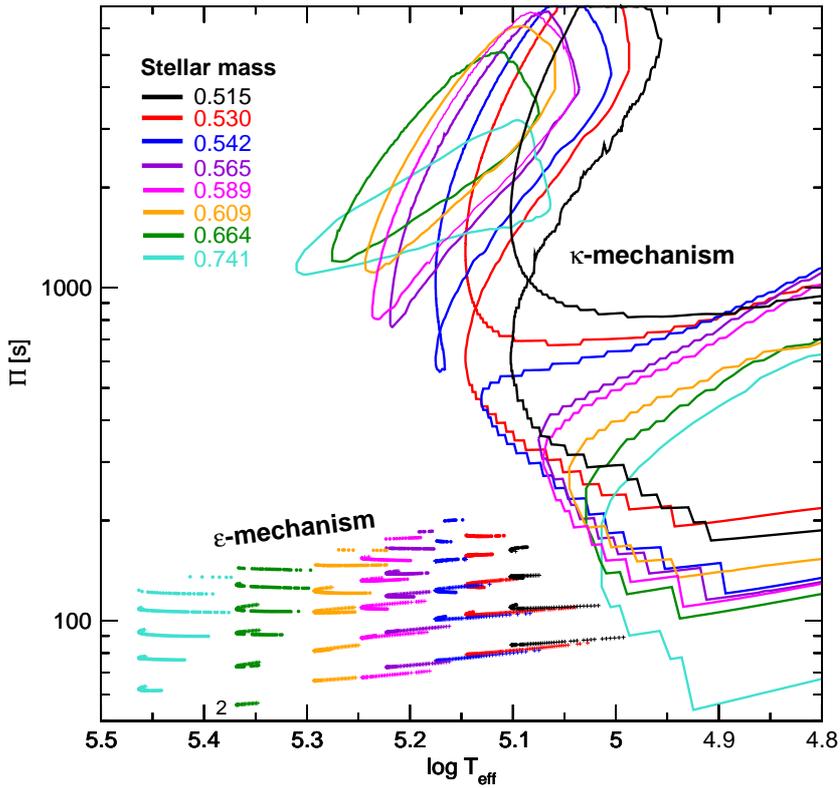}
\caption{The  dipole ($\ell=  1$) instability  domains  for overstable
         $\kappa$-destabilized   $g$-modes,  bounded  with   lines  of
         different colours  for the various  stellar masses, according
         to  C\'orsico et al.   (2006).  Short-period  dipole unstable
         $\epsilon$-destabilized  $g$-modes (see  Sect. \ref{epsilon})
         are  depicted  with  dot  (plus) symbols  for  stages  before
         (after) the evolutionary knee.}
\label{gwvir}
\end{figure}

The  work of  C\'orsico et  al.  (2006)  constitutes one  of  the most
complete and detailed studies on  pulsation stability of PG 1159 stars
performed so far.  Specifically,  C\'orsico et al. (2006) have carried
out  detailed non-adiabatic  pulsation  computations on  full PG  1159
evolutionary models  with stellar masses in the  range $0.530-0.741 \,
M_{\odot}$  resulting from the  complete evolutionary  calculations of
Althaus  et  al.  (2005a)  and  Miller  Bertolami  \& Althaus  (2006).
Numerous detailed  investigations about  pulsating PG 1159  stars have
been performed on the basis of artificial stellar models.  In spite of
the fact  that significant pulsation  damping and driving occur  in PG
1159  envelope  stars,  the  employment  of  such  simplified  stellar
configurations appear not  well justified in the case  of these stars.
This  is  in   contrast  to  the  situation  of   their  more  evolved
counterparts,  white dwarf  stars, for  which  their thermo-mechanical
structure has  relaxed to the correct  one by the time  the domains of
pulsational instability are reached. One of the main goals of the work
of C\'orsico  et al.   (2006) has  been to assess  to what  degree the
conclusions arrived  at in  previous studies on  PG 1159  stars change
when  realistic  stellar configurations  are  adopted.   The study  of
C\'orsico  et al.   (2006) confirms  many results  already  known from
previous studies, namely: (i) $g$-modes  in PG 1159 models are excited
by  the $\kappa$-mechanism  due to  partial ionization  of  carbon and
oxygen, and no abundance gradients  between the driving region and the
stellar  surface are  necessary to  drive $g$-mode  pulsations  at the
correct effective temperatures and  period ranges; (ii) there exists a
well-defined  instability domain with  a blue  edge which  is strongly
dependent   on  the   stellar   mass  (see   Figs.   \ref{gwvir}   and
\ref{epsil-strip});  (iii)  different surface  He  abundances lead  to
sizeable differences  in the precise location of  the theoretical blue
edge  of the instability  domain; (iv)  the instability  domain splits
into  two  separated  regions,   one  of  them  at  high  luminosities
characterized  by long  periods, and  the other  at  low luminosities,
corresponding to shorter periods (Fig. \ref{gwvir}); (v) all pulsating
PG 1159 stars  lay into the predicted instability  domain in the $\log
(T_{\rm eff})-\log g$ plane  (see Fig.  \ref{epsil-strip}); (vi) there
is a very good agreement  between the full period spectrum observed in
GW Vir stars and the theoretical ranges of unstable periods; and (vii)
the  pulsation  periods  of  excited modes  decrease  with  decreasing
luminosity   (increasing   surface   gravity),   in  line   with   the
observational  trend  (Fig. \ref{gwvir}).   As  for  the new  results,
C\'orsico et  al. (2006)  found that  there exists a  red edge  of the
instability domain at  the high-luminosity (low-gravity) regime.  This
red edge is mass-dependent.  The  border of the instability domains in
the  $\log  T_{\rm  eff}-\log   \Pi$  plane  at  the  high-luminosity,
long-period regime  is well delineated  (Fig.  \ref{gwvir}).  Finally,
C\'orsico et  al. (2006) found  that some non-variables  occupying the
instability strip have standard He abundances and the presence of them
between pulsators can not be explained through the argument of Quirion
et al. (2004).

The more recent  stability study on GW Vir star is  that of Quirion et
al. (2007).  These authors have carried out a impressive non-adiabatic
survey of pulsating PG 1159  stars and confirmed their previous result
that the extent of the  instability domain in the $\log g-T_{\rm eff}$
diagram is a  strong function of the C and O  content in the envelopes
of these stars. Because the chemical composition of the envelope of PG
1159  stars varies from  star to  star, the  blue edge  of the  GW Vir
instability  strip must  necessarily be  the superposition  of several
blue  edges,  i.e.,  a  fuzzy  edge.   In  the  study  of  Quirion  et
al. (2007), the effects of varying the total mass, of adding H, and of
changing the metallicity are  investigated, and the expected ranges of
excited periods under various conditions are provided.

\subsubsection{Asteroseismological inferences on GW Vir stars}
\label{astro-infer}

\begin{table*}[t]
\scriptsize
\begin{center}
\caption{Stellar masses  for all of the  intensively studied pulsating
         PG 1159  stars, including also one pulsating  [WCE] star. All
         masses are in solar units.}
\begin{tabular}{lcccclc}
\hline 
\hline
\noalign{\smallskip}
Star & $\Delta \Pi_{\ell}^{\rm a}$ &$\overline{\Delta \Pi_{\ell}}$ & Approximate  & Period fit & Pulsations    & Spectroscopy\\
     &                             &                               &  formula     &            & (other works) & \\
\noalign{\smallskip}
\hline
\ngc       & 0.571$^{\rm a}$        & 0.576$^{\rm a}$ & 0.530$^{\rm a}$ & ---             & 0.55$^{\rm j}$ ($\Delta \Pi_{\ell}^{\rm a}$) & 0.56 \\ 
\rxj       & 0.568$^{\rm b}$        & 0.560$^{\rm b}$ & 0.525$^{\rm a}$ & 0.565$^{\rm b}$ & 0.56$^{\rm h}$ ($\Delta \Pi_{\ell}^{\rm a}$) & 0.72 \\ 
\pp        & 0.577--0.585$^{\rm d}$ & 0.561$^{\rm d}$ & 0.570$^{\rm a}$ & 0.565$^{\rm d}$ & 0.59$^{\rm i}$ ($\Delta \Pi_{\ell}^{\rm a}$) & 0.54 \\ 
\pr        & 0.627$^{\rm a}$        & 0.578$^{\rm a}$ & 0.609$^{\rm a}$ & 0.589$^{\rm a}$ & 0.61$^{\rm e}$ (period fit)                  & 0.55 \\ 
\pt        & 0.597$^{\rm a}$        & 0.566$^{\rm a}$ & 0.587$^{\rm a}$ & 0.542$^{\rm a}$ & 0.57$^{\rm g}$ ($\Delta \Pi_{\ell}^{\rm a}$) & 0.53 \\ 
\pg        & 0.625$^{\rm c}$        & 0.567$^{\rm c}$ & 0.593$^{\rm a}$ & 0.556$^{\rm c}$ & 0.69$^{\rm f}$ ($\Delta \Pi_{\ell}^{\rm a}$) & 0.53 \\ 
\hline
\hline
\end{tabular} 
\label{tabla-masas}
\end{center}
{\footnotesize  References: 
$^{\rm a}$C\'orsico et al. (2009b),
$^{\rm b}$C\'orsico et al. (2007a),  
$^{\rm c}$C\'orsico et al. (2007b),  
$^{\rm d}$C\'orsico et al. (2008), 
$^{\rm e}$Reed et al. (2000), 
$^{\rm f}$Fu et al. (2007),
$^{\rm g}$Kawaler et al. (2004),   
$^{\rm h}$Vauclair et al. (2002),
$^{\rm i}$Costa et al. (2008), 
$^{\rm j}$Bond et al. (1996).}
\end{table*}

In recent  years, considerable observational effort  has been invested
to  study pulsating  PG 1159  stars. Particularly  noteworthy  are the
works of Vauclair et al. (2002) on \rxj, Fu et al.  (2007) on \pg, and
Costa et al.   (2008) and Costa \& Kepler (2008)  on \pp.  These stars
have been  monitored through  long-term observations carried  out with
the WET.  On  the theoretical front, recent important  progress in the
numerical modeling  of PG  1159 stars (Althaus  et al.   2005a; Miller
Bertolami  \&  Althaus  2006)  has  paved the  way  for  unprecedented
asteroseismological inferences  for the mentioned  stars (C\'orsico et
al.  2007a,  2007b, 2008, 2009b).   In these studies,  three different
methods have been  employed to infer the stellar mass  of GW Vir stars
(see  Sect.  \ref{mass} and  \ref{period-to-period}).  With  the first
method, the stellar mass is  obtained by comparing the observed period
spacing with the asymptotic period  spacing of the models.  The second
method consists  in a comparison  of the observed period  spacing with
the average of  the computed period spacings. Finally,  with the third
method, models that best  reproduce the individual observed periods of
each star are sought,  and, if possible, an unique asteroseismological
model is found.  The period-to-period fits (third method) performed in
these studies  are of unprecedented quality, and  no artificial tuning
of the  value of  structural parameters such  as the thickness  of the
outer envelope, the  surface chemical abundances, or the  shape of the
core chemical  profile is  made. Instead, they  are kept fixed  at the
values  predicted by  the  evolutionary calculations.   Also, for  the
first time, in this series of studies the same evolutionary tracks are
used   for  both   the  asteroseismological   and   the  spectroscopic
derivations  of  the stellar  mass.   This  remarkable refinement  (as
compared  with previous  works) leads  to a  better  agreement between
asteroseismological and spectroscopic masses of GW Vir stars.

\begin{figure}
\centering
\includegraphics[width=0.9\columnwidth,clip]{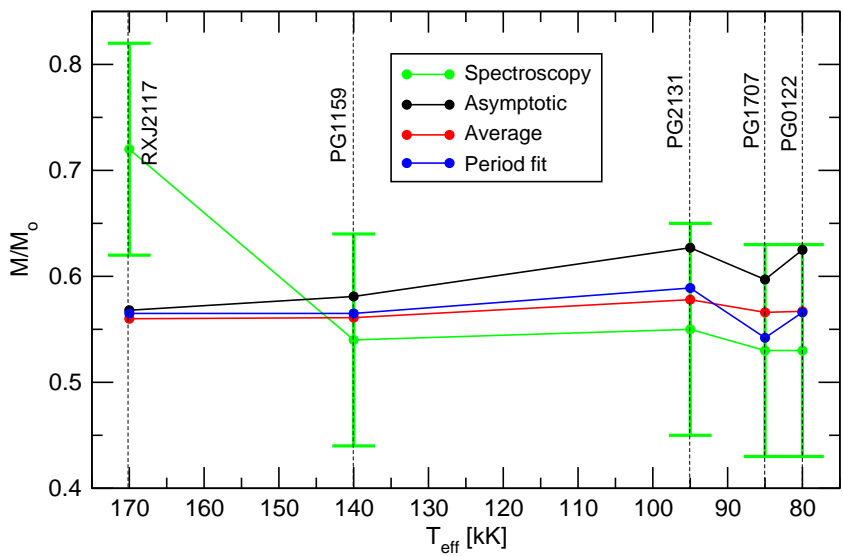}
\caption{A  comparison  between  the  stellar  mass of  GW  Vir  stars
         inferred from different approaches.}
\label{masas}
\end{figure}

Table  \ref{tabla-masas}  summarizes  the  asteroseismological  masses
obtained  for each  star,  along  with the  associated  values of  the
spectroscopic  mass. A  graphical representation  of these  results is
presented  in Fig.   \ref{masas}.  The  internal (formal)  errors (not
plotted) of the  asteroseismological mass determinations are $\lesssim
0.01 \, M_{\odot}$.   Note that there is a  good agreement between the
asteroseismological and  spectroscopic masses, except for  the case of
\rxj,  where  the  spectroscopic  mass  is  too  large  ($\sim  0.72\,
M_{\odot}$) to  be compatible  with all the  three asteroseismological
mass determinations ($\sim  0.56-0.57\, M_{\odot}$).  This discrepancy
could be due to large errors in the spectroscopic $T_{\rm eff}$ and/or
$g$  values.   We  also   note  that  the  asymptotic  period  spacing
overestimates  the stellar  mass  of  the studied  PG  1159 stars,  as
expected on the grounds that these  stars (except \rxj) are not in the
asymptotic  limit  of $g$-mode  pulsations,  and  then the  asymptotic
theory is not completely valid in such cases.

The  main conclusion  of the  series of  studies by  C\'orsico  et al.
(2007b, 2008, 2009b) is that  for most well-observed pulsating PG 1159
stars (\pg, \pp, \pr, and \pt)  it is possible to find a stellar model
(the  asteroseismological  model) with  $M_*$  near the  spectroscopic
determinations  to a  high internal  accuracy.   The next  step is  an
assessment  of  the question  if  the  asteroseismological models  can
provide more accurate masses for  these objects.  At first glance, the
scatter in  the masses derived from  the different asteroseismological
methods (see  Table \ref{tabla-masas}) suggests  that it could  not be
the  case.    In  fact,  when  all   asteroseismological  methods  are
considered, the uncertainty in  the seismological determination of the
mass  amounts   to  $\sim  0.05  \,  M_{\odot}$,   comparable  to  the
spectroscopic one --- which is $\sim 0.05-0.1 \, M_{\odot}$ (Werner et
al. 2008).   However, it is worth  noting that, when  results based on
the asymptotic  period spacing  --- an approach  that is  not entirely
correct for the  high-gravity regime of PG 1159  stars, see Althaus et
al.  (2008b)  --- are  not taken into  account, the scattering  in the
derived masses is of only $\sim 0.02\, M_{\odot}$. It can be concluded
that  asteroseismology of  PG 1159  stars  is a  precise and  powerful
technique which  can determine  the masses of  GW Vir stars  even more
accurately than spectroscopy.

\subsubsection{The rate of period change of \pp}

Recently,  Costa \&  Kepler (2008)  have measured  the rate  of period
change  for  a   large  number  of  modes  present   in  the  star  PG
1159$-$035. This  study demonstrates that some  $\dot{\Pi}$ values are
positive (periods  increase with  time), and other  $\dot{\Pi}$ values
are negative (periods  decrease with time).  This is  at odds with the
asteroseismological model for \pp\ derived by C\'orsico et al. (2008).
Also, the rates  of period changes measured by  Costa \& Kepler (2008)
are 10  times larger than  those predicted by  the asteroseismological
model.  We stress that the PG 1159 evolutionary models employed in the
series  of asteroseismological  studies of  C\'orsico et  al.  (2007a,
2007b,  2008,   2009b)  are  characterized  by   thick  He-rich  outer
envelopes,  as they  are  predicted  by the  standard  theory for  the
formation of  PG 1159  stars.  However, Althaus  et al.   (2008b) have
demonstrated  that  thinner He-rich  envelopes  ($2-3$ times  thinner)
solve  the severe  discrepancy between  the measured  rates  of period
change in \pp\ and the  theoretical values.  In view of this important
result,  and to  place it  on a  solid basis,  it would  be  useful to
recompute the asteroseismological models for all the pulsating PG 1159
stars  (and, in  particular, for  \pp)  analyzed in  C\'orsico et  al.
(2007a, 2007b, 2008, 2009b) with non-canonical models characterized by
thinner He-rich envelopes.

\subsubsection{Short period $g$-modes driven by the $\epsilon$-mechanism?}
\label{epsilon}

\begin{figure}
\centering
\includegraphics[width=0.9\columnwidth,clip]{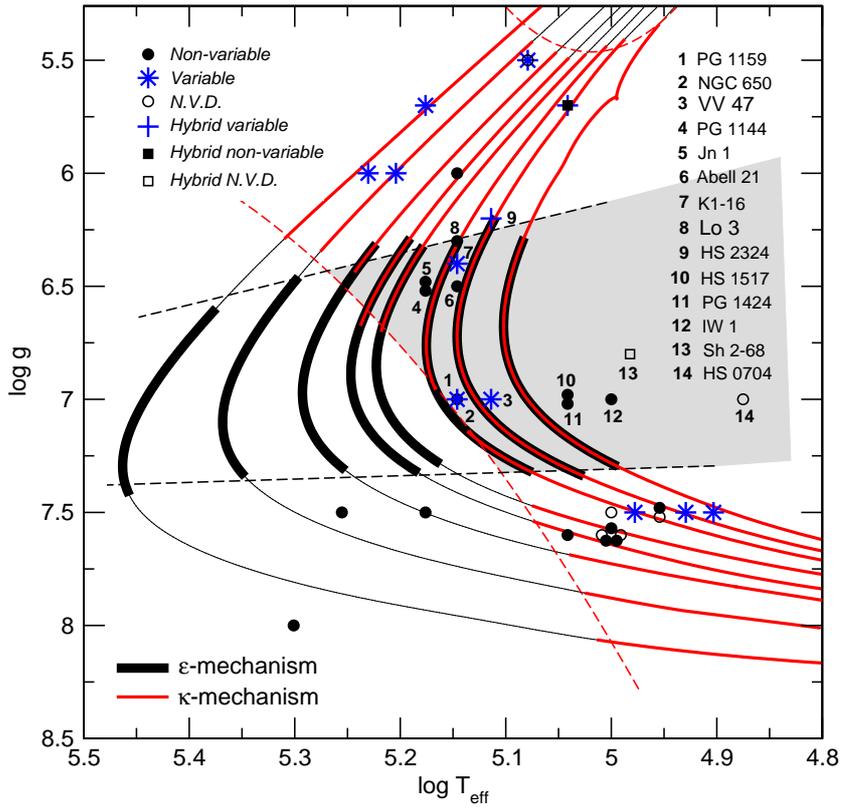}
\caption{The PG 1159 evolutionary tracks (thin solid lines) of Althaus
         et  al.   (2005a), Miller  Bertolami  \&  Althaus (2006)  and
         C\'orsico et al.  (2006),  with stellar masses of (from right
         to  left): $M_*=  0.515, 0.530,  0.542, 0.565,  0.589, 0.609,
         0.664, 0.742  \, M_{\odot}$.   The location of  models having
         $\ell= 1$  (dipole) $\kappa$-destabilized modes  according to
         C\'orsico et  al. (2006) is  displayed with solid  red curves
         along   the  tracks,   and   models  harboring   short-period
         $\epsilon$-destabilized modes  according to C\'orsico  et al.
         (2009c) are shown with thick solid curves.  The superposition
         of both  types of curves  corresponds to stellar  models with
         both  $\epsilon$-  and  $\kappa$-destabilized  modes  (shaded
         area).   The location  and  designation of  relevant PG  1159
         stars is also shown.}
\label{epsil-strip}
\end{figure}

A  recent  stability study  on  PG 1159  star  models  carried out  by
C\'orsico  et  al.  (2009c)  has  revealed the  presence  of  unstable
short-period  $g$-modes destabilized by  the He-burning  shell through
the $\epsilon$-mechanism.   This work, which  covers a broad  range of
stellar masses  and effective  temperatures, confirms and  extends the
pioneering studies of Kawaler et  al. (1986), Saio (1996) and Gautschy
(1997) on  this  topic.   The  main  results  of  this  study  can  be
summarized   as  follows:  there   exists  a   separate,  well-defined
theoretical PG 1159 instability strip  in the $\log T_{\rm eff} - \log
g$  diagram characterized  by  short-period $g$-modes  excited by  the
$\epsilon$-mechanism  due   to  the  presence   of  active  He-burning
shells. Notably, this instability strip partially overlaps the already
known GW Vir instability strip due to the $\kappa$-mechanism acting on
the partial ionization of carbon  and/or oxygen in the envelope of the
PG  1159  stars, as  can  be seen  in  Fig.   \ref {epsil-strip}.   At
variance  with the classical  $\kappa$-mechanism, responsible  for the
intermediate/long-period  GW Vir pulsations,  the $\epsilon$-mechanism
should be efficient  even in PG 1159 stars with  low carbon and oxygen
content in  their envelopes. The  $\epsilon$-driven $g$-modes probably
have time enough to reach observable amplitudes before the star leaves
the  instability  strip. \vv\  is  the only  PG  1159  star for  which
observational  evidence  of  the  presence of  short-period  $g$-modes
exists (Gonz\'alez  P\'erez et al.  2006).  The  stability analysis of
C\'orsico et al. (2009c) provides very strong support to the idea that
the physical  origin of the short  periodicities of \vv\  could be the
$\epsilon$-mechanism powered  by the active  He-burning shell, whereas
the long-period branch  of the period spectrum of  this star should be
due  to  the  $\kappa$-mechanism  acting  on  the  region  of  partial
ionization of  carbon and oxygen.   If true, this could  be indicating
that \vv\ should have a  thick He-rich envelope able to support stable
He-shell burning.  However, the real existence of the short periods of
\vv\ must be confirmed by  follow-up observations.  If this were done,
this star could  be the first known pulsating  PG 1159 star undergoing
non-radial $g$-modes  destabilized by the  $\epsilon$-mechanism.  Even
more, \vv\ could  be the first known pulsating star  in which both the
$\kappa$-mechanism  and the $\epsilon$-mechanism  of mode  driving are
operating simultaneously.


\section{Conclusion}

There  are several  reasons  that make  white  dwarfs reliable  cosmic
clocks and laboratories to  study different kind of problems.  Indeed,
they constitute the final evolutionary  stage for the vast majority of
stars.   They are  a homogeneous  class of  stars with  a  narrow mass
distribution and well known structure with slow rotation rates and low
magnetic fields.   And most importantly,  their evolution, as  we have
shown, is relatively simple and can be described in terms of a cooling
problem.   Additionally,  their  pulsational  properties can  be  well
described,  and the  excellent  observational data  gathered over  the
years provide strong constraints to the theory, allowing to refine the
theoretical  models.  However,  there  are still  some open  questions
regarding  the evolutionary  history of  white dwarf  progenitors, the
physics of  white dwarf cooling and their  pulsational properties that
still  need to be  addressed.  Once  this is  done, white  dwarfs will
constitute  the  best  studied  class  of stars  and  will  allow  the
astronomical community to use them  as tracers of the evolution of the
Galaxy.

\begin{acknowledgements}
Part of this  work was supported by the  AGENCIA grant BID 1728/OC-AR,
by  MCINN grants AYA2008--04211--C02--01  and AYA08--1839/ESP,  and by
AGAUR  grants  SGR1002/2009 and  SGR15\-/2009  of  the Generalitat  de
Catalunya.   LGA also  acknowledges the  AGAUR of  the  Generalitat de
Catalunya for a PIV grant.
\end{acknowledgements}

\end{document}